\RequirePackage{lineno}
\documentclass[aps,prd,twocolumn,superscriptaddress,floatfix]{revtex4}
\usepackage{epsfig}
\usepackage{graphicx}
\usepackage{dcolumn}
\usepackage{amsmath}
\usepackage{xspace}

\begin{document}
\title{Measurement of the Cross Section for Prompt Isolated Diphoton
       Production in $p\bar p$ Collisions at $\sqrt{s}=1.96$ TeV}

\affiliation{Institute of Physics, Academia Sinica, Taipei, Taiwan 11529, Republic of China} 
\affiliation{Argonne National Laboratory, Argonne, Illinois 60439, USA} 
\affiliation{University of Athens, 157 71 Athens, Greece} 
\affiliation{Institut de Fisica d'Altes Energies, ICREA, Universitat Autonoma de Barcelona, E-08193, Bellaterra (Barcelona), Spain} 
\affiliation{Baylor University, Waco, Texas 76798, USA} 
\affiliation{Istituto Nazionale di Fisica Nucleare Bologna, $^z$University of Bologna, I-40127 Bologna, Italy} 
\affiliation{University of California, Davis, Davis, California 95616, USA} 
\affiliation{University of California, Los Angeles, Los Angeles, California 90024, USA} 
\affiliation{Instituto de Fisica de Cantabria, CSIC-University of Cantabria, 39005 Santander, Spain} 
\affiliation{Carnegie Mellon University, Pittsburgh, Pennsylvania 15213, USA} 
\affiliation{Enrico Fermi Institute, University of Chicago, Chicago, Illinois 60637, USA}
\affiliation{Comenius University, 842 48 Bratislava, Slovakia; Institute of Experimental Physics, 040 01 Kosice, Slovakia} 
\affiliation{Joint Institute for Nuclear Research, RU-141980 Dubna, Russia} 
\affiliation{Duke University, Durham, North Carolina 27708, USA} 
\affiliation{Fermi National Accelerator Laboratory, Batavia, Illinois 60510, USA} 
\affiliation{University of Florida, Gainesville, Florida 32611, USA} 
\affiliation{Laboratori Nazionali di Frascati, Istituto Nazionale di Fisica Nucleare, I-00044 Frascati, Italy} 
\affiliation{University of Geneva, CH-1211 Geneva 4, Switzerland} 
\affiliation{Glasgow University, Glasgow G12 8QQ, United Kingdom} 
\affiliation{Harvard University, Cambridge, Massachusetts 02138, USA} 
\affiliation{Division of High Energy Physics, Department of Physics, University of Helsinki and Helsinki Institute of Physics, FIN-00014, Helsinki, Finland} 
\affiliation{University of Illinois, Urbana, Illinois 61801, USA} 
\affiliation{The Johns Hopkins University, Baltimore, Maryland 21218, USA} 
\affiliation{Institut f\"{u}r Experimentelle Kernphysik, Karlsruhe Institute of Technology, D-76131 Karlsruhe, Germany} 
\affiliation{Center for High Energy Physics: Kyungpook National University, Daegu 702-701, Korea; Seoul National University, Seoul 151-742, Korea; Sungkyunkwan University, Suwon 440-746, Korea; Korea Institute of Science and Technology Information, Daejeon 305-806, Korea; Chonnam National University, Gwangju 500-757, Korea; Chonbuk National University, Jeonju 561-756, Korea} 
\affiliation{Ernest Orlando Lawrence Berkeley National Laboratory, Berkeley, California 94720, USA} 
\affiliation{University of Liverpool, Liverpool L69 7ZE, United Kingdom} 
\affiliation{University College London, London WC1E 6BT, United Kingdom} 
\affiliation{Centro de Investigaciones Energeticas Medioambientales y Tecnologicas, E-28040 Madrid, Spain} 
\affiliation{Massachusetts Institute of Technology, Cambridge, Massachusetts 02139, USA} 
\affiliation{Institute of Particle Physics: McGill University, Montr\'{e}al, Qu\'{e}bec, Canada H3A~2T8; Simon Fraser University, Burnaby, British Columbia, Canada V5A~1S6; University of Toronto, Toronto, Ontario, Canada M5S~1A7; and TRIUMF, Vancouver, British Columbia, Canada V6T~2A3} 
\affiliation{University of Michigan, Ann Arbor, Michigan 48109, USA} 
\affiliation{Michigan State University, East Lansing, Michigan 48824, USA}
\affiliation{Institution for Theoretical and Experimental Physics, ITEP, Moscow 117259, Russia}
\affiliation{University of New Mexico, Albuquerque, New Mexico 87131, USA} 
\affiliation{Northwestern University, Evanston, Illinois 60208, USA} 
\affiliation{The Ohio State University, Columbus, Ohio 43210, USA} 
\affiliation{Okayama University, Okayama 700-8530, Japan} 
\affiliation{Osaka City University, Osaka 588, Japan} 
\affiliation{University of Oxford, Oxford OX1 3RH, United Kingdom} 
\affiliation{Istituto Nazionale di Fisica Nucleare, Sezione di Padova-Trento, $^{aa}$University of Padova, I-35131 Padova, Italy} 
\affiliation{LPNHE, Universite Pierre et Marie Curie/IN2P3-CNRS, UMR7585, Paris, F-75252 France} 
\affiliation{University of Pennsylvania, Philadelphia, Pennsylvania 19104, USA}
\affiliation{Istituto Nazionale di Fisica Nucleare Pisa, $^{bb}$University of Pisa, $^{cc}$University of Siena and $^{dd}$Scuola Normale Superiore, I-56127 Pisa, Italy} 
\affiliation{University of Pittsburgh, Pittsburgh, Pennsylvania 15260, USA} 
\affiliation{Purdue University, West Lafayette, Indiana 47907, USA} 
\affiliation{University of Rochester, Rochester, New York 14627, USA} 
\affiliation{The Rockefeller University, New York, New York 10065, USA} 
\affiliation{Istituto Nazionale di Fisica Nucleare, Sezione di Roma 1, $^{ee}$Sapienza Universit\`{a} di Roma, I-00185 Roma, Italy} 

\affiliation{Rutgers University, Piscataway, New Jersey 08855, USA} 
\affiliation{Texas A\&M University, College Station, Texas 77843, USA} 
\affiliation{Istituto Nazionale di Fisica Nucleare Trieste/Udine, I-34100 Trieste, $^{ff}$University of Trieste/Udine, I-33100 Udine, Italy} 
\affiliation{University of Tsukuba, Tsukuba, Ibaraki 305, Japan} 
\affiliation{Tufts University, Medford, Massachusetts 02155, USA} 
\affiliation{University of Virginia, Charlottesville, VA  22906, USA}
\affiliation{Waseda University, Tokyo 169, Japan} 
\affiliation{Wayne State University, Detroit, Michigan 48201, USA} 
\affiliation{University of Wisconsin, Madison, Wisconsin 53706, USA} 
\affiliation{Yale University, New Haven, Connecticut 06520, USA} 
\author{T.~Aaltonen}
\affiliation{Division of High Energy Physics, Department of Physics, University of Helsinki and Helsinki Institute of Physics, FIN-00014, Helsinki, Finland}
\author{B.~\'{A}lvarez~Gonz\'{a}lez$^v$}
\affiliation{Instituto de Fisica de Cantabria, CSIC-University of Cantabria, 39005 Santander, Spain}
\author{S.~Amerio}
\affiliation{Istituto Nazionale di Fisica Nucleare, Sezione di Padova-Trento, $^{aa}$University of Padova, I-35131 Padova, Italy} 

\author{D.~Amidei}
\affiliation{University of Michigan, Ann Arbor, Michigan 48109, USA}
\author{A.~Anastassov}
\affiliation{Northwestern University, Evanston, Illinois 60208, USA}
\author{A.~Annovi}
\affiliation{Laboratori Nazionali di Frascati, Istituto Nazionale di Fisica Nucleare, I-00044 Frascati, Italy}
\author{J.~Antos}
\affiliation{Comenius University, 842 48 Bratislava, Slovakia; Institute of Experimental Physics, 040 01 Kosice, Slovakia}
\author{G.~Apollinari}
\affiliation{Fermi National Accelerator Laboratory, Batavia, Illinois 60510, USA}
\author{J.A.~Appel}
\affiliation{Fermi National Accelerator Laboratory, Batavia, Illinois 60510, USA}
\author{A.~Apresyan}
\affiliation{Purdue University, West Lafayette, Indiana 47907, USA}
% ...
\collaboration{CDF Collaboration\footnote{With visitors from $^a$University of Massachusetts Amherst, Amherst, Massachusetts 01003,
$^b$Istituto Nazionale di Fisica Nucleare, Sezione di Cagliari, 09042 Monserrato (Cagliari), Italy,
$^c$University of California Irvine, Irvine, CA  92697, 
$^d$University of California Santa Barbara, Santa Barbara, CA 93106
$^e$University of California Santa Cruz, Santa Cruz, CA  95064,
$^f$CERN,CH-1211 Geneva, Switzerland,
$^g$Cornell University, Ithaca, NY  14853, 
$^h$University of Cyprus, Nicosia CY-1678, Cyprus, 
$^i$University College Dublin, Dublin 4, Ireland,
$^j$University of Fukui, Fukui City, Fukui Prefecture, Japan 910-0017,
$^k$Universidad Iberoamericana, Mexico D.F., Mexico,
$^l$Iowa State University, Ames, IA  50011,
$^m$University of Iowa, Iowa City, IA  52242,
$^n$Kinki University, Higashi-Osaka City, Japan 577-8502,
$^o$Kansas State University, Manhattan, KS 66506,
$^p$University of Manchester, Manchester M13 9PL, England,
$^q$Queen Mary, University of London, London, E1 4NS, England,
$^r$Muons, Inc., Batavia, IL 60510,
$^s$Nagasaki Institute of Applied Science, Nagasaki, Japan, 
$^t$National Research Nuclear University, Moscow, Russia,
$^u$University of Notre Dame, Notre Dame, IN 46556,
$^v$Universidad de Oviedo, E-33007 Oviedo, Spain, 
$^w$Texas Tech University, Lubbock, TX  79609, 
$^x$Universidad Tecnica Federico Santa Maria, 110v Valparaiso, Chile,
$^y$Yarmouk University, Irbid 211-63, Jordan,
$^{gg}$On leave from J.~Stefan Institute, Ljubljana, Slovenia. 
}}
\noaffiliation

\date{\today} 
\begin{abstract}
This article reports a measurement of the production cross section of prompt
isolated photon pairs in proton-antiproton collisions at $\sqrt{s}=1.96$ TeV
using the CDF II detector at the Fermilab Tevatron collider. The data
correspond to an integrated luminosity of 5.36 fb$^{-1}$. The cross section
is presented as a function of kinematic variables sensitive to the reaction
mechanisms. The results are compared with three perturbative QCD calculations:
(1) a leading order parton shower Monte Carlo, (2) a fixed next-to-leading
order calculation and (3) a next-to-leading order/next-to-next-to-leading-log
resummed calculation. The comparisons show that, within their known
limitations, all calculations predict the main features of the data, but no
calculation adequately describes all aspects of the data.
\end{abstract}

\maketitle

\section{Introduction}
\label{sec:introduction}

The measurement of the production cross section of two energetic isolated
central photons (diphotons) in high energy hadron collisions is important
for testing standard model (SM) predictions in the domain of searches for
undiscovered particles and new physics. Understanding the reaction mechanisms
in the complicated environment formed in such collisions is a challenge for
perturbative Quantum ChromoDynamics (pQCD) calculations. Photons orginating
from hard collisions of hadrons (``direct'' or ``prompt'' photons) are an
ideal probe for testing these calculations because they do not interact with
other final state particles, and their energies and directions can be measured
with high precision in modern electromagnetic calorimeters. Prompt diphoton
production creates an irreducible background to the diphoton decay channel
of proposed new particles, such as low mass Higgs bosons or Randall-Sundrum
gravitons in models of extra spatial dimensions \cite{higgs,grav}. An improved
knowledge of the SM background will help the development of more powerful
search strategies for these particles.

\begin{figure}[!ht]
\centering
\includegraphics[width=1.0\linewidth]{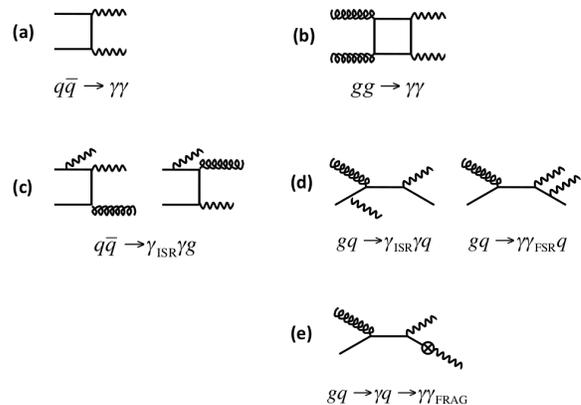}
\caption{Basic diagrams for prompt diphoton production: (a-b) direct,
     (c-d) one-photon radiation from an initial- (ISR) or final-state quark
     (FSR), (e) fragmentation where one photon is emitted along the direction
     of a final-state quark taking almost all of its energy. The symbol
     $\otimes$ denotes the non-perturbative mechanism of the fragmentation
     process (FRAG).}
\label{fig:diagrams}
\end{figure}

The basic mechanisms of prompt diphoton production in hadron collisions are
quark-antiquark annihilation $q\bar q\rightarrow\gamma\gamma$, quark-gluon
scattering $gq\rightarrow\gamma\gamma q$, and gluon-gluon fusion
$gg\rightarrow\gamma\gamma$. The respective basic diagrams are shown
in Fig. \ref{fig:diagrams}. At the Tevatron, the dominant mechanism is
quark-antiquark annihilation. In quark-gluon scattering, most of the time at
least one of the two photons is emitted almost parallel to the scattered quark.
Contributions from this mechanism are therefore suppressed by requiring
isolated prompt photons. Each mechanism can be modeled by calculating the
respective matrix element for the specific event kinematics. Matrix element
calculations of Leading Order (LO) in the strong coupling are relatively
simple and are thus implemented in advanced parton shower Monte Carlo (MC)
event generators \cite{pythia,herwig,sherpa}, which allow for gluon and
photon radiation as well as multiple interactions in the colliding beams.
By including radiation before and after the hard scattering, parton shower
generators take into account soft gluon and photon emissions, thus resulting
in an effective resummation of all of the Leading Logarithmic (LL) terms in
the cross section to all orders of the strong and electromagnetic couplings
constants. Next-to-Leading Order (NLO) calculations \cite{diphox,g2mc,resbos}
additionally include one-loop corrections at the cost of not featuring
realistic multi-particle event representations as the LO generators do.
Recent NLO calculations include an analytical resummation of the cross
section for initial-state gluon radiation to all orders in the strong
coupling constant \cite{resbos}, reaching a higher logarithmic accuracy
than in the parton shower Monte Carlo generators. By this method, all soft
gluon emissions in the initial state are taken into account, and reliable
predictions for the low diphoton transverse momentum region are possible.
A fixed-order NLO calculation implemented by the {\sc diphox} program
\cite{diphox} also accounts for the case where a final state quark loses
almost all of its energy to the photon detected in the event \cite{frag}.
This process is called ``fragmentation'' and, in contrast to final state
photon radiation in parton showering, it involves non-perturbative
calculations. One or both photons in the event may come from fragmentation.
The case where both photons come from fragmentation of a single quark is
also possible, but is not included in calculations, as in this case the
photons are nearly collinear and non-isolated most of the time.

The prompt diphoton cross section has been previously measured by the CDF
Collaboration using 200 pb$^{-1}$ of data \cite{cdf6312}, but the large
statistical uncertainties did not allow for a precise comparison with
theoretical calculations. The nearly 30--times larger CDF II data set
currently available presents an opportunity to significantly extend the
kinematic range and perform a detailed study of diphoton kinematic
distributions. A recent measurement of the diphoton cross section using
4.2 fb$^{-1}$ has been reported by the D0 Collaboration \cite{d0}. The
reported differential cross sections were only partly reproduced by
theoretical calculations \cite{pythia,diphox,resbos}, although the
discrepancies between the NLO calculations \cite{resbos,diphox} and the
data were less important in kinematic regions where the Higgs boson or new
heavy particles are expected.

This article is organized as follows. An overview of the detector is given
in Section \ref{sec:detector}. The event selection is presented in Section
\ref{sec:data}. Section \ref{sec:analysis} deals with extracting the cross
section from the selected diphoton sample. The results are 
presented and discussed in Section \ref{sec:results}. The conclusions are
given in Section \ref{sec:summary}. Appendix \ref{sec:likelihood} explains
details of the non-prompt photon subtraction technique introduced in Section
\ref{sec:analysis}. Finally, tables of the measured cross section, differential
in various kinematic quantities, are given in Appendix \ref{sec:xsec-tab}.

\section{Detector Overview}
\label{sec:detector}

The CDF II detector is a cylindrically--symmetric apparatus \cite{eta}
designed to study $p\bar{p}$ collisions at the Fermilab Tevatron. The
detector has been described in detail elsewhere \cite{CDF}; only the
detector components that are relevant to this analysis are briefly
discussed here. The magnetic spectrometer consists of tracking devices
inside a 3-m diameter, 5-m long superconducting solenoid magnet which
provides an axial magnetic field of 1.4 T. A set of silicon microstrip
detectors (L00, SVX, and ISL) \cite{L00,SVX,ISL} and a 3.1-m long
drift chamber (COT) \cite{COT} with 96 layers of sense wires measure
momenta and trajectories (tracks) of charged particles in the
pseudorapidity regions of $\left|\eta\right|$$<$2 and
$\left|\eta\right|$$<$1 \cite{eta}, respectively. Surrounding the
magnet coil is the projective-tower-geometry sampling calorimeter,
which is used to identify and measure the energy and direction of
photons, electrons, and jets. The calorimeter consists of lead-scintillator
electromagnetic and iron-scintillator hadron compartments and it is divided
into a central barrel ($\left|\eta\right|$$<$1.1) and a pair of ``end plugs''
that cover the region 1.1$<$$\left|\eta\right|$$<$3.6. The central
calorimeter is composed of towers with a segmentation of
$\Delta\eta\times\Delta\phi\simeq 0.1\times 15^o$. The energy
resolution of the central electromagnetic calorimeter for electrons is
$\sigma(E_{\rm T})/E_{\rm T}=13.5\%/ \sqrt{E_{\rm T}({\rm GeV})}\oplus 1.5\%$
\cite{cem}, while the energy resolution of the central hadron calorimeter
for charged pions that do not interact in the electromagnetic section is
$\sigma(E_{\rm T})/E_{\rm T}=50\%/\sqrt{E_{\rm T}({\rm GeV})}\oplus 3\%$
\cite{cwha}. Multiwire proportional chambers with cathode-strip readout
(the CES system), located at the depth of six radiation lengths (near
shower maximum) in the central electromagnetic calorimeter, are used for
identification and precise position measurement of photons and electrons.
Cathode strips and anode wires, with a channel spacing between 1.5~cm and
2~cm, running along the azimuthal (strips) and the beam line (wires)
direction provide location and two-dimensional profiles of electromagnetic
showers. The position resolution of the CES is 2~mm for a 50~GeV photon.
The electromagnetic compartments of the calorimeter are also used to
measure the arrival time of particles depositing energy in each tower
\cite{EMtiming}. A system of Cherenkov luminosity counters (CLC) \cite{CLC},
located around the beam pipe and inside the plug calorimeters, is used to
measure the number of inelastic $p\bar p$ collisions per bunch crossing,
and thereby the luminosity.

The online event selection at CDF is done by a three-level
trigger \cite{trigger} system with each level providing a rate
reduction sufficient to allow for processing at the next level with
minimal deadtime. Level-1 uses custom--designed hardware to find
physics objects based on a subset of the detector information.
Level-2 does limited event reconstruction. 
Level-3 uses the full detector information
and consists of a farm of computers that reconstruct the data and apply
selection criteria similar to the offline requirements.

\section{Data Selection and Event Reconstruction}
\label{sec:data}

Inclusive $\gamma\gamma$ events are selected online by a three-level
trigger that requires two isolated electromagnetic (EM) clusters with
$E_{\rm T}^{\gamma}$$>$12 GeV (diphoton-12 trigger) or two  electromagnetic
clusters with $E_{\rm T}^{\gamma}$$>$18 GeV and no isolation requirement
(diphoton-18 trigger). The transverse energy of the clusters is calculated
with respect to the nominal center of the detector at $z$=0 cm. The trigger
requirements at each level are briefly described below.

\begin{table*}[tbp]
\begin{center}
\caption{Summary of the diphoton trigger requirements.}
\label{tab:trig}
\begin{tabular}{lccc}
\hline\hline
Trigger Level & Diphoton-12 & Diphoton-18 \\  \hline
        & EM $E_{\rm T}>8$ GeV & same \\
Level-1 & $E^{HAD}/E^{EM}<0.125$ & same \\
        & $N_{cluster}=2$ & same \\ \hline
        & EM $E_{\rm T}>10$ GeV & EM $E_{\rm T}>16$ GeV \\
Level-2 & $E^{HAD}/E^{EM}<0.125$ & same \\
        & $E_{\rm T}^{ISO}<3$ GeV or $E_{\rm T}^{ISO}/E_{\rm T}<0.15$ & not applied \\
        & $N_{cluster}=2$ & same \\ \hline
        & EM $E_{\rm T}>12$ GeV & EM $E_{\rm T}>18$ GeV \\
Level-3 & $E^{HAD}/E^{EM}<0.055+0.00045$$\times$$E$/GeV if $E<200$ GeV & same \\
        & $E_{\rm T}^{ISO}<2$ GeV or $E_{\rm T}^{ISO}/E_{\rm T}<0.1$ & not applied \\
	& shower profile: $\chi^2_{CES}<20$  & same \\ 
        & $N_{cluster}=2$ & same \\
\hline\hline
\end{tabular}
\end{center}
\end{table*}

At Level-1, events having two towers with EM $E_{\rm T}$$>$8 GeV each are
required. For each trigger tower, the amount of energy in the hadronic
compartment of the calorimeter ($E^{HAD}$) has to be consistent with
that of an electromagnetic object. A trigger tower consists of two
adjacent towers in the same calorimeter wedge, so that the granularity
is approximately $\Delta\eta\times\Delta\phi\simeq 0.2\times 15^o$.

The Level-2 requirements are different for the two triggers. The diphoton-12
trigger selects events if there are two isolated seeds with EM
$E_{\rm T}$$>$10~GeV each. The isolation (ISO) energy is calculated as a sum
of the transverse energy in the towers nearby the seed tower. The ISO energy
for both photons has to be less than 3 GeV or 15$\%$ of the seed energy,
whatever is larger. The diphoton-18 trigger requires two towers with EM
$E_{\rm T}$$>$16 GeV each at Level-2.

Events are fully reconstructed at Level-3. At this level, for all photons
in both triggers, the energy profile at the shower maximum of each photon
candidate has to be consistent with that of a single photon. The diphoton-12
trigger selects events with two isolated photon candidates with
$E_{\rm T}$$>$12 GeV. The isolation energy at level-3 is calculated as the sum
of $E_{\rm T}$ in all towers (except for photon towers) within the cone of
$\Delta R=\sqrt{(\Delta\eta)^2+(\Delta\phi)^2}<0.4$ centered around the
photon candidate. This ISO energy has to be less than 2 GeV or 10$\%$ of
the photon energy, whatever is larger. The diphoton-18 trigger has no
isolation requirement and accepts events with two photon candidates with
$E_{\rm T}$$>$18 GeV. Table~\ref{tab:trig} gives a summary of all trigger
requirements for events with EM objects in the central calorimeter and
with $E_{\rm T}$ calculatated with respect to the event vertex.

The triggered $\gamma\gamma$ candidate events are then subject to the
offline selection. Each event is required to have two central photon
candidates inside a well-instrumented region of the calorimeter 
(approximately 0.05$<$$|\eta|$$<$1.05) with $E_{\rm T}$$>$17 GeV for one
candidate and $E_{\rm T}$$>$15 GeV for the other. This asymmetric cut helps
to avoid instabilities in fixed NLO calculations \cite{diphox}. Photon
candidates must satisfy strict (referred to as ``tight'') photon
identification requirements. The EM cluster has to be located inside the
well--instrumented region of the CES chamber, away from the $\phi$-boundary
of a calorimeter tower \cite{CESfid}. The energy deposition pattern in
both transverse profiles at CES has to be consistent with that of a single
electromagnetic object. The ratio of the energy measured in the hadron
(HAD) calorimeter to the EM energy, $E^\mathrm{HAD}/E^\mathrm{EM}$, has to
satisfy the requirement $E^{HAD}/E^{EM}$$<$0.055+0.00045$\times$$E^{\gamma}$.
To distinguish photons from electrons, no high-$p_{\rm T}$ charged--particle
track should point into the cluster ($N_{track}$$\leq$1 with track
$p_{\rm T}$$<$1.0+0.005$\times$$E_{\rm T}$). The main sources of ``fake''
photons are energetic $\pi^{0}$ and $\eta^{0}$ mesons produced in jets.
These mesons are usually produced in association with other particles. To
reduce this contamination from jets, the photon candidate must be isolated
in the calorimeter. To calculate the calorimeter isolation (cal-ISO), the
$E_{\rm T}$ deposited in the calorimeter towers within the cone of
$\Delta R<0.4$ around the EM cluster is summed, and the $E_{\rm T}$ of the
EM cluster is subtracted. Cal-ISO is then corrected for the photon's energy
leakage into towers in the neighboring wedge and for the contribution
from multiple interactions in the same bunch crossing \cite{diphoAnom}.
Cal-ISO must be consistent with the amount of energy expected from the
underlying event (see Table \ref{tab:phoID}). In addition to the calorimeter
isolation, there should be no other significant energy ($E_{\rm T}$ of 2$^{nd}$
CES cluster) deposited in the CES chamber containing the photon candidate.
Table \ref{tab:phoID} provides a summary of the photon identification
requirements described above. To reduce contamination due to cosmic-ray,
beam-related, and other non-collision backgrounds, the event must contain a
well-reconstructed vertex, formed from tracks, with $|z|$$<$60 cm. If multiple
vertices are reconstructed, the vertex with the largest$\sum p_{\rm T}$ of the
associated tracks is selected. The transverse energy of the photon candidates
is calculated with respect to this primary vertex.

\begin{table*}[tbp]
\begin{center}
\caption{Summary of the standard (``tight'') photon identification
         requirements for the $\gamma\gamma$ sample.}
\label{tab:phoID}
\begin{tabular}{lc}
\hline\hline
Cuts & ``Tight'' photon ID \\ \hline \hline
Calorimeter fiduciality & central \\ \hline
$E_{\rm T}^{\gamma}$ & $\geq 15$ GeV (1$^{\rm st}$$\gamma$),~\!
                 $\geq 17$ GeV (2$^{\rm nd}$$\gamma$) \\ \hline
Shower profile in CES: $\chi^{2}$ & $\leq$20 \\ \hline
$E^{HAD}/E^{EM}$ & $\leq$0.055+0.00045$\times$$E$/GeV \\ \hline
cal-ISO & $\leq$0.1$\times$$E_{\rm T}$ if $E_{\rm T}$$<$20 GeV or \\
        & $\leq$2.0 GeV+0.02$\times$$(E_{\rm T}-20$ GeV$)$ \\ \hline
$N_{tracks}$ in cluster & $\leq 1$ \\ \hline
track $p_{\rm T}$ if $N_{tracks}=1$ & $\leq$1.0 GeV+0.005$\times$$E_{\rm T}$ \\
\hline
$E_{\rm T}$ of $2^{nd}$ CES &
 $\leq$0.14$\times$$E_{\rm T}$ if $E_{\rm T}$$<$18 GeV \\
cluster & $\leq$2.4 GeV+0.01$\times$$E_{\rm T}$ if $E_{\rm T}$$\geq$18 GeV \\
\hline\hline
\end{tabular}
\end{center}
\end{table*}

Inclusive $\gamma\gamma$ events satisfying the above criteria form the
baseline $\gamma\gamma$ sample used in the analysis. Due to the presence
of fakes, this sample consists of real $\gamma\gamma$, {\it jet-}$\gamma$
and {\it jet-jet} events. (An object misidentified as a photon is referred
to as a ``fake'' photon.) Events with one or two fake photons are classified
as background. The baseline signal plus background $\gamma\gamma$ sample
consists of roughly 60,000 events in data corresponding to 5.36 fb$^{-1}$
of integrated luminosity. Signal and background samples were simulated
with the {\sc pythia} event generator which includes simulation of the
underlying event and multiple hadron interactions, as well as initial-
(ISR) and final-state radiation (FSR) and a hadronization model of the final
state partons \cite{pythia}. The {\sc pythia} events were processed through
a {\sc geant}-based detector simulation \cite{cdfsim} and trigger emulation,
followed by the same reconstruction program as that for the data.

\section{Cross Section Measurement}
\label{sec:analysis}

This section describes the steps of the cross section measurement. Kinematic
variables of interest are histogrammed to measure the corresponding
differential cross section. The background is subtracted from each histogram
bin. The signal histograms are normalized to the integrated luminosity and to
the size of each bin to obtain uncorrected differential cross section
histograms. These are then corrected for the reconstruction efficiency,
acceptance, and resolution effects.

\subsection{Background Subtraction}
\label{subsec:bkg-sub}

\begin{figure}[!ht]
\centering
\includegraphics[width=1.0\linewidth]{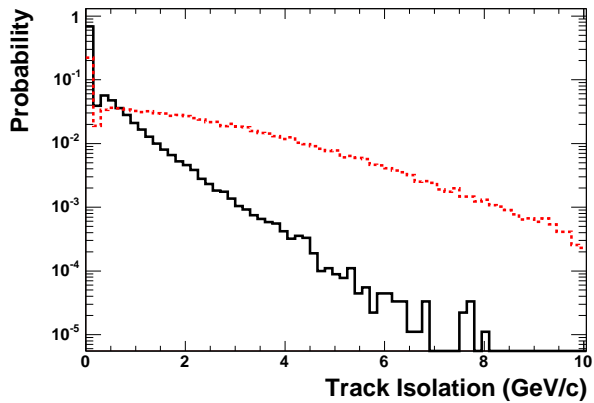}
\caption{The track-ISO distribution in signal (black--solid) and background
     (red--dashed) events.}
\label{fig:trk_ISO}
\end{figure}

The fake photon background subtraction is based on the use of the track
isolation (track-ISO), which is calculated as the $\sum{p_{\rm T}}$ of tracks
with $\Delta$$R$ to the photon $<$0.4 and 
$|z_{vertex}$-$z_{track}|$$<$5 cm. The concept of this technique is similar
to the one used in the earlier measurement of the inclusive photon cross
section \cite{cdf6312}. The main idea behind the method is that true and
fake photons have very different isolation distributions (see Fig.
\ref{fig:trk_ISO}). Therefore, one expects different efficiencies for signal
(true photons) and background (fake photons) for a given isolation cut. In a
single-photon sample this property can be used to extract the number of true
photons:

\begin{equation}
w=\sum_{i=1}^{N}\frac{\epsilon_{i}-\epsilon_{b}(E_{{\rm T}i})}
{\epsilon_{s}(E_{{\rm T}i})-\epsilon_{b}(E_{{\rm T}i})}
\label{eq:phofr}
\end{equation}

\noindent
where $\epsilon_i$=1 if track-ISO$<$$cut$ and $\epsilon_i$=0 if
track-ISO$>$$cut$, $\epsilon_s(E_{\rm T})$ is the signal efficiency for
track-ISO$<$$cut$, $\epsilon_b(E_{\rm T})$ is the background efficiency for
track-ISO$<$$cut$ and $N$ is the total number of candidate photons in the
sample. This technique can be generalized in the case of the $\gamma\gamma$
sample and is based on a maximum likelihood approach which is described in
detail in Appendix \ref{sec:likelihood}.

As mentioned in Section \ref{sec:data}, two types of isolation can be
defined for central photons: calorimeter and track isolation. Cal-ISO is
sensitive to the following contributions: underlying event (UE), multiple
interactions (MI), leakage from the photon cluster (LE), and fragmentation
contribution (FR) from jets (for fakes). Track-ISO, on the other hand,
is only a measure of UE and FR. Therefore, it can potentially offer a better
separation between true and fake photons. Using track-ISO for the fake photon
background subtraction also has additional advantages that low-$P_{\rm T}$
tracks are very well measured (unlike the calorimeter energy) and jet
fragmentation studies \cite{jetsfrag} indicate that track observables are
well described by {\sc pythia} both for the UE and for the jets.

\begin{figure*}[!ht]
\centering
\includegraphics[width=0.45\linewidth]
{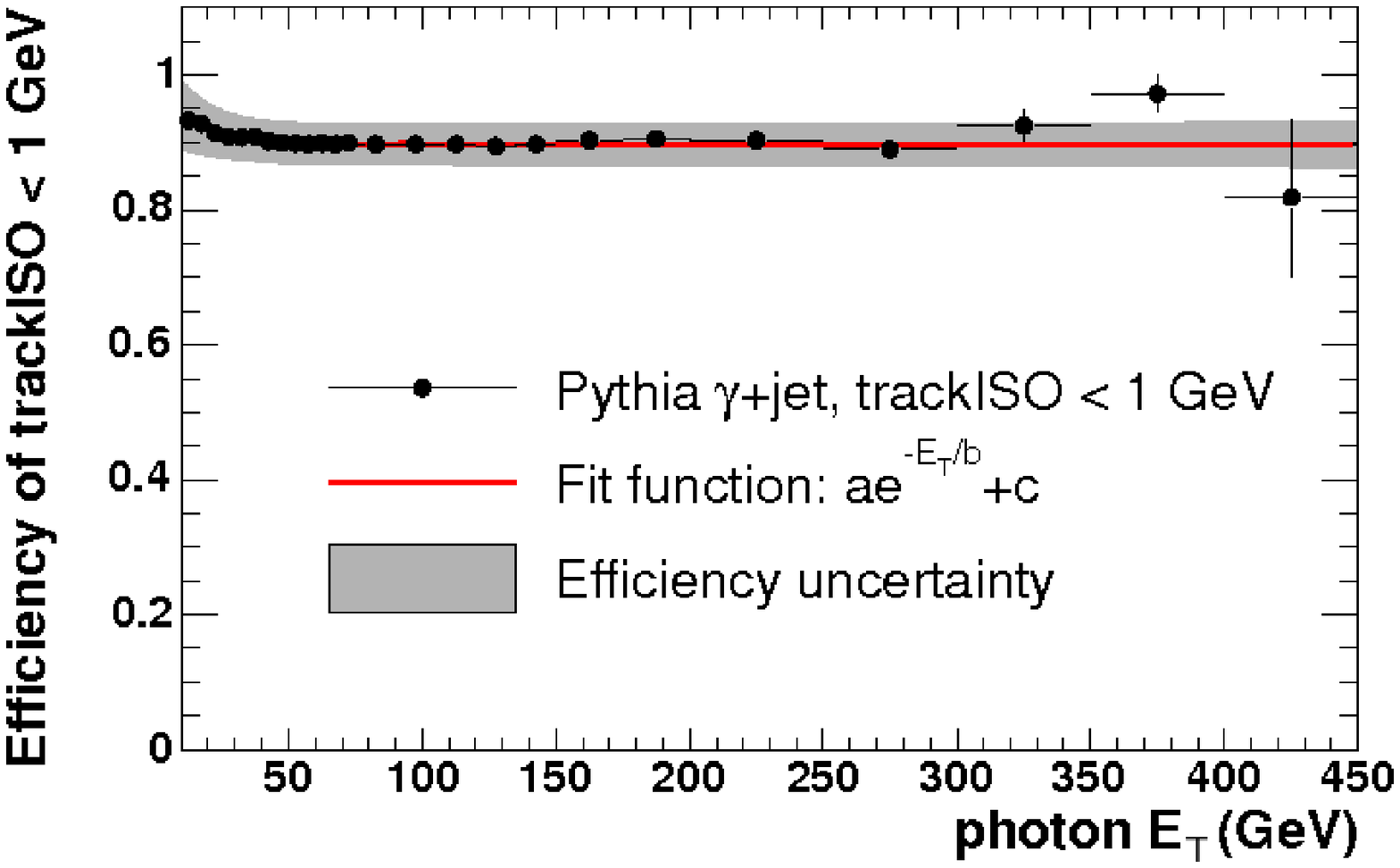}
\includegraphics[width=0.45\linewidth]{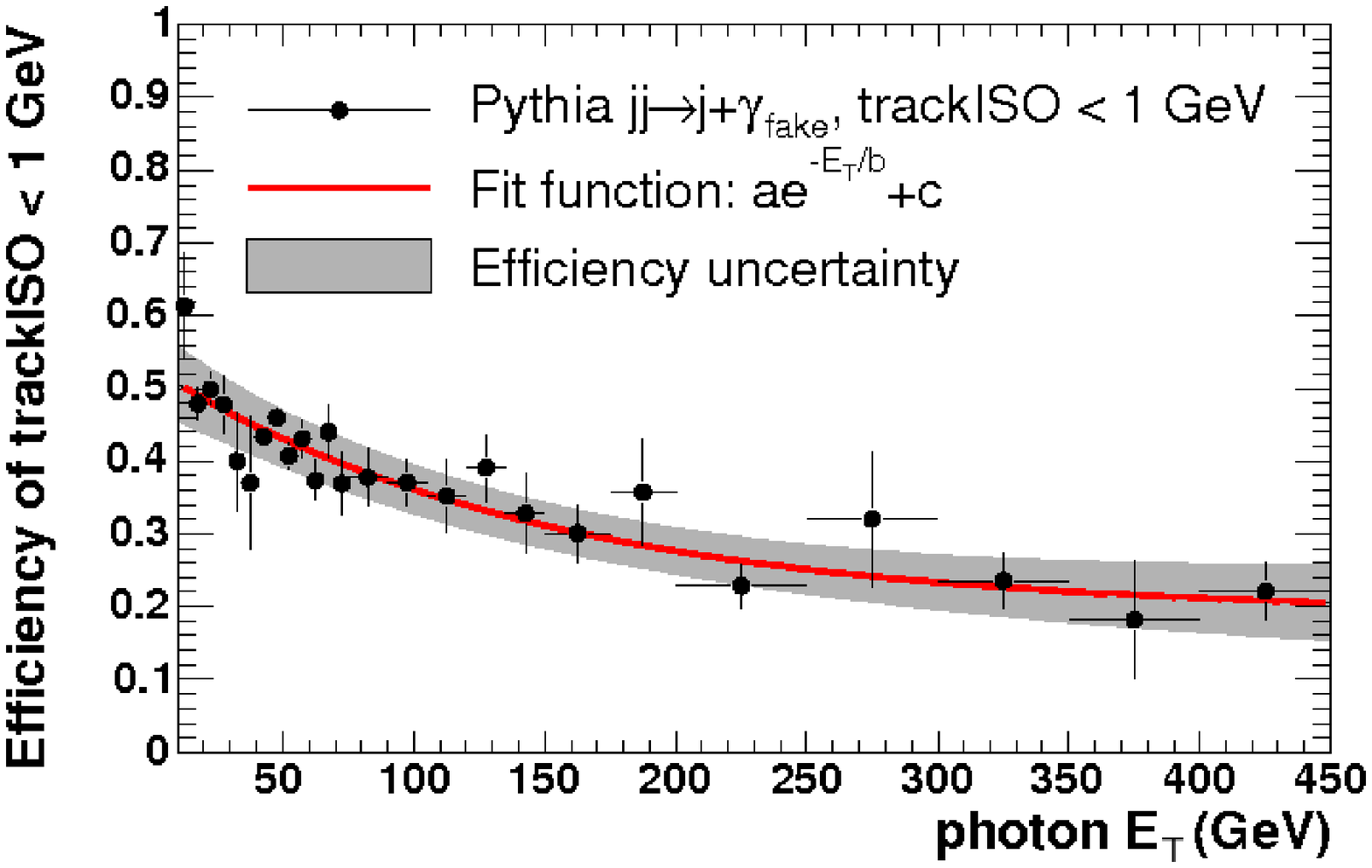}
\caption{Signal (left) and background (right) efficiencies for
   track-ISO$<$1 GeV. The shaded area is the total systematic uncertainty.}
\label{fig:eff_systematics1}
\end{figure*}

To perform the background subtraction, signal ($\epsilon_s$) and background
($\epsilon_b$) efficiencies are needed for a certain cut on track-ISO. The
form of Eq. \ref{eq:phofr} suggests that the best accuracy in photon purity
can be achieved when the absolute value of the denominator is maximum.
When this happens the terms in the sum of Eq. \ref{eq:phofr} are minimized
in magnitude and thus the purity is less sensitive to the statistical
uncertainty of the number of events in the sample. Therefore, a scan of
$\epsilon_s-\epsilon_b$ as a function of the track-ISO cut is performed using
MC samples of true and fake photons. The difference peaks at track-ISO$\sim$1
GeV. The threshold of the track-ISO cut for the signal and background
efficiency functions is thus chosen at 1 GeV.

The signal track-ISO efficiency is obtained from {\sc pythia} photon--plus--jet
samples. The background efficiency is obtained from {\sc pythia} dijet
samples. All {\sc pythia} samples used in this work are derived from version
6.2.16 of the program using the CTEQ5L PDF set for \cite{cteq5l} and the
``tune A'' for UE parameters \cite{tuneA}. Background events are filtered out
if a detector photon is matched to a generator level photon originating from
quark ISR or FSR. This ensures that the background track-ISO efficiency
function is obtained for neutral hadrons (mostly $\pi^0$ or $\eta^0$) faking
a photon signature. Similarly, for the signal events detector photons are
required to match generator level photons from the hard scattering (thus
fragmentation photons are removed). The signal efficiency $\epsilon_s$ and
the background efficiency $\epsilon_b$ are shown in Fig.
\ref{fig:eff_systematics1} as functions of the photon $E_{\rm T}$. Both
functions are parameterized by a linear combination of an exponential and
a constant.

The isolation cones for the two photon candidates are not entirely independent.
For example, if a particular event has a higher (lower) than average underlying
event activity, then it is likely that both isolation cones will simultaneously
have more (less) energy. In addition, the ordering in $E_{\rm T}$ of the two
photons also introduces some bias. The signal (for $E_{\rm T}$$<$50 GeV) and
the background track-ISO efficiencies drop with increasing $E_{\rm T}$ (see Fig.
\ref{fig:eff_systematics1}). Therefore, the lower $E_{\rm T}$ threshold for the
first photon in the event relative to the second photon implies, on average,
that the $E_{\rm T}$ of the first photon will be systematically lower than the
$E_{\rm T}$ of the second photon, thus introducing some bias due to the
$E_{\rm T}$ dependence of the efficiencies. This effect is negligible for
$E_{\rm T}$$>$50 GeV, where the signal track-ISO efficiency is flat and the
background is weak, but it becomes significant at low $E_{\rm T}$. It is a
small effect in the single photon purity, but it is at least a factor of two
more important for diphoton events. These correlations must be taken into
account when calculating a probability of two photon candidates to pass-pass,
pass-fail, fail-pass or fail-fail the 1~GeV isolation cut described above.
{\sc pythia} diphoton events were used to obtain ``per event'' track-ISO
efficiencies for these combinations. Correlations are much less important
for events with one or two fake photons because they are diluted by a much
larger contribution from jet fragmentation.

\begin{figure*}[!ht]
\centering
\includegraphics[width=0.45\linewidth]
{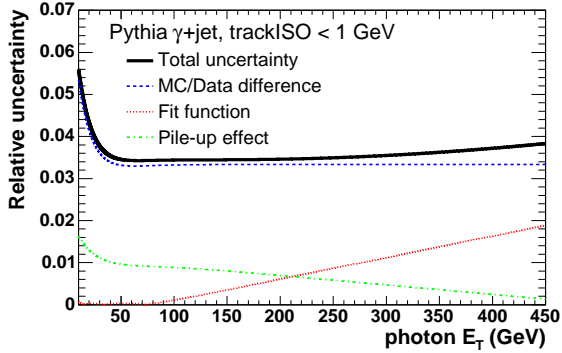}
\includegraphics[width=0.45\linewidth]
{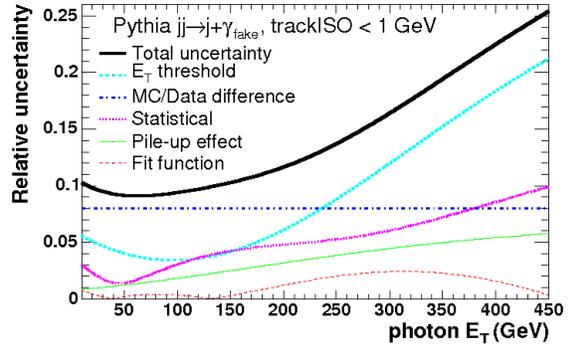}
\caption{Relative systematic uncertainties on the signal (left) and
   background (right) efficiencies for track-ISO$<$1 GeV.}
\label{fig:eff_systematics2}
\end{figure*}

The systematic uncertainties in the signal and background track-ISO
efficiencies are estimated and propagated into the final estimate of
photon purity. Correlations between different sources of systematics are
taken into account. The following sources of uncertainties are considered:
1) mis-modeling of the distribution of the number of vertices ($N_{vx}$) in
MC (pile-up effect); 2) statistical uncertainties in the fit parameters;
3) choice of the fit function for the efficiency; 4) generator--related
data-MC differences; 5) effect of the $E_{\rm T}$ threshold for selected
photon candidates (only for the background). These uncertainties are
presented in Fig. \ref{fig:eff_systematics2} and discussed below.

The MC simulation does not describe accurately the distribution of the
number of reconstructed vertices. This effect can be either removed by
re-weighting the MC to match the data or the associated uncertainty can
be assigned for the effect of mis-modeling. The latter approach is
chosen in this analysis because the track-ISO, to leading order, is
not sensitive to the presence of multiple interactions and the effect
is very small. This uncertainty is conservatively estimated as the
difference between the extreme cases of track-ISO efficiencies obtained
in events with $N_{vx}$=1 and track-ISO efficiencies obtained in events
with $N_{vx}$$>$1. For the photon energies relevant to this analysis, the
relative effect is $<$1$\%$ for the signal and $<$3$\%$ for the background.

\begin{figure}[!ht]
\centering
\includegraphics[width=1.0\linewidth]
{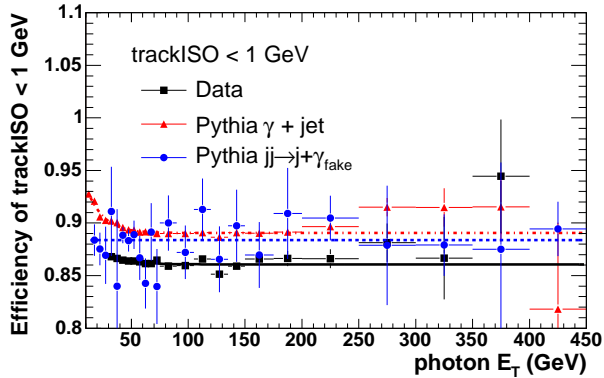}
\caption{Comparison of track-ISO$<$1 GeV efficiencies in complementary
   cones from data (squares), signal MC (triangles) and background MC
   (circles).}
\label{fig:compl_eff}
\end{figure}

The fit statistical uncertainties are included in the estimation of
systematic uncertainties. Correlations between fit parameters are
properly taken into account. The relative effect is negligible for
the signal and 1$\%$-3$\%$ for the background in the range of photon
energies relevant to this analysis.

The default fit function choice is an exponential plus a constant term.
As one can see from the track-ISO efficiency plots shown in Fig.
\ref{fig:eff_systematics1}, the quality of the fit is very good. The studies
of the efficiency dependence on the track-ISO cut indicated that, for some
cut values, the exponential plus a linear function can be a better fit to
the signal efficiency. This function was thus chosen as an alternative
track-ISO efficiency parameterization and the difference with the default
function was taken as the associated uncertainty. The relative effect is
$<$1$\%$ for both signal and background with $E_{\rm T}$$<$200 GeV.

\begin{figure}[!ht]
\centering
\includegraphics[width=1.0\linewidth]{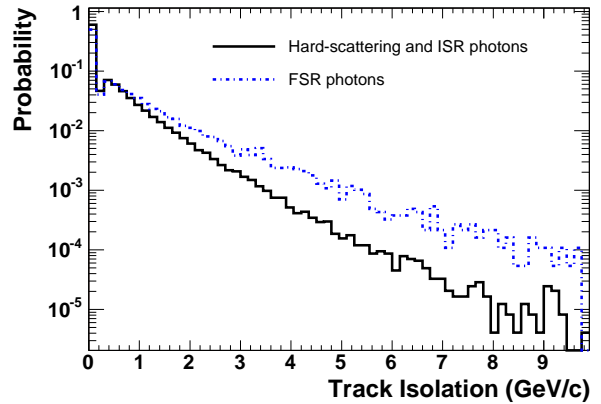}
\caption{The track-ISO distribution in hard scattering and ISR (blue) and
     FSR (black) events.}
\label{fig:trk_ISO_FSR}
\end{figure}

The modeling of both signal and background relies on the MC. Therefore,
it is necessary to assign a systematic uncertainty on possible data-MC
differences both for signal and background photons. In the case of signal,
it is necessary to check the modeling of the underlying event in the MC.
This is done by means of complementary cones. The complementary cones are
chosen such that their axes have the same angle $\theta$ with respect to
the beam line as the photon candidate and are rotated by $\pm$$\pi$/2 in
$\phi$. These cones are assumed, on average, to collect the same amount
of the underlying event as cones of the same size around true photons.
This assumption is tested and confirmed  in the MC. It is also checked
that complementary cones for signal and background look very similar.
Finally, the signal track-ISO efficiency is obtained from complementary
cones in data and signal MC and the difference between the two is taken
as the associated systematic uncertainty in the signal track-ISO efficiency.
The comparison of track-ISO efficiencies for complementary cones in data,
MC signal and background is shown in Fig. \ref{fig:compl_eff}. The relative
effect is $\sim$3.5$\%$ for most of the photon energies. This is the largest
systematic uncertainty for the signal track-ISO efficiency. The systematic
uncertainty from this source decreases from 5.5$\%$ at $E_{\rm T}$=10 GeV to
$\sim$3.5$\%$ at $E_{\rm T}$=40 GeV and then it stays at roughly the same level
for $E_{\rm T}$$>$40 GeV. An additional uncertainty arises from the fact that
the signal efficiency is derived from true photons generated only by
direct $gq\rightarrow\gamma q$, $q\bar q\rightarrow\gamma g$ and
$gg\rightarrow\gamma g$ production, omitting photons radiated from initial
or final state quarks. Fig. \ref{fig:trk_ISO_FSR} shows that the track
isolation of photons radiated from final state quarks is somewhat different
than that of photons produced by hard scattering or radiated from initial
state quarks. This difference is estimated to have a constant 2$\%$ effect
on the signal track-ISO efficiency which is added to its total systematic
uncertainty.

The systematic uncertainty due to data-MC differences in the track-ISO
background efficiency is estimated by comparing the track-ISO cut efficiency
for a leading track in dijet events from data and MC. This method assumes
that jets with a leading neutral particle (e.g., $\pi^0/\eta$) have the same
or very similar fragmentation properties as jets with a leading charged
particle (e.g., $\pi^{\pm}$ or K$^{\pm}$). The following procedure is applied
to both data and MC. Events with two well-balanced and back-to-back jets are
used, satisfying
$\vert$$E_{\rm T}$(jet1)$-$$E_{\rm T}$(jet2)$\vert$/[$E_{\rm T}$(jet1)+$E_{\rm T}$(jet2)]$<$0.3
and $\vert$$\phi$(jet1)$-$$\phi$(jet2)$\vert$$>$2.7 rad. The event is rejected
if there is a third jet with
$E_{\rm T}$$>$0.1$\times$[$E_{\rm T}$(jet1)+$E_{\rm T}$(jet2)]. One of the jets
(a probe jet) is required to be in the central detector region,
$\vert$$\eta$$\vert$$<$1.1, thus matching the pseudorapidity requirement for
photons. In the next step, a well-reconstructed track (a probe track) is
selected with the largest $p_{\rm T}$ inside the probe jet, i.e. inside a cone
of $\Delta R$=0.4 around the jet direction. For this track, an analog of the
cal-ISO is calculated as the $\sum E_{\rm T}$ of all towers inside a cone of
$\Delta R$=0.4 around the track direction. Towers associated with the track
(up to 3 towers in $\eta$) are excluded from the sum. The cal-ISO for the
probe track has to satisfy exactly the same requirements as the isolation for
a photon with $E_{\rm T}$=$p_{\rm T}$. An analog of the track-ISO for the probe
track is also calculated by following exactly the same procedure as for
photons, with the only exception being that the track itself is excluded from
the sum. Finally, the efficiency of the track-ISO$<$1 GeV cut for the probe
track is compared in data and MC. The observed relative difference of 8\%,
independent of the track $p_{\rm T}$, is taken as an estimate of the systematic
uncertainty due to data-MC differences in the track-ISO background efficiency.

\begin{figure}[!ht]
\centering
\includegraphics[width=1.0\linewidth]
{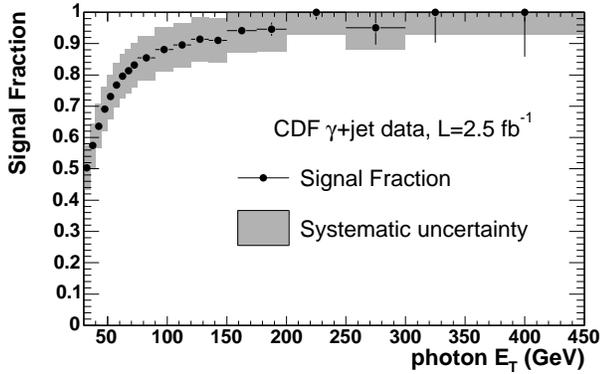}
\caption{The estimated signal fraction in the inclusive photon data.
   The shaded area is the total systematic uncertainty in the signal fraction.}
\label{fig:signal_fr_1pho}
\end{figure}

Finally, the last source of systematic uncertainty in the background
track-ISO efficiency is associated with the choice of an $E_{\rm T}$ threshold
for selecting fake photons from a particular jet sample. The fake rate for
jets is very small and, as a consequence, the MC dijet samples do not have
enough statistics to yield a sufficient number of fake photons after the
selection cuts. To maximize the statistics, fake photons are accepted
from each dijet sample if $E_{\rm T}$$>$$\hat{p}_{\rm T}$ where $\hat{p}_{\rm T}$
is the parton transverse momentum cutoff used in the event generation. Ideally,
events with $E_{\rm T}$$>$($\hat{p}_{\rm T}$$+$offset) should have been selected
to avoid a bias due to the $\hat{p}_{\rm T}$ threshold effect. This is necessary
because fake photons carry, on the average, only $~$90$\%$ of the energy
of the original parton. Therefore, the procedure is biased toward selecting
fakes originating from gluon jets produced by radiation, which are not
limited by the $\hat{p}_{\rm T}$ threshold of hard scattering, and as a
consequence toward lower background efficiencies. To obtain a conservative
estimate of this effect, the threshold was lowered even more, thus accepting
fake photons with $E_{\rm T}$$>$$F$$\times$$\hat{p}_{\rm T}$ where
$F$$\sim$0.8-0.9, depending on $\hat{p}_{\rm T}$. By decreasing the threshold,
the effect is overestimated, but this gives a conservative estimate of the
associated uncertainty. The total systematic uncertainty of the background
track-ISO efficiency is at the level of 10$\%$-12$\%$ in the range of photon
$E_{\rm T}$ from 15~GeV to 200~GeV, the range relevant to this analysis.

\begin{figure*}[!ht]
\centering
\includegraphics[width=0.4\linewidth]{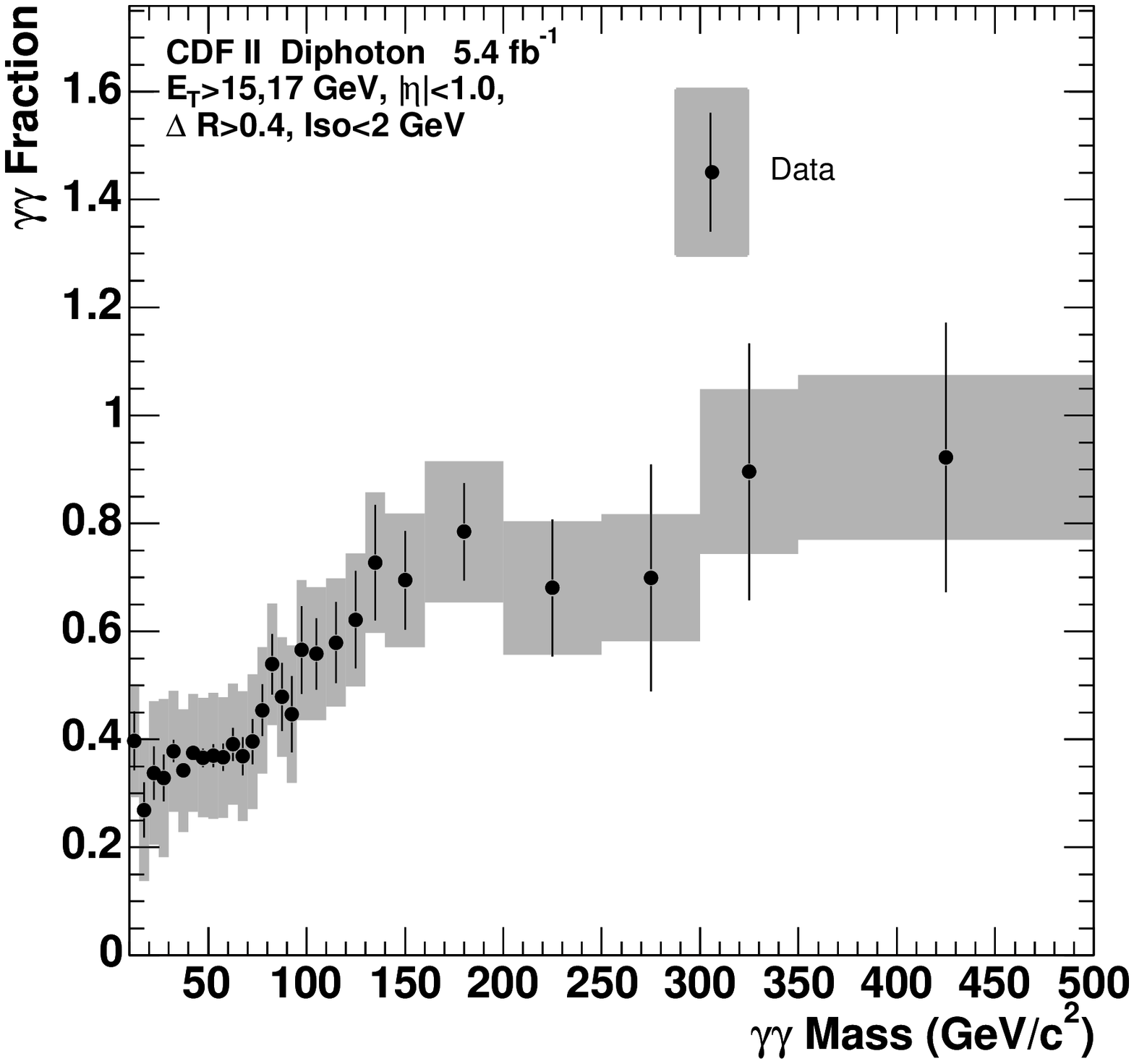}
\includegraphics[width=0.4\linewidth]{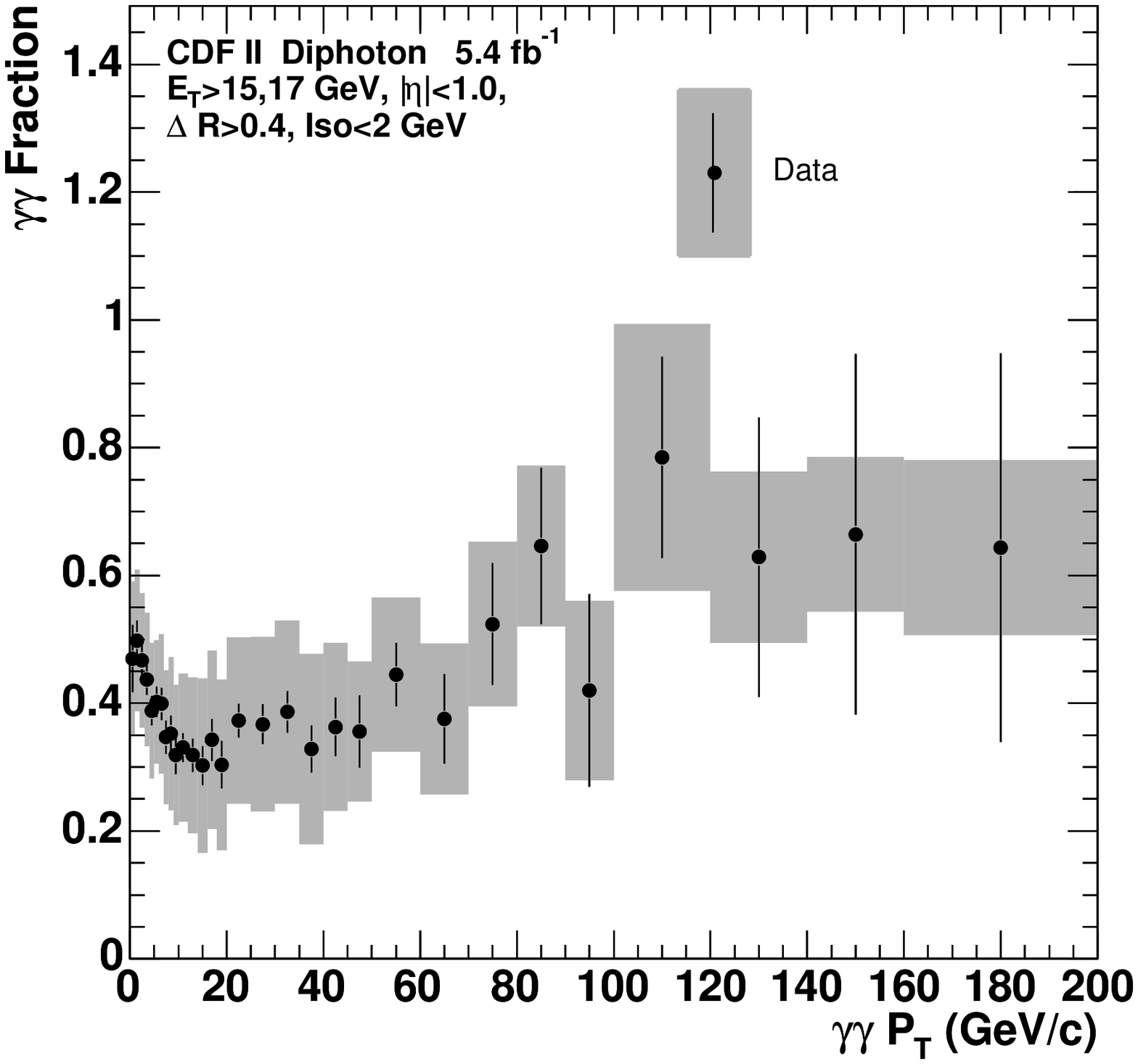}
\includegraphics[width=0.4\linewidth]{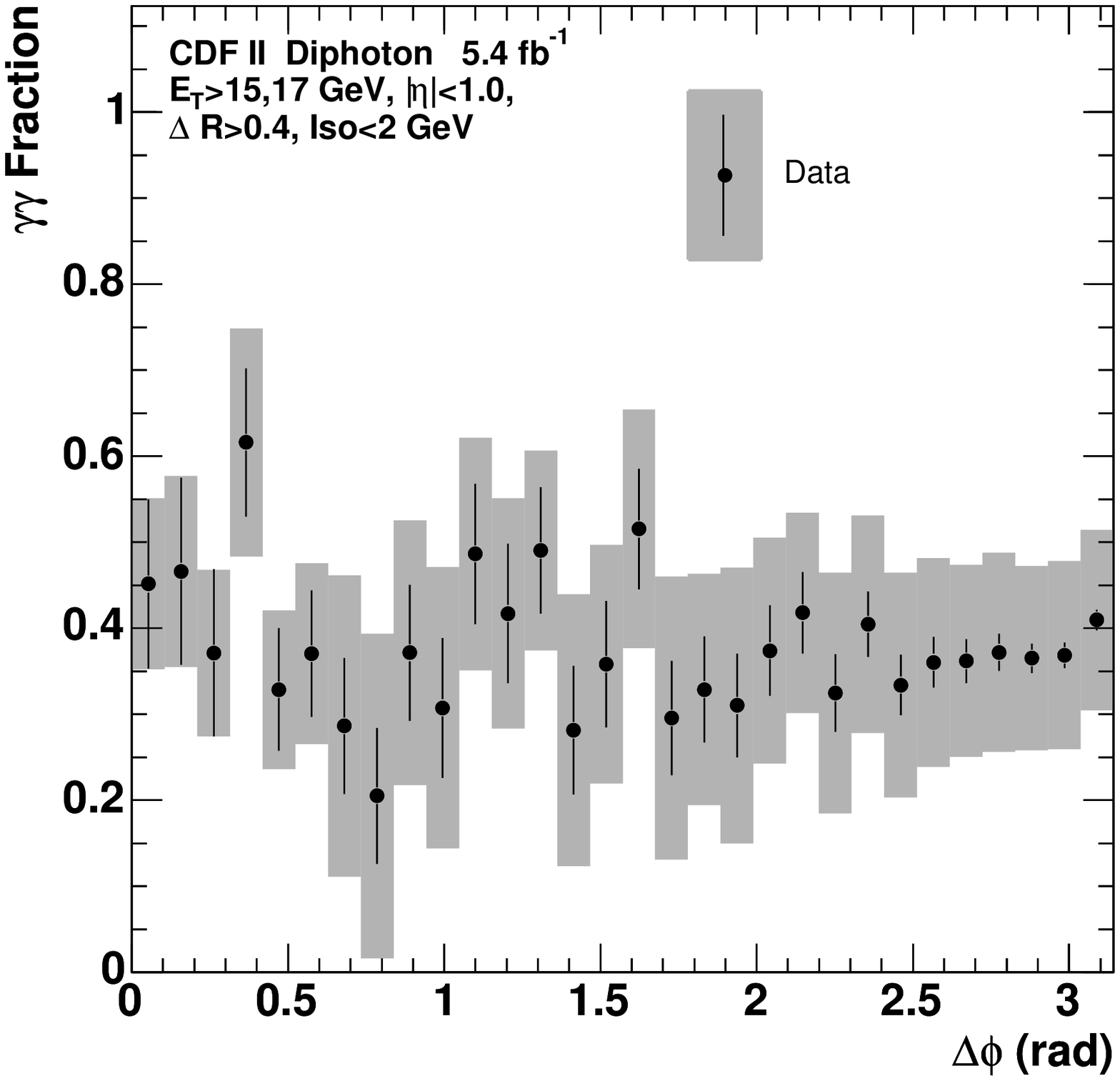}
\includegraphics[width=0.4\linewidth]{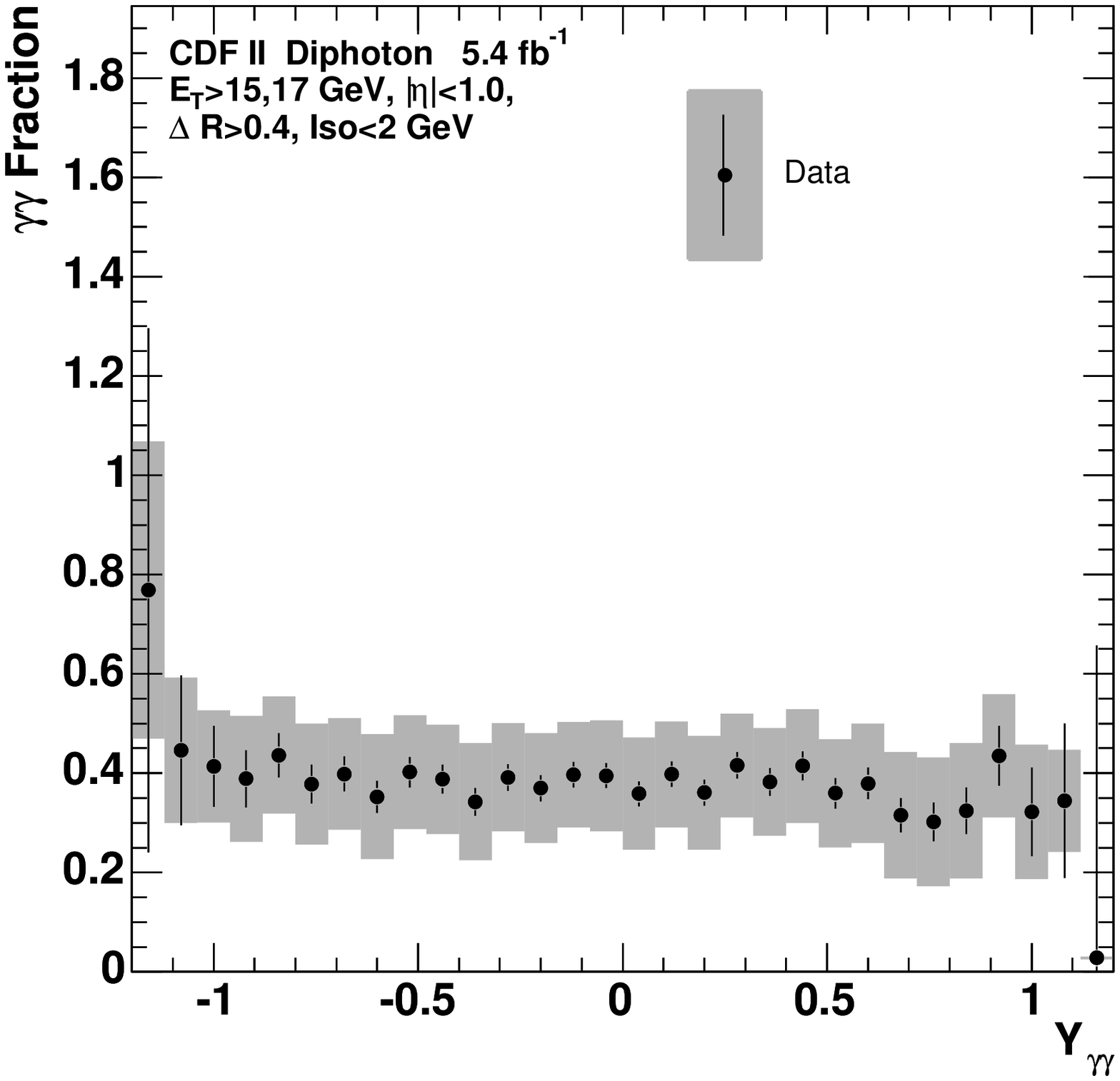}
\includegraphics[width=0.4\linewidth]{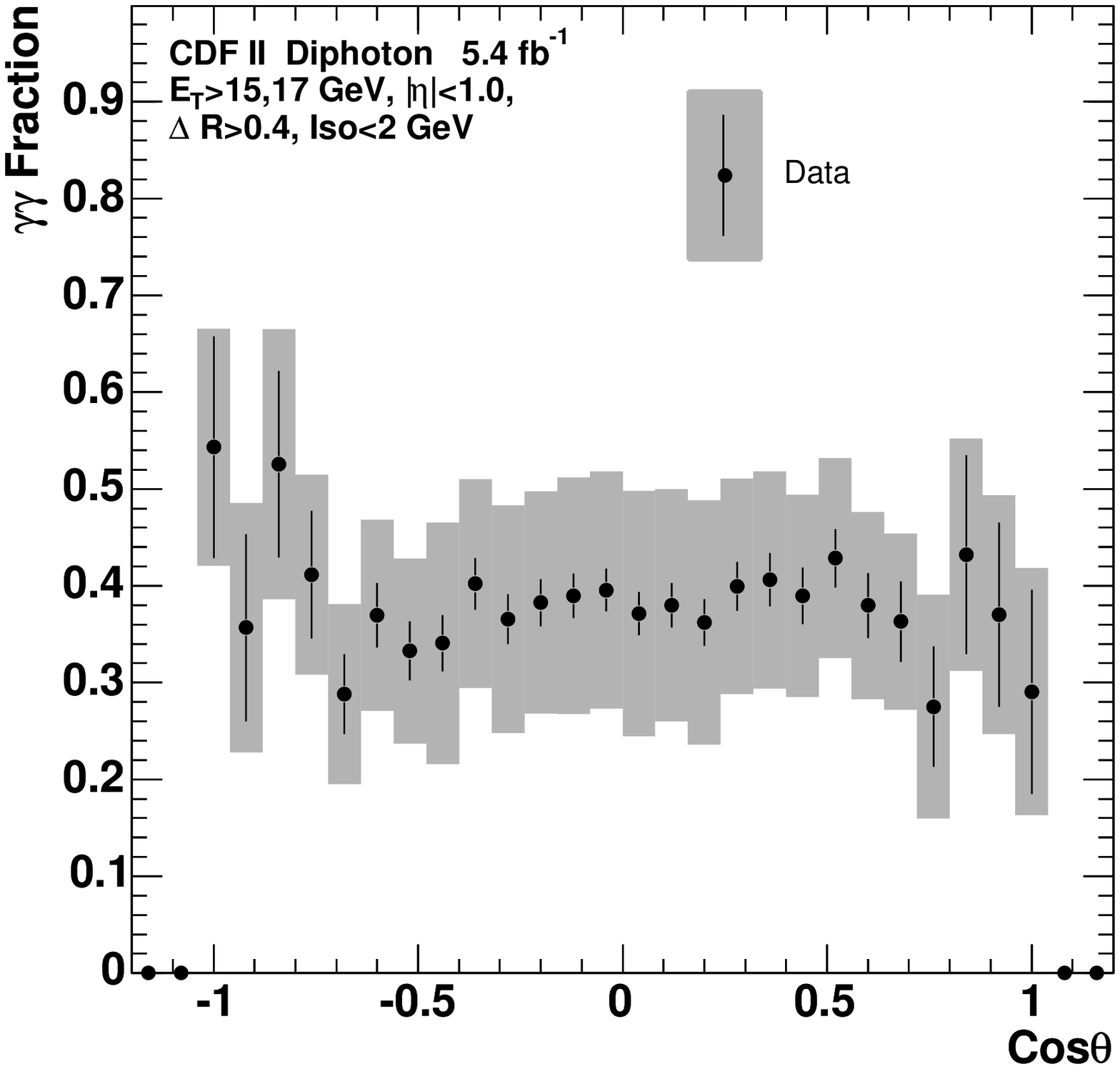}
\includegraphics[width=0.4\linewidth]{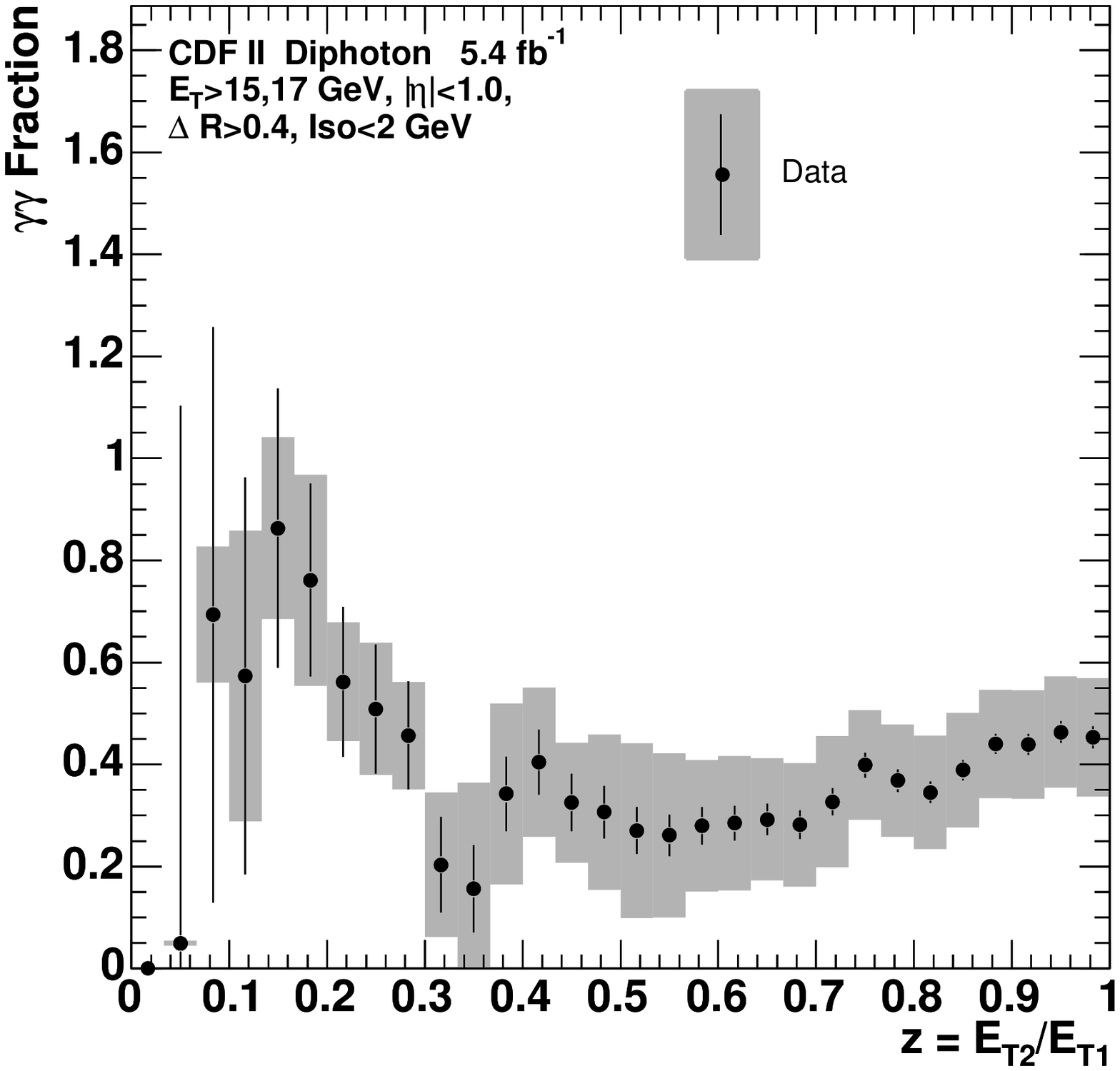}
\caption{The estimated signal fraction in the diphoton data as a function of
   several kinematic variables. The shaded area is the total systematic
   uncertainty in the signal fraction.}
\label{fig:signal_fr_2pho}
\end{figure*}

The background subtraction procedure has been tested with MC signal
and fake events as well as with inclusive photon data. Tests with MC
provide closure checks: the returned purity was 100$\%$ for signal
events and 0$\%$ for fakes (within the corresponding uncertainties).
The estimated photon purity for inclusive photon data as a function
of the photon transverse energy is shown in Fig. \ref{fig:signal_fr_1pho}
and is similar to the purity obtained in the inclusive photon cross
section analysis \cite{cdf6312}. The uncertainty in the signal fraction
of the inclusive photon sample achieved with the track-ISO method is
between $\sim$11$\%$ at low $E_{\rm T}$ and $\sim$5$\%$ at very high
$E_{\rm T}$. Fig. \ref{fig:signal_fr_2pho} shows the estimated purity
for the diphoton data as a function of the kinematic variables defined
in Subsection \ref{subsec:kin-var}.

\subsection{Event Reconstruction and Selection Efficiency}
\label{subsec:effi}

The corrections for event reconstruction and selection efficiency were
derived primarily from {\sc pythia} diphoton MC samples. The numerator of
the efficiency is the number of events with two photons that pass all of
the trigger criteria and selection cuts listed in Tables \ref{tab:trig}
and \ref{tab:phoID}. The definition of the cross section measurement is
determined by the definition of the denominator of the efficiency. The
denominator cuts are summarized in Table \ref{tab:MCaccept}. This work
reports a cross section for isolated photons, so the selection of denominator
events includes isolation. This isolation is found by summing over all
generated hadrons and photons originating from the primary vertex within
a cone of $\Delta R$$=$$0.4$ around each photon.

\begin{table}[tbp]
\begin{center}
\caption{Summary of the requirements applied to the generated MC events
     to define the denominator of the selection efficiency and
     the meaning of cross section measurement itself. $y^{\gamma}$
     is the photon rapidity \cite{eta}.}
\label{tab:MCaccept}
\begin{tabular}{lc}
\hline\hline
Cuts & Selected $\gamma\gamma$ events \\ \hline \hline
$E_{\rm T}^{\gamma}$ & $\geq 15$ GeV (1$^{\rm st}$$\gamma$),~\!
                 $\geq 17$ GeV (2$^{\rm nd}$$\gamma$) \\ \hline
$|y^{\gamma}|$ & $\leq$1.0 for both photons \\ \hline
Isolation & $\leq$2.0 GeV for both photons \\ \hline
$\Delta R(\gamma\gamma)$ & $\geq$$0.4$ \\ \hline
~~~~~~~~~~~~~~~ & No matching requirement between \\
~~~~~~~~~~~~~~~ & generated and reconstructed objects \\
\hline\hline
\end{tabular}
\end{center}
\end{table}

\begin{figure*}[!ht]
\centering
\includegraphics[width=0.4\linewidth]{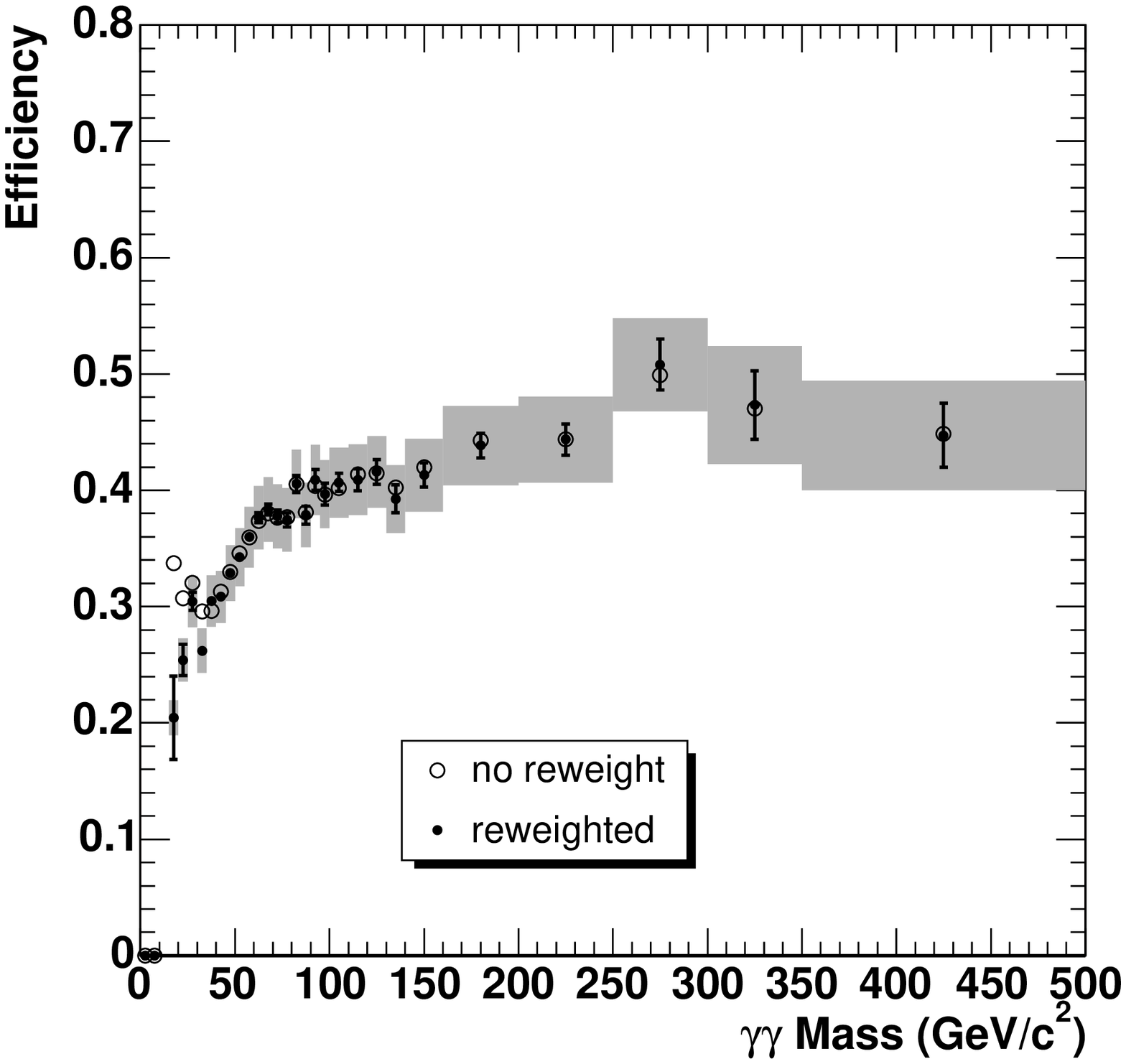}
\includegraphics[width=0.4\linewidth]{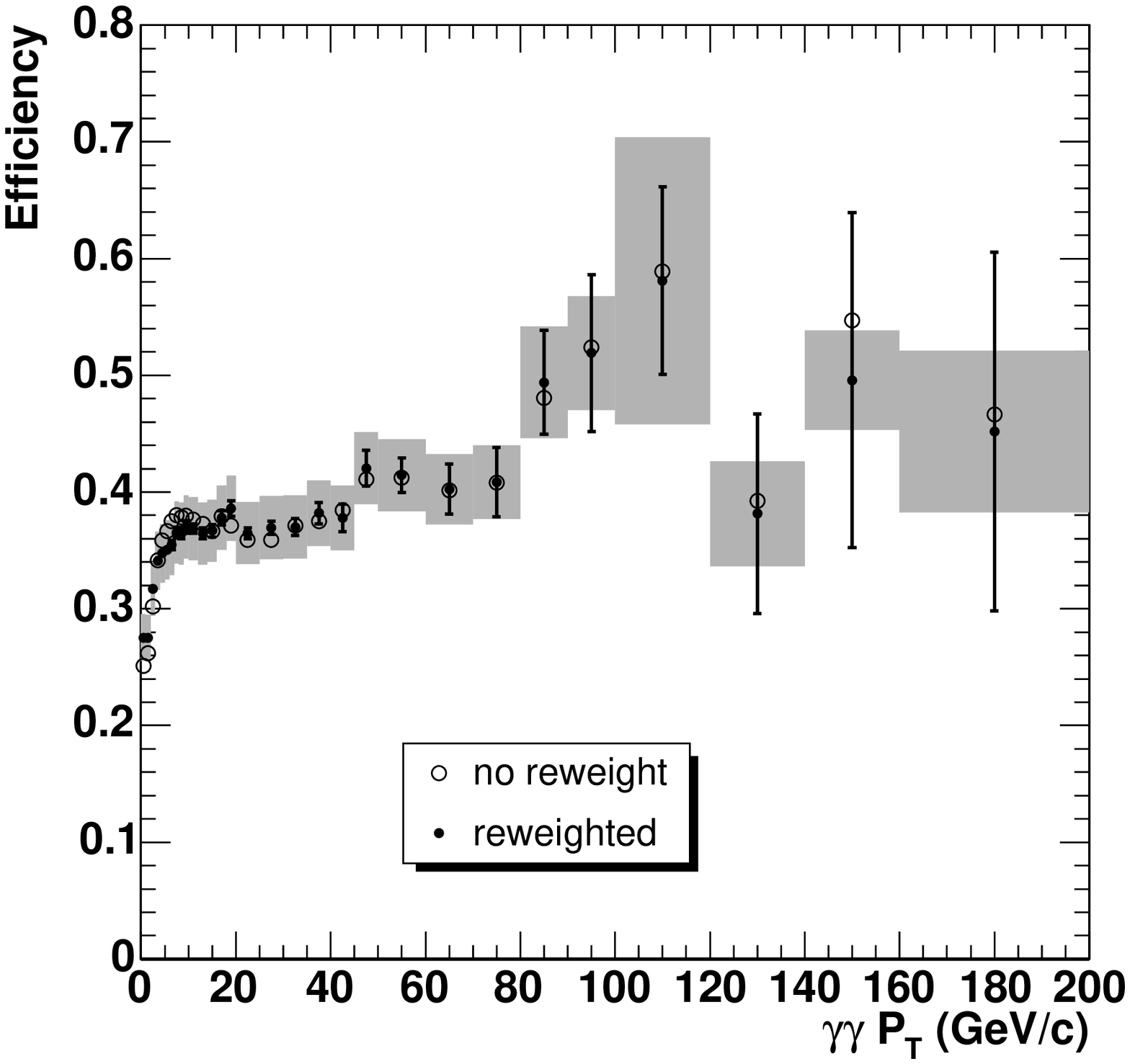}
\includegraphics[width=0.4\linewidth]{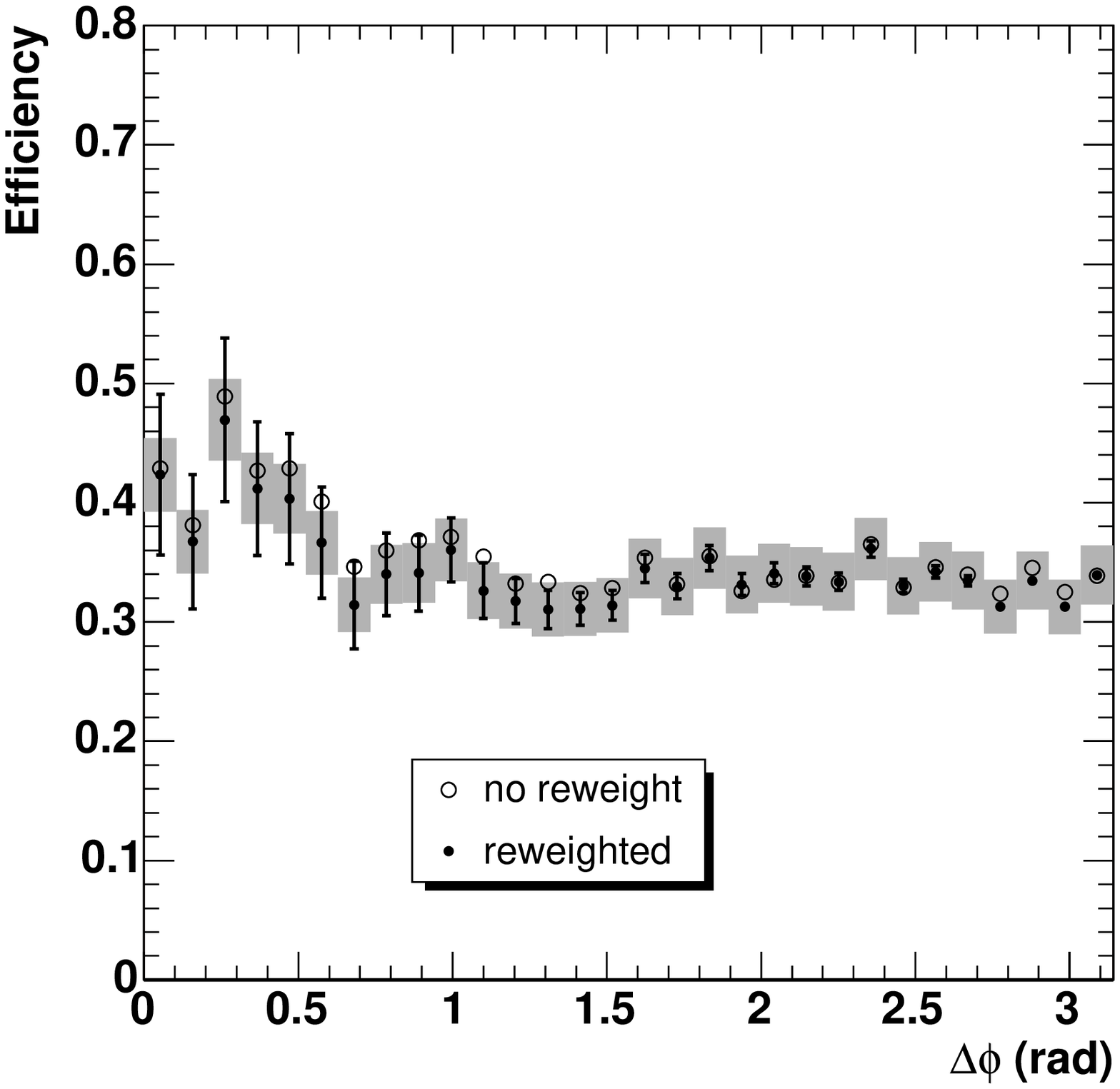}
\includegraphics[width=0.4\linewidth]{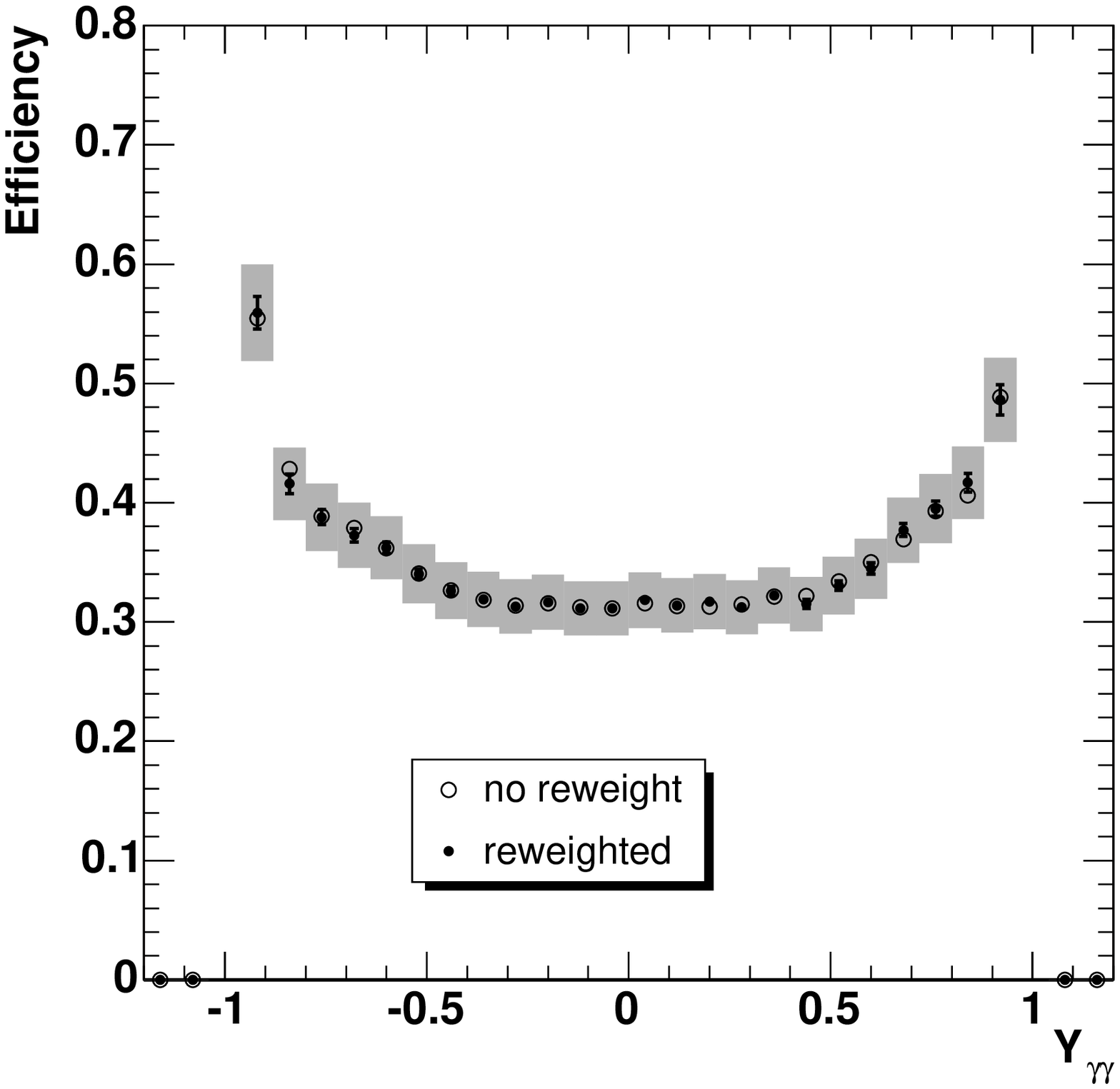}
\includegraphics[width=0.4\linewidth]{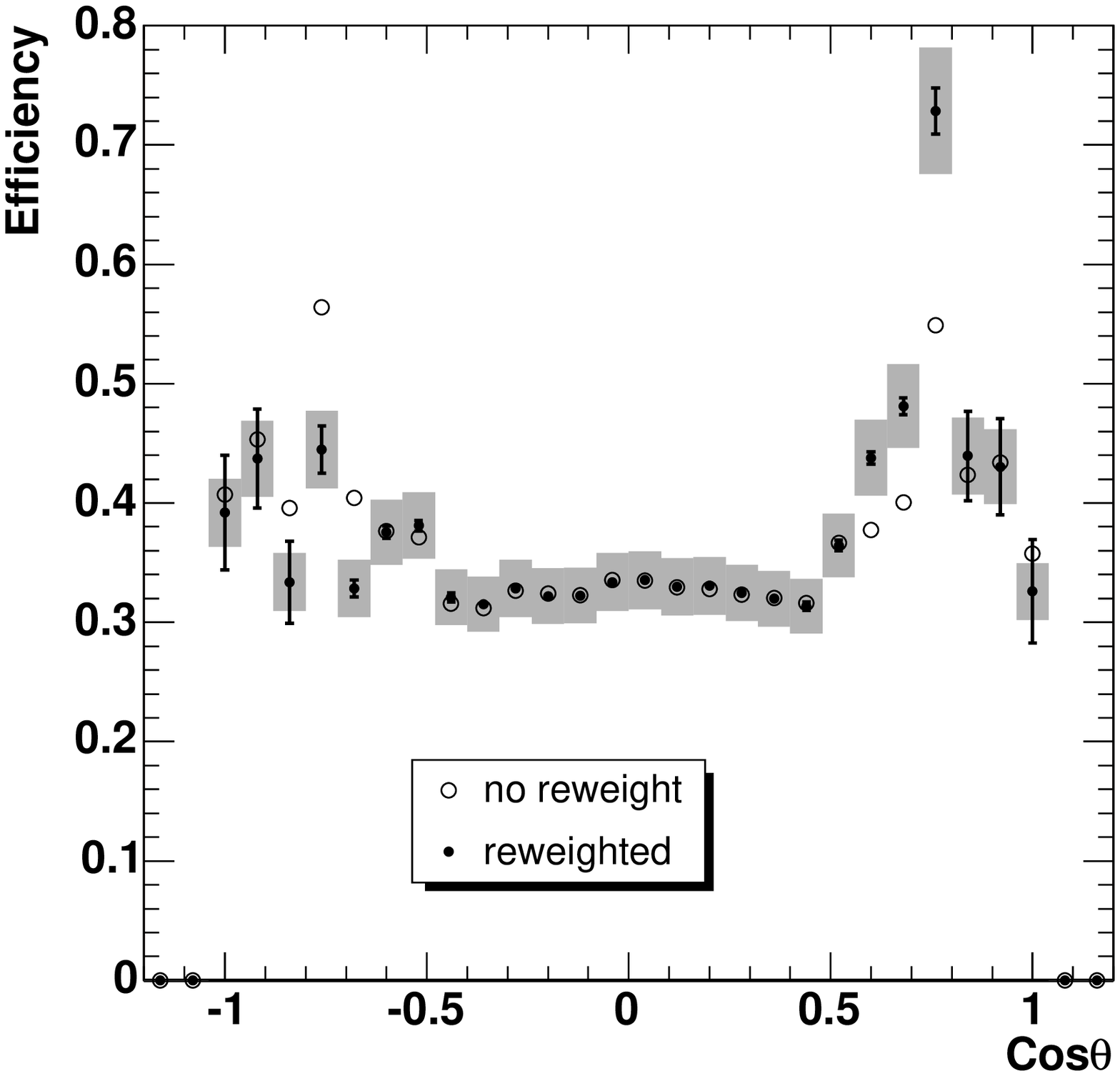}
\includegraphics[width=0.4\linewidth]{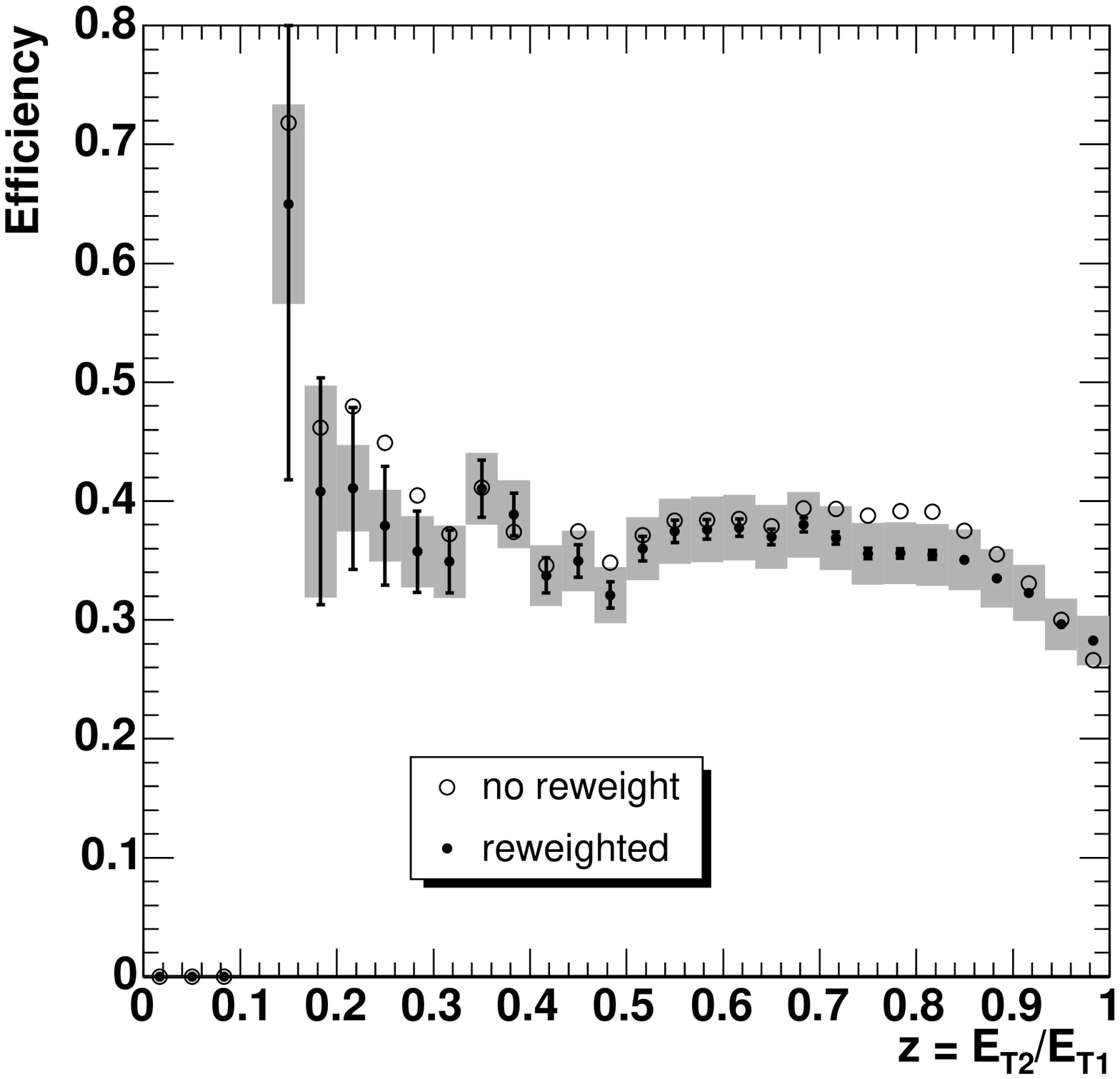}
\caption{The estimated efficiency as a function of several kinematic
   variables. The shaded area is the total systematic uncertainty in the
   efficiency.}
\label{fig:effi}
\end{figure*}

For each kinematic quantity, one histogram of the reconstructed quantity and
one of the quantity derived from generator variables are constructed. In the
first iteration, the efficiency is computed as the ratio of these histograms.
This ratio also corrects event migration in neighboring bins due to finite
resolution. Events which pass the denominator cuts and have a reconstructed
value for the histogrammed quantity but not a generator level value are
assigned the reconstructed values as the best approximation to the generator
level values. To improve accuracy, the efficiency calculation is iterated a
second time. Once all corrections are applied to the data, including the
efficiency, it is the best available representation of the true distribution.
Then the {\sc pythia} events are reweighted so that the second iteration of
the denominator histogram agrees with the corrected data. The purpose is to
correct the {\sc pythia} distribution closer to the true distribution, making
the efficiency more accurate. In practice, this does not have a large effect
on any distribution (see Fig. \ref{fig:effi}). A third iteration changes the
efficiency at the level of 1\% or less and, therefore, only two iterations
are applied.

The following corrections are applied to the efficiency:
\begin{itemize}
\item $Z^0\rightarrow e^+e^-$ events in data and MC are compared to derive a
correction to the photon ID efficiency reported by the MC. The correction
is reported as a function of $N_{vx}$ and of run periods. The correction
is weighted by the period luminositites, and the observed $N_{vx}$
distributions to find an overall multiplicative efficiency correction
of 0.967 per photon. In addition, there is some indication of an
$E_{\rm T}$-dependence, so the factor 0.967 is allowed to vary linearly up
to 1.0 between 40 and 80 GeV, and then is held constant at 1.0 above 80 GeV.
\item A small correction is included near the $E_{\rm T}$ cut threshold due
to the trigger turn-on curve. This is implemented as a factor of 0.98 at
$E_{\rm T}=15$ GeV, going linearly up to 1.0 at $E_{\rm T}=18$ GeV. Ref.
\cite{cdf6312} concludes that there is no need for other corrections for
the trigger. \item {\sc pythia} includes the underlying event, but NLO
calculations do not. This makes the {\sc pythia}-based efficiency correction
too large when comparing the isolated cross section to NLO predictions. It
is too large since the UE causes events to be removed from the isolated
denominator of the efficiency. A correction is derived by convoluting the
{\sc pythia} UE isolation energy with the {\sc diphox} energy in the
isolation cone \cite{diphox}. This reduces the probability for the
{\sc diphox} event to pass the isolation cuts. This effect is measured
to be a factor of 0.88 per event which is then applied as a correction to
the data.
\end{itemize}
The efficiency obtained for the kinematic quantities defined in Subsection
\ref{subsec:kin-var} is shown in Fig. \ref{fig:effi}. The typical efficiency
is 40$\%$.

\begin{figure*}[!ht]
\centering
\includegraphics[width=0.4\linewidth]{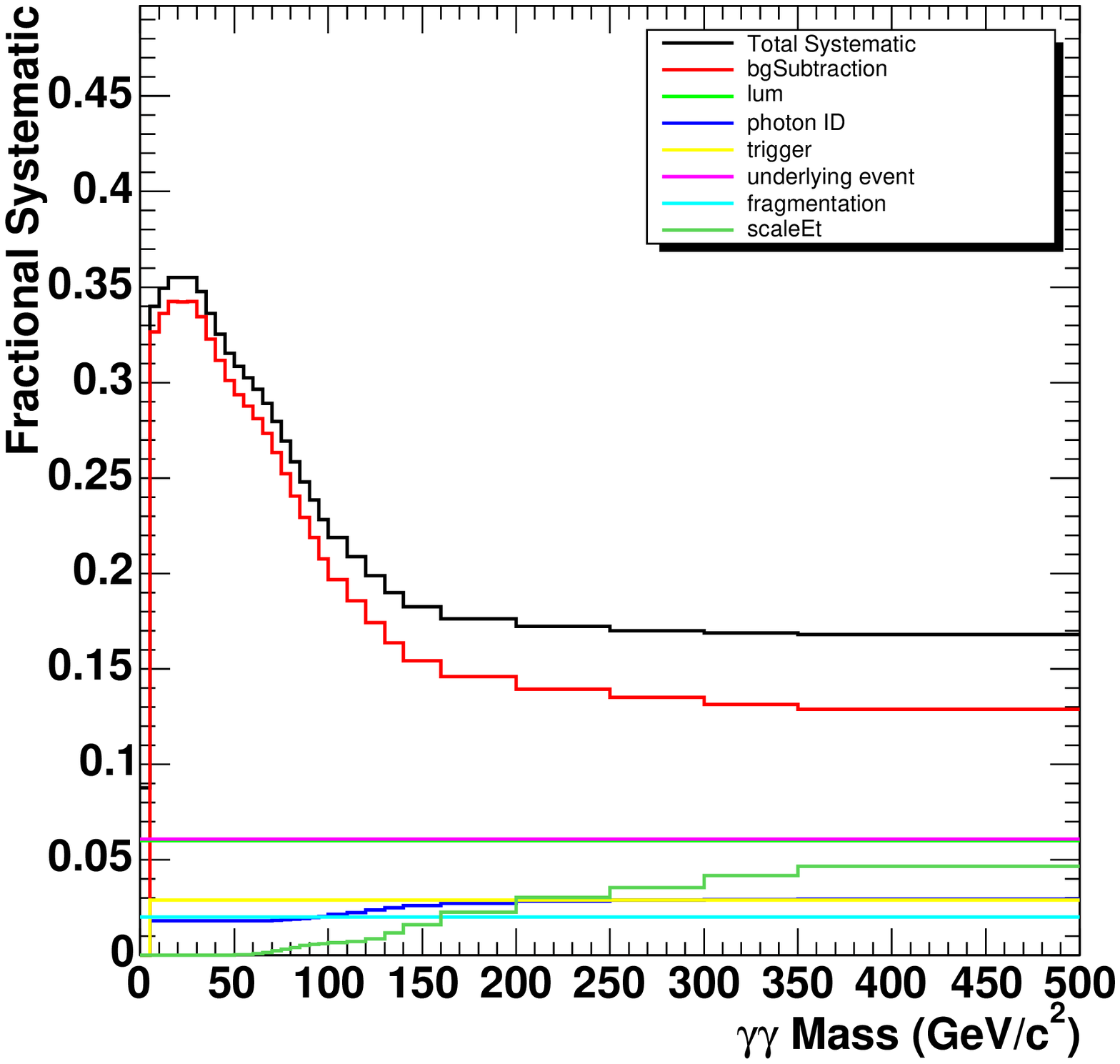}
\includegraphics[width=0.4\linewidth]{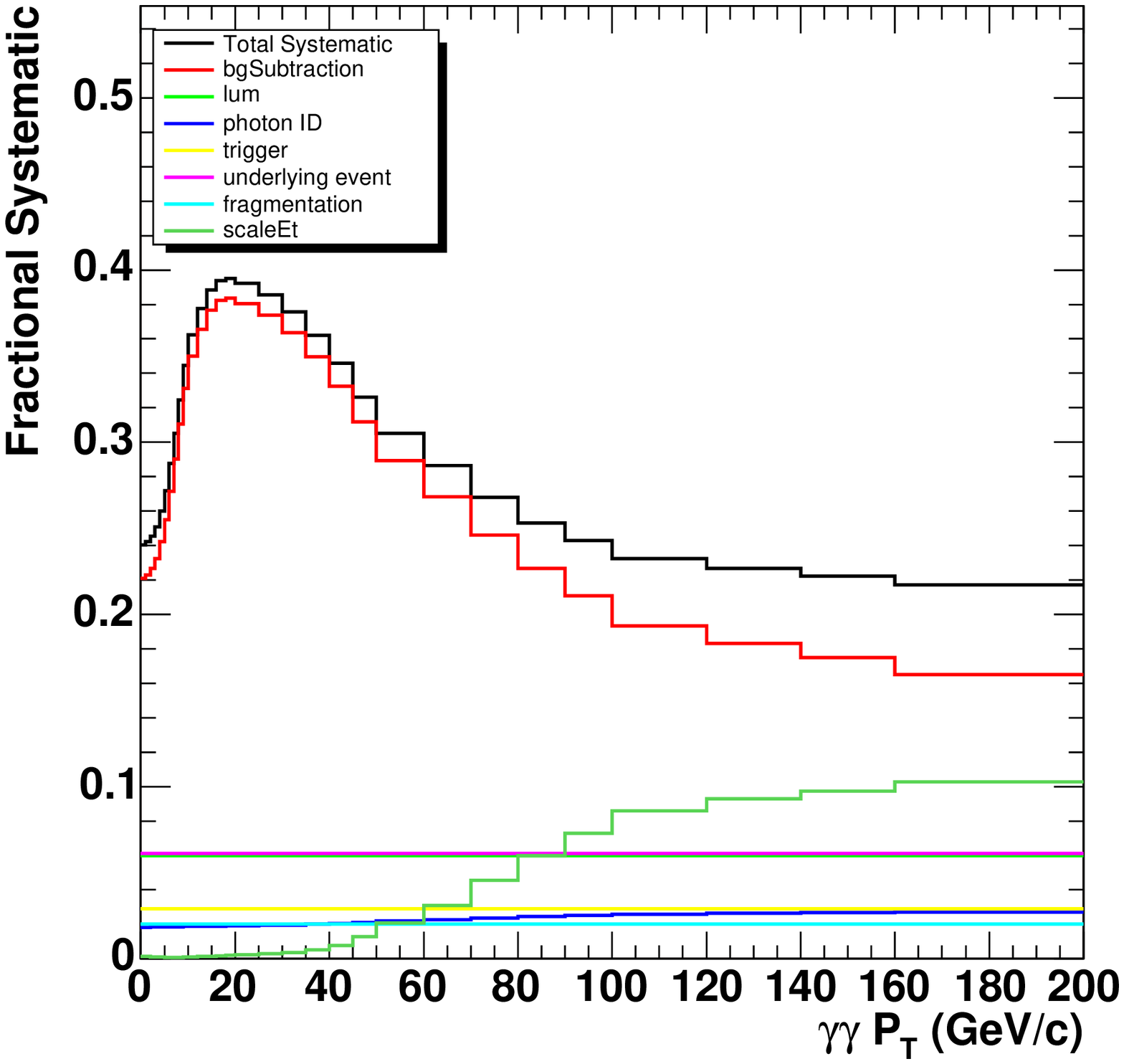}
\includegraphics[width=0.4\linewidth]{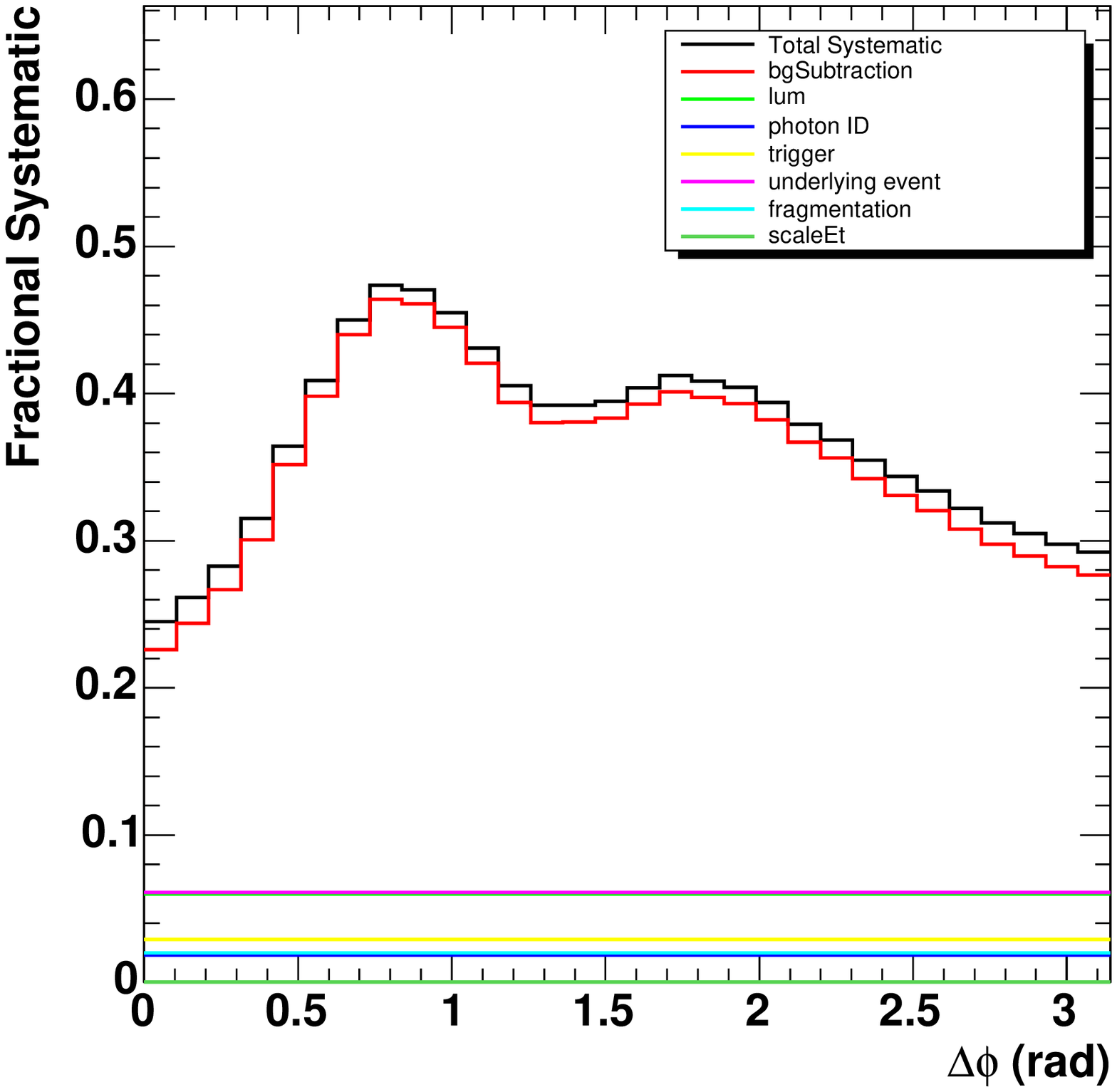}
\includegraphics[width=0.4\linewidth]{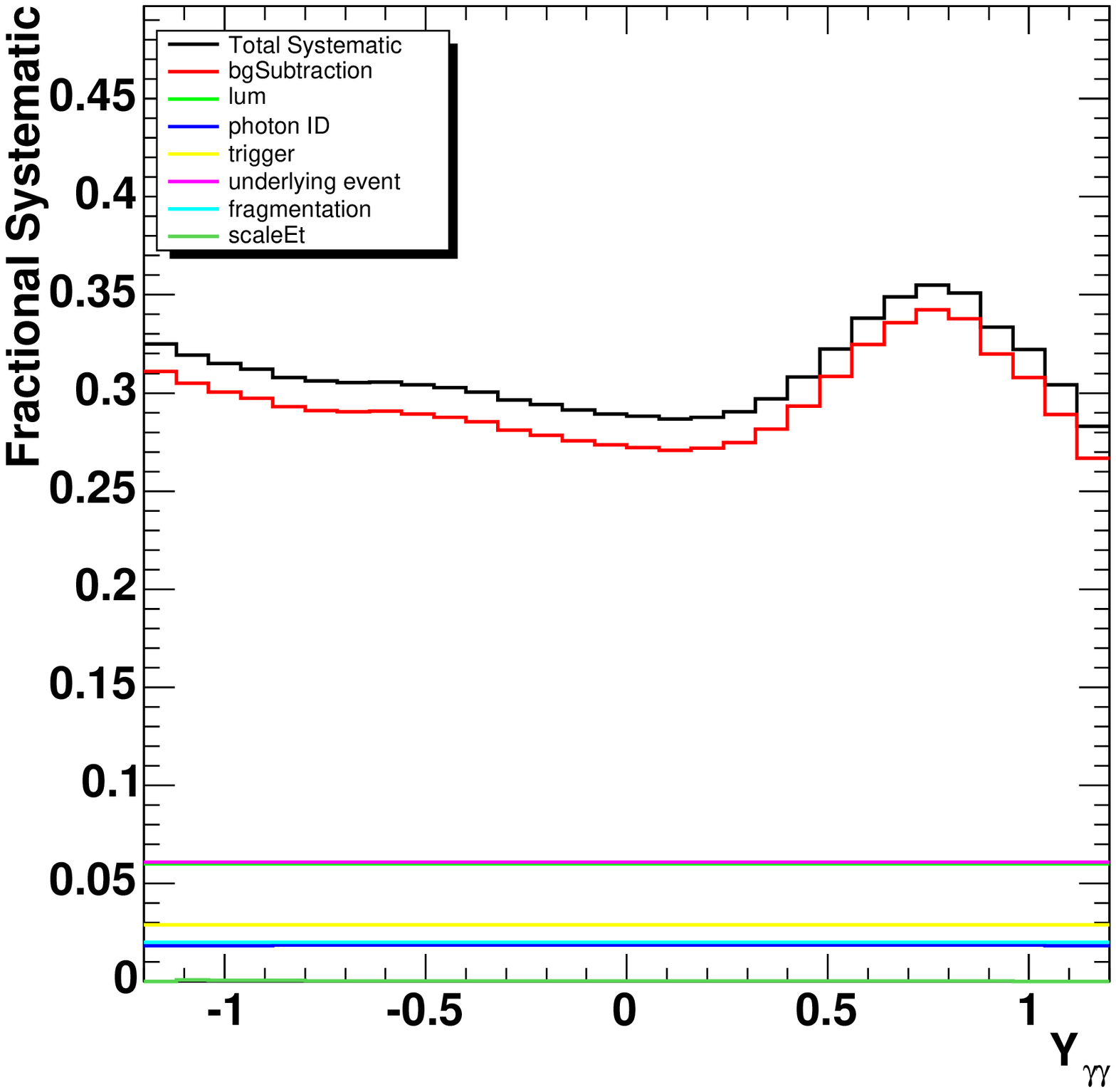}
\includegraphics[width=0.4\linewidth]{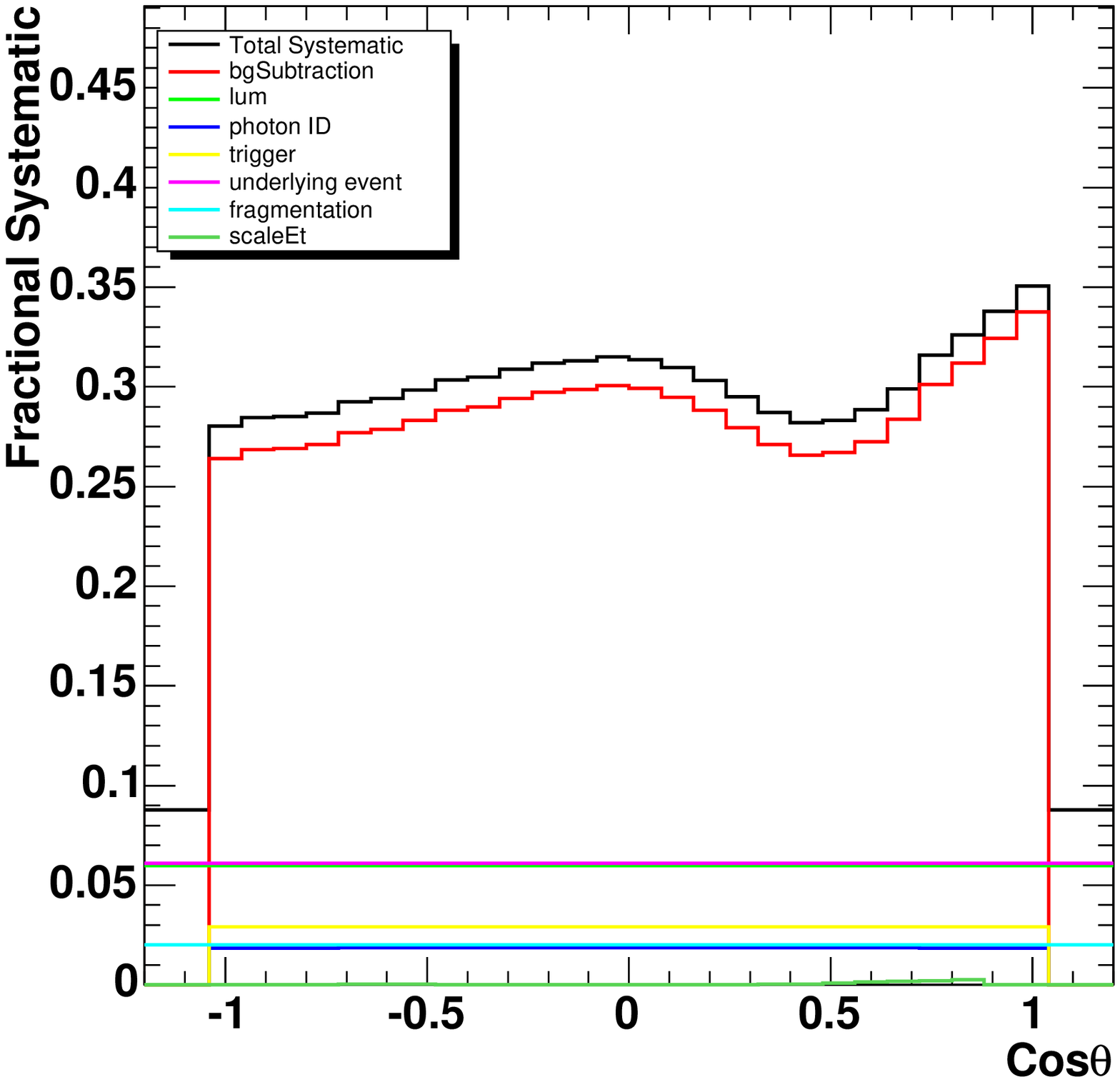}
\includegraphics[width=0.4\linewidth]{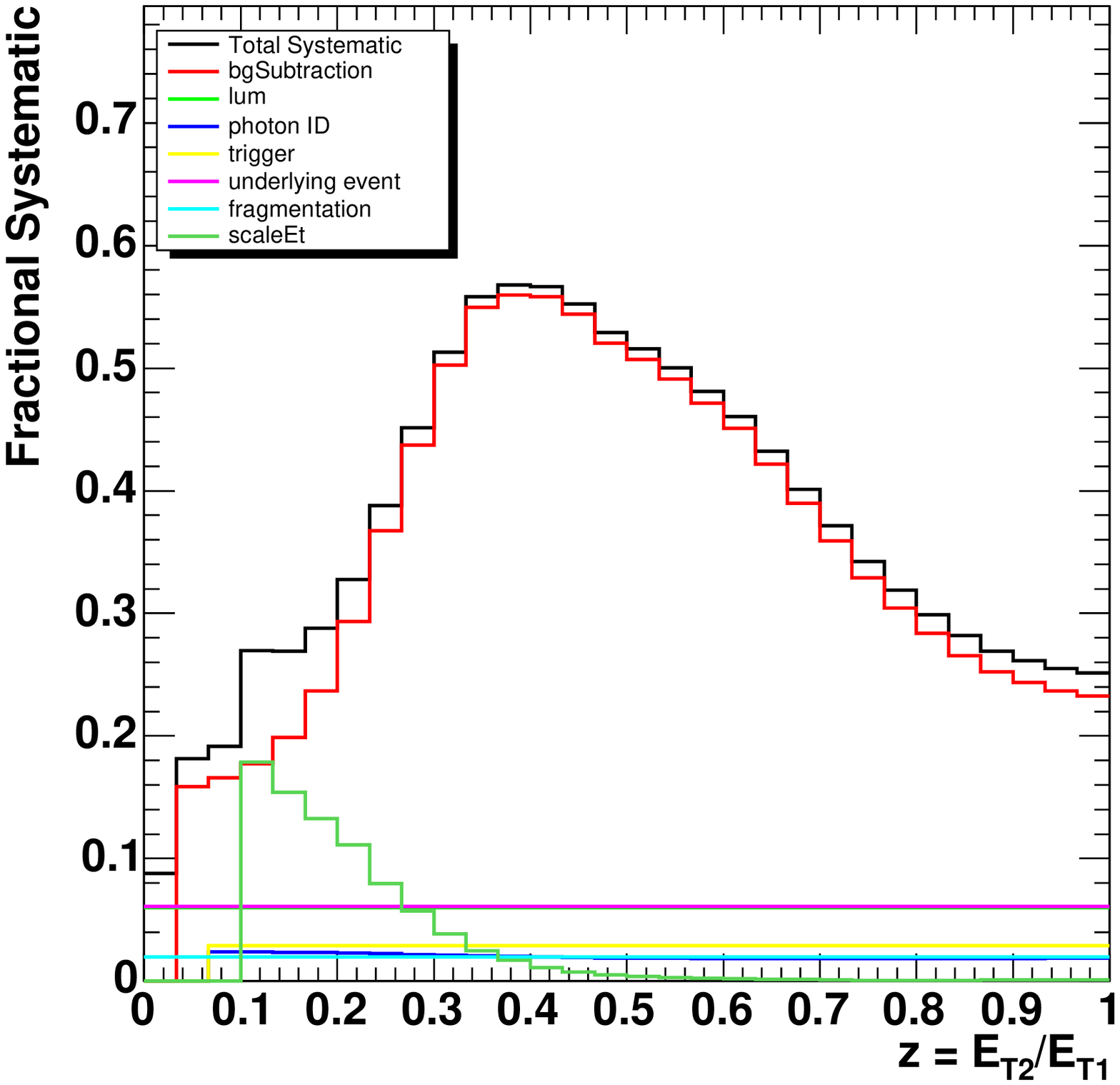}
\caption{The estimated systematic uncertainties in the cross section as a
   function of several kinematic variables.}
\label{fig:syst}
\end{figure*}

In addition to the total systematic uncertainty arising from the background
subtraction, whose details are discussed in Subsection \ref{subsec:bkg-sub},
the following systematic uncertainties are included in the cross section
measurement.
\begin{itemize}
\item The $Z^0$-based efficiency correction has an uncertainty from several
sources, including uncertainty in the amount of material leading to 
conversion events, which are rejected, and the difference between the 
electron and photon response to cuts. These are summarized as 1.8\% below
$E_{\rm T}=40$ GeV, rising linearly to 3\% at 80 GeV and fixed above that point.
This increase completely covers the $E_{\rm T}$--dependence in the photon ID
efficiency mentioned above.
\item The photon energy scale is varied and the change in the kinematic
distribution is reported as an uncertainty. For the diphoton mass, the
variation is 0 at $E_{\rm T}=40$ GeV, rising linearly up to 1.5\% at 80 GeV,
then fixed above 80 GeV. These uncertainties are based on energy scale
studies in the inclusive photon cross section measurement \cite{cdf9590}.
\item A 3\% uncertainty due to trigger efficiency is taken from Ref.
\cite{cdf6312}.
\item A 6\% uncertainty (3\% per photon) for underlying event correction
is taken from Ref. \cite{cdf9590}.
\item No uncertainty in the acceptance from variations in the ISR/FSR
model is included since the primary mechanism for the effect is extra
jets interfering with isolation. Both the numerator and denominator photons
in the efficiency calculation are isolated, therefore the efficiency is
immune to this effect, to leading order.
\item No uncertainty in the efficiency due to the choice of the $Q^2$
scale is included because the primary mechanism of this effect is through
the boosting of the final state. Since the efficiency's numerator and
denominator are calculated with full kinematic requirements, the efficiency
is immune to this effect to leading order.
\end{itemize}
The breakdown of the systematic uncertainties for the kinemaric quantities
defined in Subsection \ref{subsec:kin-var} is shown in Fig. \ref{fig:syst}.
In all distributions, the dominant uncertainty comes from the background
subtraction. The total systematic uncertainty is obtained by adding all
individual components quadratically and averages near 30$\%$.

\subsection{Corrections and Tests with the $Z^0\rightarrow e^+e^-$ Sample}
\label{subsec:corr}

The $Z^0\rightarrow e^+e^-$ sample is used in this analysis for two purposes: 
1) to set the energy scale in the CEM calorimeter and 2) to check the overall
cross section normalization. The $Z^0\rightarrow e^+e^-$ data sample is
derived from the diphoton trigger dataset. The same global event selection
as for the diphoton sample is applied. Two objects are required to pass a
``photon-like electron'' selection. The cuts applied are those of the
standard photon selection, with modifications to allow for the electron
track. The modifications are:
\begin{itemize}
\item The number of allowed tracks in the cluster is increased by 1.
\item The leading track $p_{\rm T}$ cut is applied on the second-highest
$p_{\rm T}$ track instead of the highest one.
\item The track isolation is corrected by subtracting the leading track
$p_{\rm T}$.
\item $0.8 < E/p < 1.2$ is required for the energy-to-momentum ratio of the
leading track. (Events which fail this cut also tend to fail the CES $\chi^2$
cut.)
\end{itemize}
The electrons are required to match EM objects passing the Level-1 and
Level-2 trigger criteria (or trigger simulation for the MC). 
A high--luminosity sample of fully simulated and reconstructed {\sc pythia}
$Z^0\rightarrow e^+e^-$ events is used for MC.

The electromagnetic energy scale is set by tuning the reconstructed $Z^0$
mass to the world average \cite{PDG} in both the data and the MC samples.
The correction is applied as a function of time. It is applied before final
event selection to account for a few events slightly below the energy
threshold which the correction pushes above the threshold. The correction
can only have a noticable effect on kinematic variables with rapidly falling
spectra, related to the photon $E_{\rm T}$, such as the diphoton mass. A 1.5\%
systematic uncertainty due to energy scale is included, as mentioned in
Subsection \ref{subsec:effi}.

The $Z^0\rightarrow e^+e^-$ cross section is measured in order to check the
cross section measurement procedures. This measurement tests the trigger
efficiency, the ability for the MC to predict the event selection efficiency,
the efficiency corrections, and the luminosity. The cross section is measured
for events with $e^+e^-$ invariant mass between 65 and 115 GeV/$c^2$. 
The ``photon-like electron'' selection is applied and the efficiency from the
{\sc pythia} MC is used. The same photon efficiency corrections as in the
diphoton cross section (see Subsection \ref{subsec:effi}) are applied. Since
the photon--like electron cuts are used, it is assumed that the response
is similar to photon response. The resulting $Z^0\rightarrow e^+e^-$ 
cross section is found to be consistent with previous dedicated measurements
and expectation from theory \cite{dy}.

\section{Results}
\label{sec:results}

This Section presents the results of the cross section measurement.
A brief description of the theoretical calculations is given first,
then the comparisons for selected kinematic variables are shown and
discussed. Tables with the measured cross section values are given
in Appendix \ref{sec:xsec-tab}.

\subsection{Theoretical Calculations}
\label{subsec:models}

\begin{figure}[!ht]
  \centering
    \includegraphics[width=1.0\linewidth]
                    {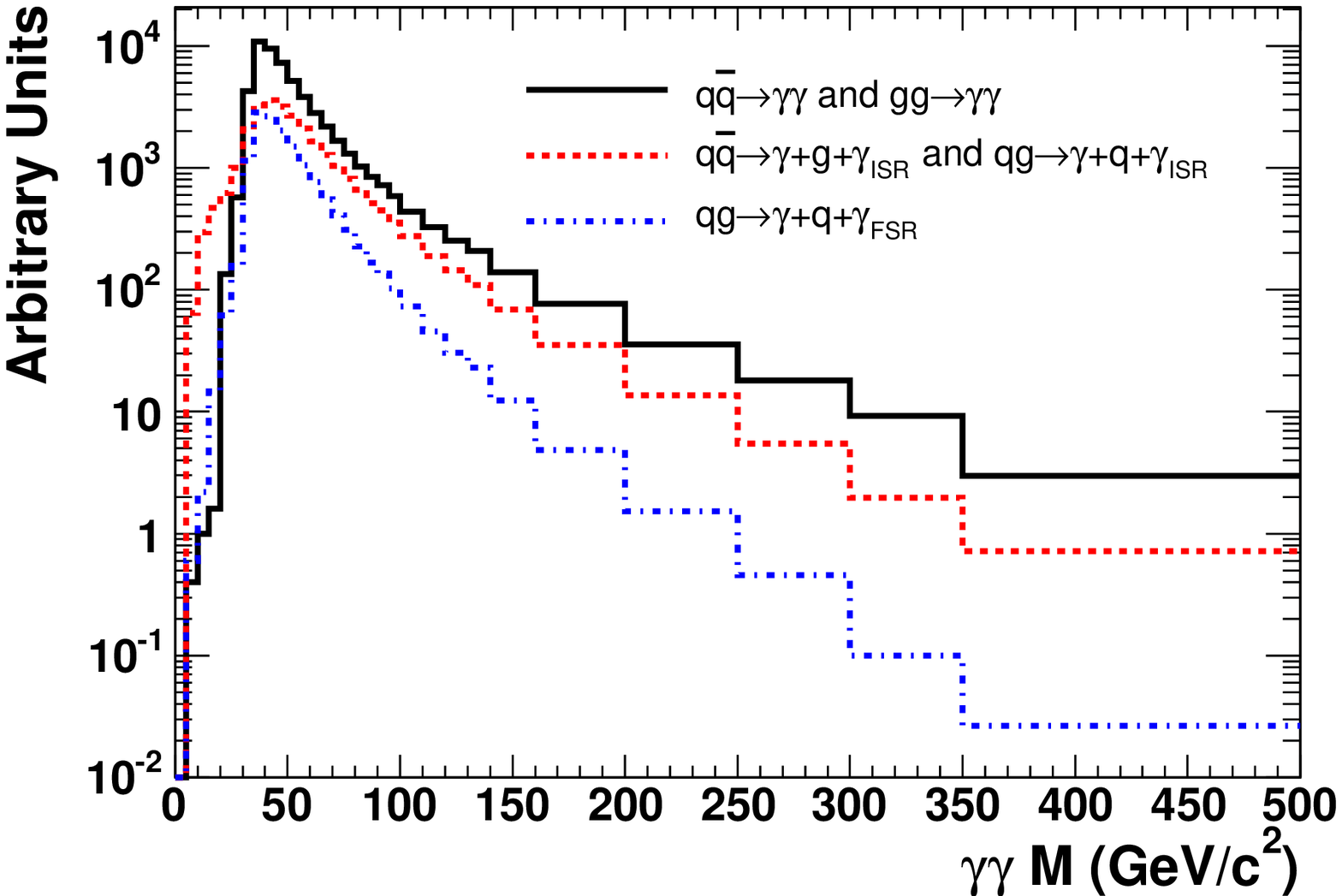}
    \includegraphics[width=1.0\linewidth]
                    {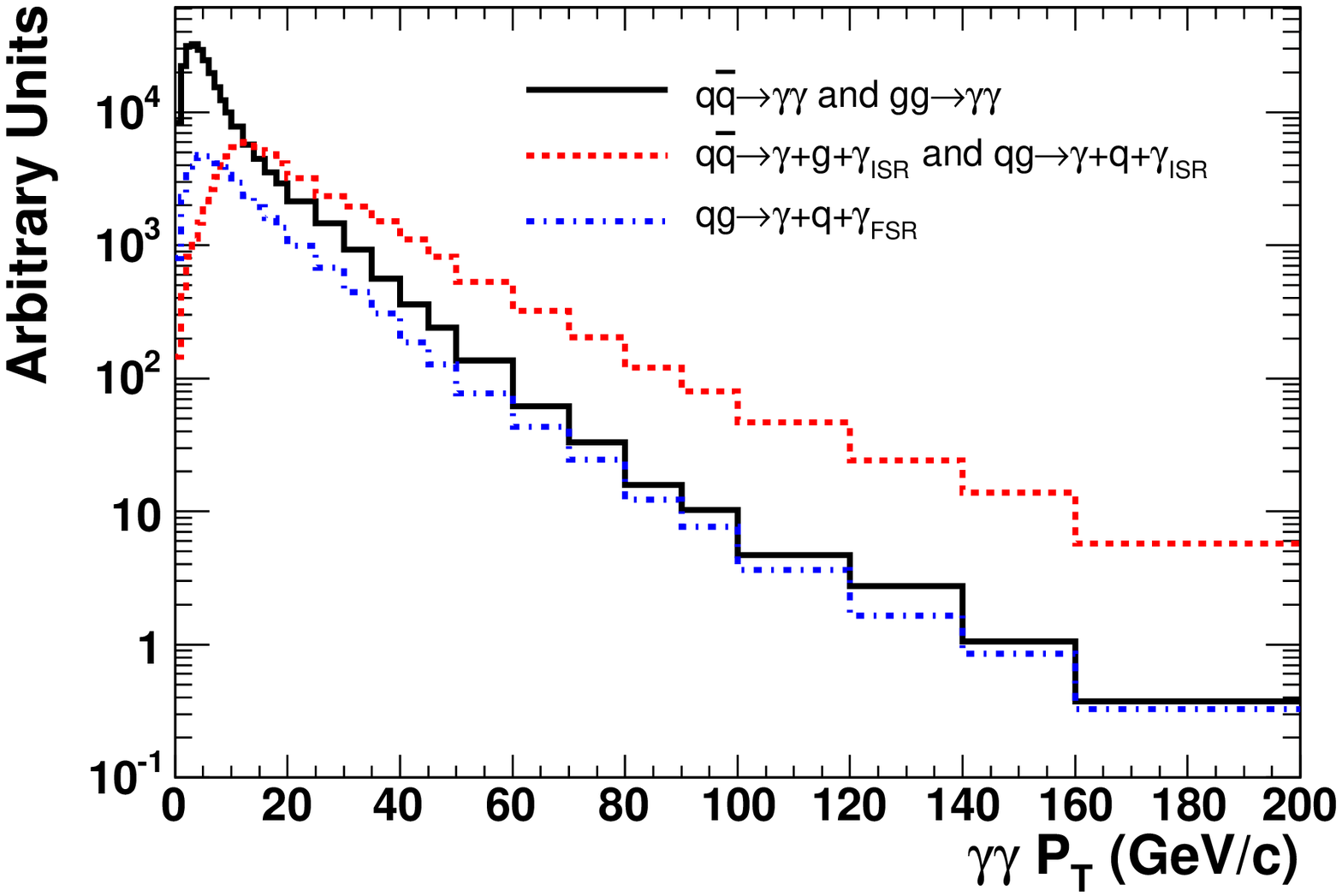}
    \includegraphics[width=1.0\linewidth]
                    {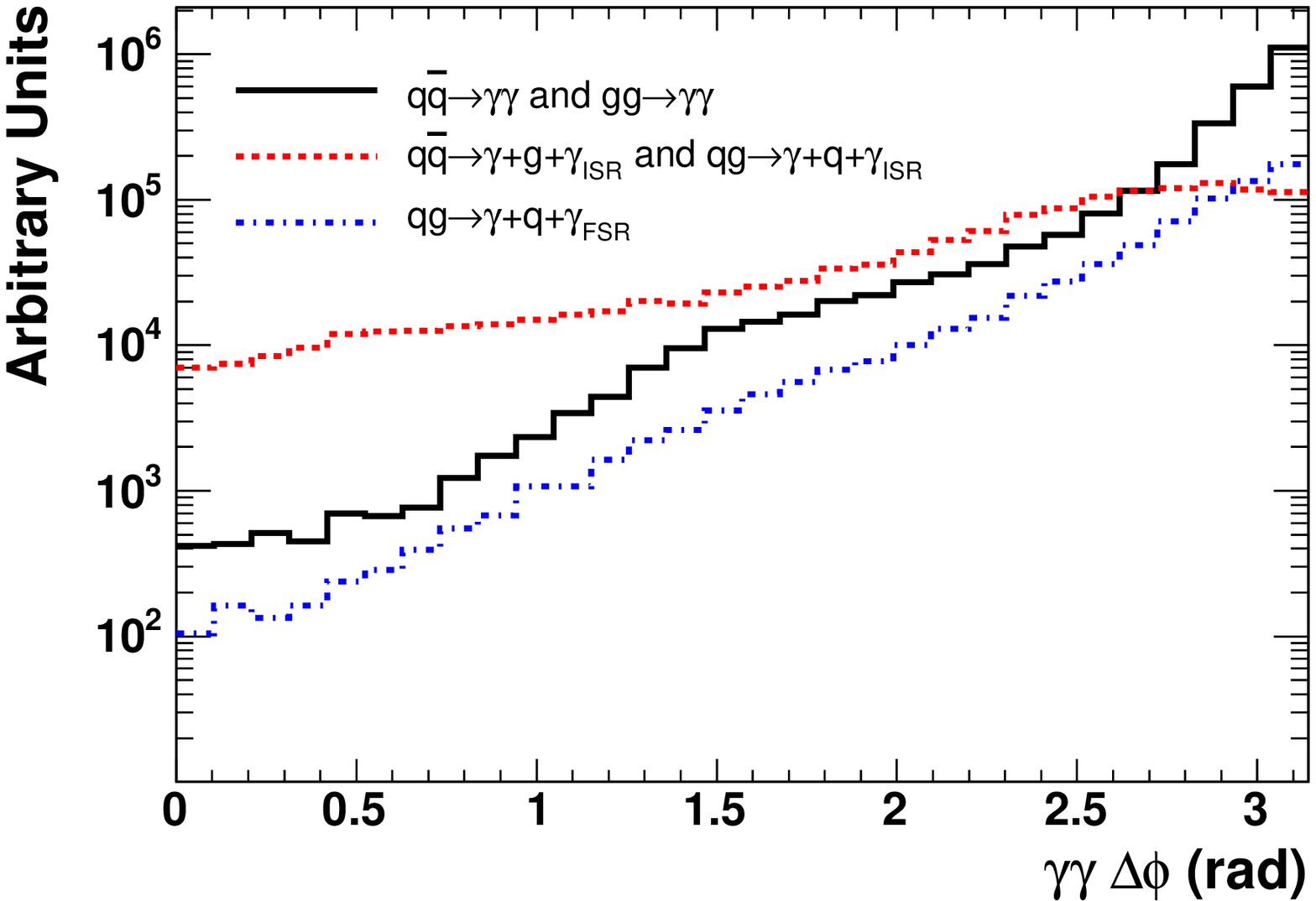}
    \caption{The individual contributions to the cross section from events
             where both photons are generated according to the {\sc pythia}
             diphoton matrix element and from events where one photon
             originates from initial or final state radiation, as functions
             of the diphoton mass (top), transverse momentum (middle) and
             azimuthal difference (bottom).}
\label{fig:pythia_breakdown}
\end{figure}

\begin{figure}[!ht]
  \centering
    \includegraphics[width=1.0\linewidth]{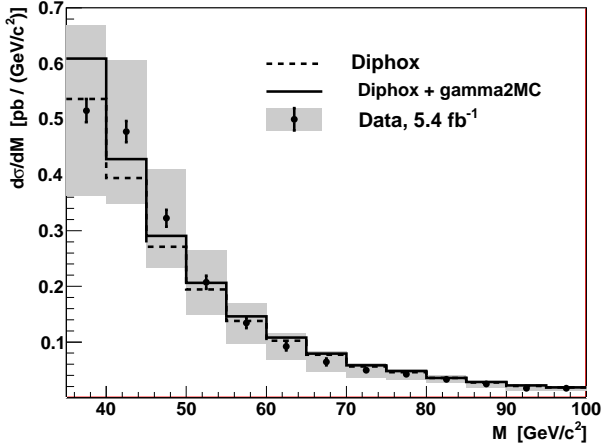}
    \caption{The measured cross section as a function of the diphoton mass
             in comparison with the {\sc diphox} predictions without and with
             the NLO $gg\rightarrow\gamma\gamma$ correction, calculated
             by the {\sc gamma2mc} program. The shaded area is the total
             systematic uncertainty in the data.}
\label{fig:diphox_g2mc}
\end{figure}

The results of this measurement are compared with three theoretical
predictions:
\begin{itemize}
\item A calculation using the {\sc pythia} program \cite{pythia}. This is a
parton--showering generator which features a realistic representation of the
physics events in terms of observable particles. It includes initial and final
state radiation and an underlying event model. {\sc pythia} implements a
Leading-Order (LO) matrix element (ME) for direct diphoton production which
includes the $q\bar q\rightarrow\gamma\gamma$ and $gg\rightarrow\gamma\gamma$
LO processes described, respectively, by diagrams (a) and (b) of Fig.
\ref{fig:diagrams}. Significant contributions also arise from the processes
$q\bar q\rightarrow\gamma\gamma g$ (diagrams (c) of Fig. \ref{fig:diagrams})
and $gq\rightarrow\gamma\gamma q$ (diagrams (d) of Fig. \ref{fig:diagrams})
where the second photon is emitted from an initial or final state quark
according to the {\sc pythia} radiation model. These contributions were
included in the calculation by running the program with a filter selecting
diphoton events from inclusive $\gamma$$+$X events, where X is either a photon
or a jet, with an efficiency of 0.025$\%$. Fig. \ref{fig:pythia_breakdown}
shows the individual contributions to the cross section as a function of the
diphoton invariant mass, transverse momentum and azimuthal difference.
Initial-state radiation (ISR) photons, in particular, produce substantially
different distributions than ME and final-state radiation (FSR) photons, having
a harder transverse momentum spectrum and stronger low--$\Delta\phi$ tail in
the azimuthal difference spectrum. In leading order, this can be attributed
to the fact that FSR occurs in quark-gluon scattering [diagram (d) of Fig.
\ref{fig:diagrams}], whereas ISR occurs both in $q\bar q$ annihilation
[diagram (c) of Fig. \ref{fig:diagrams}] and quark-gluon scattering, and
the luminosity of quark-gluon states falls off more rapidly with the parton
momenta than the luminosity of $q\bar q$ states \cite{diphox,resbos}. The
diphoton ME contributes 56$\%$ to the cross section, the processes
$q\bar q\rightarrow g\gamma\gamma_{_{\rm ISR}}$ and
$gq\rightarrow q\gamma\gamma_{_{\rm ISR}}$ 29$\%$, and the process
$gq\rightarrow q\gamma\gamma_{_{\rm FSR}}$ 15$\%$. Double radiation processes
in minimum bias dijet events, such as
$qq\rightarrow qq\gamma_{_{\rm ISR/FSR}}\gamma_{_{\rm ISR/FSR}}$,
$q\bar q\rightarrow q\bar q\gamma_{_{\rm ISR/FSR}}\gamma_{_{\rm ISR/FSR}}$,
$gq\rightarrow gq\gamma_{_{\rm ISR/FSR}}\gamma_{_{\rm ISR/FSR}}$,
$q\bar q\rightarrow gg\gamma_{_{\rm ISR}}\gamma_{_{\rm ISR}}$ and
$gg\rightarrow q\bar q\gamma_{_{\rm FSR}}\gamma_{_{\rm FSR}}$, were also examined
but their overall contribution was estimated to only $\sim 3\%$ of the total,
having no significant effect to any kinematical distribution. Therefore,
these processes were not included in the {\sc pythia} calculation.
\item A fixed Next-to-Leading-Order (NLO) calculation using the {\sc diphox}
program \cite{diphox}. This generator explictly includes parton fragmentation
into photons \cite{frag}, {\it i.e.} processes in which nearly all the energy
of a parton is transformed into a photon. Direct production contributes 85$\%$
to the cross section and fragmentation 15$\%$. The {\sc diphox} matrix element
accounts for the $q\bar q\rightarrow\gamma\gamma$ and
$gq\rightarrow\gamma\gamma q$ processes up to NLO, and LO for the
$gg\rightarrow\gamma\gamma$ process, since this is already a second order
process in the strong coupling. The NLO $gg\rightarrow\gamma\gamma$
contributions were examined with the {\sc gamma2mc} program \cite{g2mc}.
Fig. \ref{fig:diphox_g2mc} shows an example of the uncorrected and corrected
{\sc diphox} predictions in comparison with the measured cross section as a
function of the diphoton invariant mass $M$. The corrected prediction is
calculated by running {\sc diphox} without the LO $gg\rightarrow\gamma\gamma$
term and then adding the full LO+NLO $gg\rightarrow\gamma\gamma$ calculation
from {\sc gamma2mc} incoherently, since the initial state is different in
gluon fusion than in the other processes. The correction of the total cross
section for the NLO $gg\rightarrow\gamma\gamma$ contribution is nearly
10$\%$, which is comparable with the experimental and theoretical
uncertainties (see Table \ref{tab:tot-xsec}). Therefore, this correction was
not applied to the {\sc diphox} calculation.
\item A resummed NLO calculation using the {\sc resbos} program \cite{resbos}.
Here the effects of soft gluon ISR in the NLO calculation are analytically
resummed to all orders in the strong coupling and reach NNLL accuracy. The
resultant prediction is smoothly matched to the fixed-order NLO result in the
kinematic regions where the NLO matrix element is dominant. The {\sc resbos}
matrix element includes the $q\bar q\rightarrow\gamma\gamma$,
$gq\rightarrow\gamma\gamma q$ and $gg\rightarrow\gamma\gamma$ processes up to
NLO and it is adjusted so as to approximately account for fragmentation.
\end{itemize}
All calculations are done by Monte Carlo event generation and are subject to
the experimental kinematic and isolation cuts. In the fixed-order NLO
calculations the isolation cut is applied on parton variables and thus it
only approximates the isolation cut appplied in the data and in {\sc pythia}.
The {\sc resbos} predictions are restricted in the diphoton invariant mass
$M$ range from $2m_{b}=9$ GeV/$c^2$ to $2m_{t}=350$ GeV/$c^2$ and they are
shown up to $M=300$ GeV/$c^2$ in the plots of the mass distribution, where
$m_{b}$ and $m_{t}$ are the masses of the bottom and top quarks, respectively.

NLO theoretical uncertainties are estimated for the choice of scale,
representing the sensitivity to missing higher order terms, and for the PDFs.
In {\sc diphox} the default renormalization, factorization and fragmentation
scales are all set to $\mu=M/2$. In {\sc resbos} the default renormalization
and factorization scales are both set to $\mu=M$. In either case, all scales
are varied by a factor of 2 up and down relative to the default choice and
this is taken as a conservative estimate of the total scale uncertainty.
The proton PDF set is the  CTEQ6.1M set \cite{cteq6m} for both {\sc diphox}
and {\sc resbos}. The corresponding uncertainty is estimated by varying the
generated event weights within the 90\% level uncertainties given by the 20
CTEQ6.1M eigenvectors.

The measured total cross section is shown in Table \ref{tab:tot-xsec} together
with the predictions from the three theoretical calculations. All calculations
are consistent with the size of the measured cross section within the
experimental uncertainties.

\subsection{Kinematic Variables}
\label{subsec:kin-var}

The complete description of the reaction
$h_1+h_2\rightarrow\gamma_1+\gamma_2+X$, where $h_{1,2}$ are hadrons,
requires five independent kinematic variables. A suitable choice consists
of the invariant mass

\begin{equation}
M=\sqrt{2p_{\rm T\gamma 1}p_{\rm T\gamma 2}
[\cosh(y_{\gamma 1}-y_{\gamma 2})-\cos(\phi_{\gamma 1}-\phi_{\gamma 2})]}
\label{eq:mass}
\end{equation}

\noindent
the transverse momentum

\begin{equation}
P_{\rm T}=\sqrt{p_{\rm T\gamma 1}^{2}+p_{\rm T\gamma 2}^{2}
+2p_{\rm T\gamma 1}p_{\rm T\gamma 2}\cos(\phi_{\gamma 1}-\phi_{\gamma 2})}
\label{eq:moment}
\end{equation}

\noindent
the rapidity

\begin{equation}
Y_{\gamma\gamma}=\tanh^{-1}\frac{p_{\rm T\gamma 1}\sinh y_{\gamma 1}+
p_{\rm T\gamma 2}\sinh y_{\gamma 2}}{p_{\rm T\gamma 1}\cosh y_{\gamma 1}+
p_{\rm T\gamma 2}\cosh y_{\gamma 2}}
\label{eq:rapid}
\end{equation}

\noindent
the azimuthal difference

\begin{equation}
\Delta\phi=|\phi_{\gamma 1}-\phi_{\gamma 2}|\mod\pi
\label{eq:azimuth}
\end{equation}

\noindent
of the photon pair in the laboratory frame \cite{eta}, and the cosine of
the polar angle $\theta$ of the 1$^{\rm st}$ photon in the Collins-Soper
frame \cite{CS}. This is defined as the rest frame of the photon pair 
chosen so that (a) the 3-momenta $\vec{p}_{h_1}$ and $\vec{p}_{h_2}$ of the
initial hadrons lie in the $Oxz$ plane (with positive $x$) and (b) the $z$
axis bisects the angle between $\vec{p}_{h_1}$ and $-\vec{p}_{h_2}$. This
variable is generally determined by \cite{CS}

\begin{equation}
\cos\theta=\frac{2p_{\rm T\gamma 1}p_{\rm T\gamma 2}\sinh(y_{\gamma 1}-y_{\gamma 2})}
{M\sqrt{M^{2}+P_{\rm T}^{2}}}
\label{eq:cos}
\end{equation}

\noindent
For photons emitted at large angles with respect to the beam,
$\cos\theta\approx\tanh[(y_{\gamma 1}-y_{\gamma 2})/2]$ in the limit
$P_{\rm T}\rightarrow 0$. In the above equations $p_{\rm T\gamma i}$,
$y_{\gamma i}$ and $\phi_{\gamma i}$ are the transverse momentum, rapidity
and azimuth of photon $i$, respectively, with $i$=1, 2.

The set of $\{M,P_{\rm T},Y_{\gamma\gamma}\}$ describes the kinematics of the
diphoton system and, therefore, of possible heavy particles decaying into
a photon pair, such as a Higgs boson. The existence of such a particle would
manifest as a peak in the distribution of the invariant mass $M$. The
results of this analysis are presented in the form of cross sections
differential in each of the five kinematic variables
$\{M,P_{\rm T},\Delta\phi,Y_{\gamma\gamma},\cos\theta\}$ and in the variable
$z=p_{\rm T\gamma 2}/p_{\rm T\gamma 1}$, the ratio of sub-leading to leading
photon transverse momentum (0$\leq$$z$$\leq$1). Three kinematic cases are
examined:
\begin{itemize}
\item Differential cross sections without additional kinematic cuts. No
kinematic cut other than those listed in Table \ref{tab:MCaccept} is applied.
The results of this case are presented in Subsection \ref{subsec:uncon-kin}.
\item Differential cross sections for $P_{\rm T}$$<$$M$. The kinematics in this
case are similar to the diphoton decay of a heavy particle, such as a Higgs
boson, produced in events of moderate parton activity. At the Tevatron, prompt
photon pairs are almost entirely produced in this case by low-$P_{\rm T}$
quark-antiquark annihilation. The results of this case are presented in
Subsection \ref{subsec:higgs-kin}.
\item Differential cross sections for $P_{\rm T}$$>$$M$. The importance of
high-$P_{\rm T}$ contributions from gluon-gluon fusion, fragmentations and ISR
is enhanced in this case. The results of this case are presented in Subsection
\ref{subsec:box-kin}.
\end{itemize}

\begin{table}[tbp]
\begin{center}
\caption{The total diphoton production cross section obtained from the
         measurement and from the theoretical calculations. The {\sc pythia}
         $\gamma\gamma$ calculation involves only the
         $q\bar q\rightarrow\gamma\gamma$ and $gg\rightarrow\gamma\gamma$
         processes. The {\sc pythia} $\gamma\gamma +\gamma{\rm j}$ calculation
         includes also the $q\bar q\rightarrow\gamma\gamma g$ and
         $gq\rightarrow\gamma\gamma q$ processes.}
\label{tab:tot-xsec}
\begin{tabular}{lc}
\hline\hline
 & Cross section (pb) \\ \hline
Data & $12.47\pm 0.21_{\rm stat}\pm 3.74_{\rm syst}$ \\
{\sc resbos} & $11.31\pm 2.45_{\rm syst}$ \\
{\sc diphox} & $10.58\pm 0.55_{\rm syst}$ \\
{\sc pythia} $\gamma\gamma +\gamma{\rm j}$ & $9.19$ \\
{\sc pythia} $\gamma\gamma$ & $5.03$ \\
\hline\hline
\end{tabular}
\end{center}
\end{table}

The oveflow data entries are excluded from the $M$ and $P_{\rm T}$ histograms,
to keep the cross section definition consistent for the data and the theories
at the highest bins. For each kinematic variable, the following plots are
presented:
\begin{itemize}
\item The measured and calculated cross sections as functions of the
selected variable. Each of these plots includes the predictions of all three
calculations, for comparison, and shows only the uncertainties of the data.
The prediction of the {\sc pythia} $\gamma\gamma$ calculation, involving
only the $q\bar q\rightarrow\gamma\gamma$ and $gg\rightarrow\gamma\gamma$
processes, is also shown in these plots, to be compared with the {\sc pythia}
$\gamma\gamma +\gamma{\rm j}$ calculation (j$=$jet), which includes also the
$q\bar q\rightarrow\gamma\gamma g$ and $gq\rightarrow\gamma\gamma q$
processes.
\item The relative deviations of the data from each calculation, in the form
(data$-$theory)/theory, as functions of the selected variable. These plots
show the comparison of the data with each calculation separately and include
the uncertainties of the NLO predictions. No relative deviations are shown
for the {\sc pythia} $\gamma\gamma$ calculation. The benchmark
parton showering MC calculation, compared in detail with the data, is
{\sc pythia} $\gamma\gamma +\gamma{\rm j}$.
\end{itemize}

\subsection{Differential Cross Sections Without Additional Kinematic Cut}
\label{subsec:uncon-kin}

\begin{figure*}[!ht]
\centering
\includegraphics[width=0.45\linewidth]{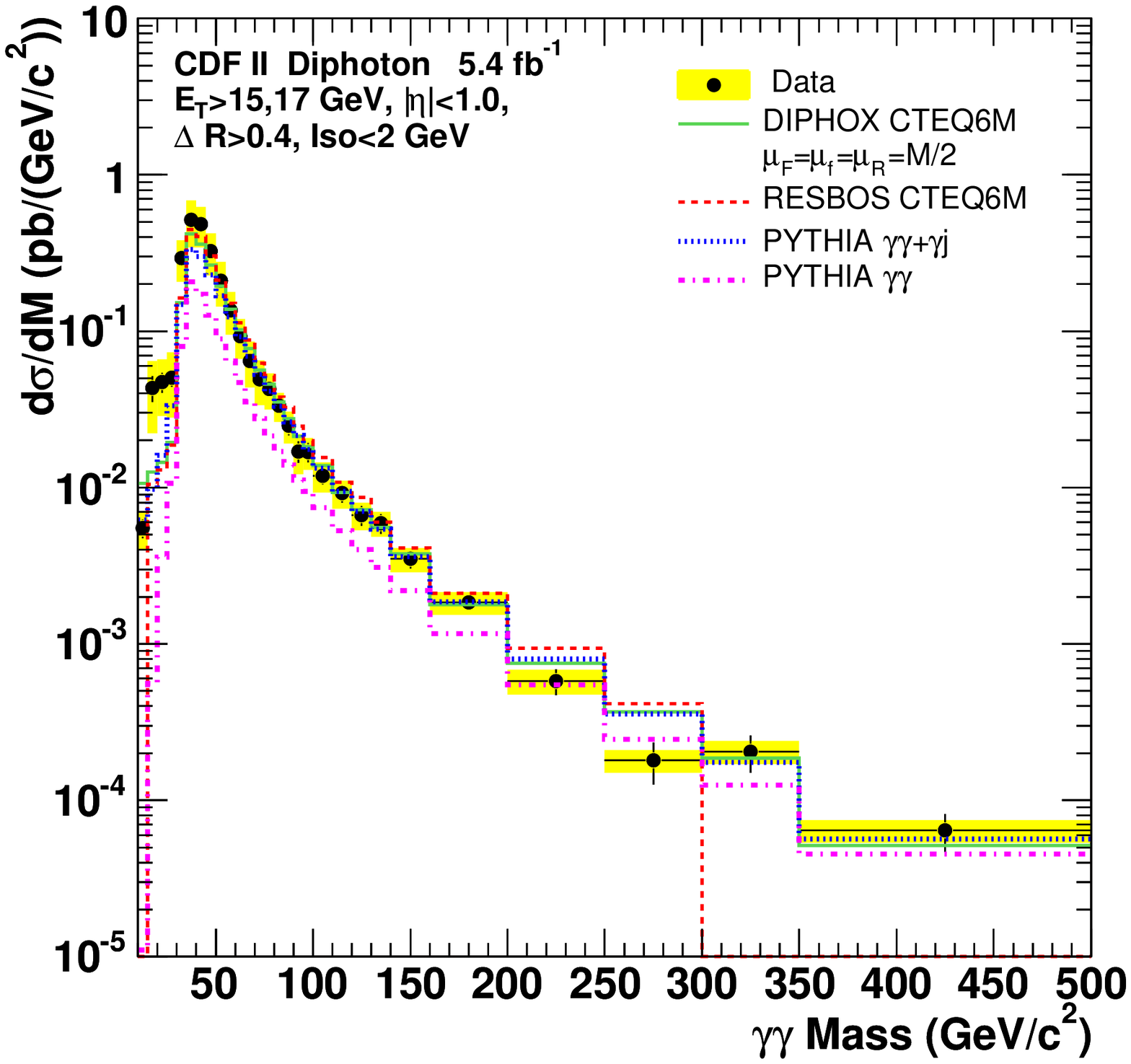}
\includegraphics[width=0.45\linewidth]{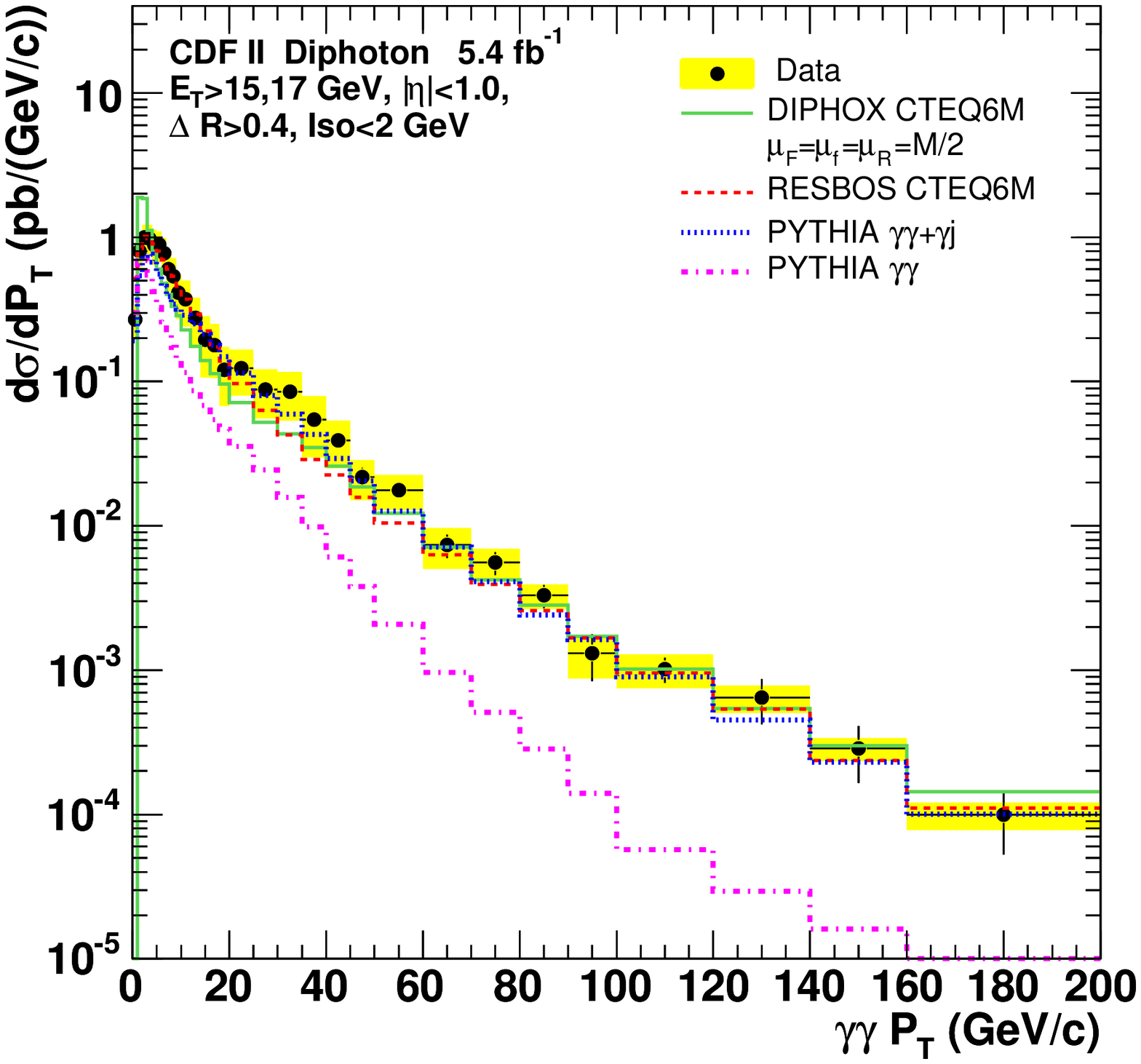}
\includegraphics[width=0.45\linewidth]
{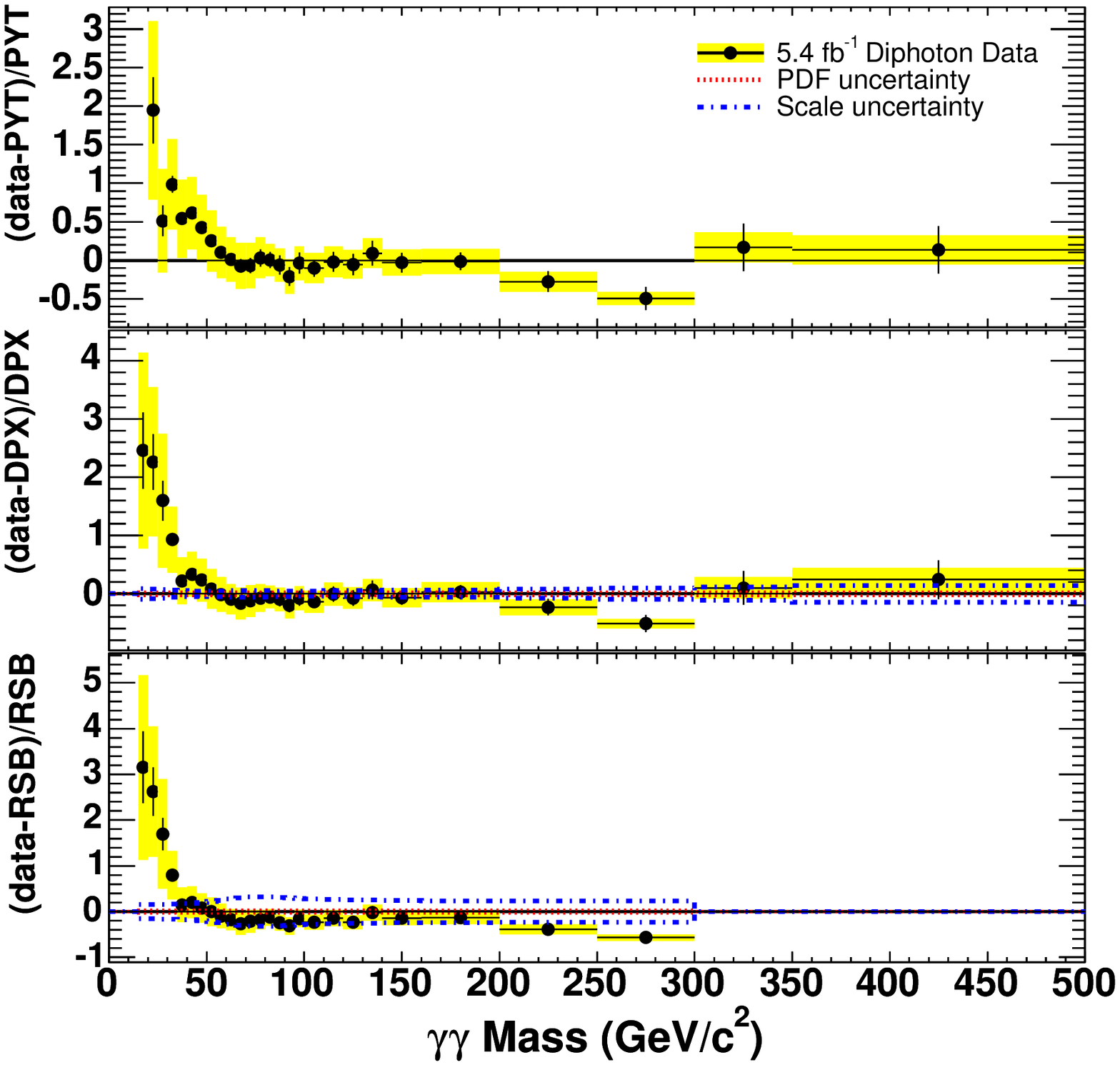}
\includegraphics[width=0.45\linewidth]{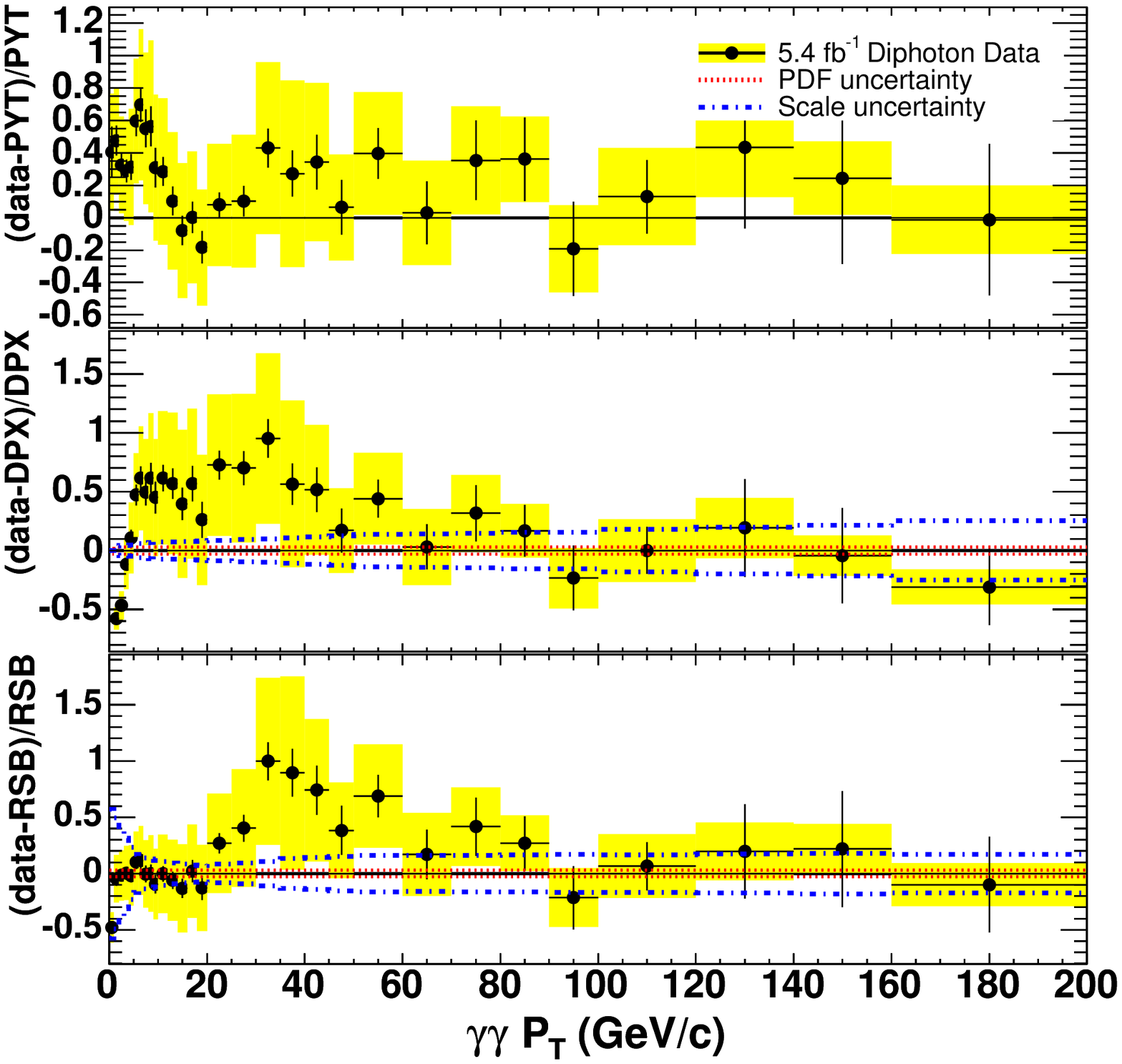}
\caption{The cross section as a function of the diphoton invariant mass (left)
     and transverse momentum (right). {\bf Top:} the absolute cross section
     values. {\bf Bottom:} the relative deviations of the data from the
     predictions. {\it Note:} the vertical axes scales differ between
     relative deviation plots. The shaded area is the total systematic
     uncertainty in the data.}
\label{fig:uncon_kin_xsec_mass}
\end{figure*}

Fig. \ref{fig:uncon_kin_xsec_mass} shows the results for $d\sigma/dM$ and
$d\sigma/dP_{\rm T}$. The mass spectrum peaks at
$M$=$2\sqrt{E_{\rm T1}^{min}E_{\rm T2}^{min}}$$\approx$32 GeV/$c^2$. All three
theoretical predictions for $d\sigma/dM$ are in reasonable agreement
with the data, within uncertainties, except in the region 6
GeV/$c^2$$<$$M$$<$32 GeV/$c^2$ below the mass peak. The low mass limit of
6 GeV/$c^2$ is set by the $\Delta R(\gamma\gamma)$ and $E_{\rm T}^{\gamma}$ cuts.
This region is rich in events coming from gluon scattering and fragmentation.
All three predictions underestimate the data in this region.

The excess of the data over all three predictions for $M$ below the
peak of the mass spectrum is reflected in the region 20
GeV/$c$$<$$P_{\rm T}$$<$50 GeV/$c$ of the $P_{\rm T}$ spectrum, which has a
shoulder around $P_{\rm T}$=$p_{\rm T2}^{min}$$+$$p_{\rm T1}^{min}$=32 GeV/$c$
(the so-called ``Guillet shoulder''). This arises from a collinear enhancement
for the two photons in the fragmentation processes which, however, is
suppressed by the $\Delta R(\gamma\gamma)$ cut. The {\sc resbos} predictions
for $d\sigma/dP_{\rm T}$ are in overall agreement with the data, within
uncertainties, except in this region. The {\sc diphox} prediction
underestimates the data, in addition, for $P_{\rm T}$$<$20 GeV/$c$, where
the resummation effects implemented in {\sc resbos} provide a better
description. The {\sc pythia} prediction underestimates the data at very
low $P_{\rm T}$, $P_{\rm T}$$<$10 GeV/$c$, showing that the LL resummation
of parton showering is less accurate than the NNLL resummation implemented
in {\sc resbos}. The {\sc pythia} prediction is in reasonably good agreement
with the data, within uncertainties, in the rest of the $P_{\rm T}$ range due to
the $q\bar q\rightarrow g\gamma\gamma_{_{\rm ISR}}$ and
$gq\rightarrow q\gamma\gamma_{_{\rm ISR}}$ processes which make the {\sc pythia}
$P_{\rm T}$ spectrum sufficiently hard (see the middle plot of Fig.
\ref{fig:pythia_breakdown}).

\begin{figure*}[!ht]
\centering
\includegraphics[width=0.45\linewidth]{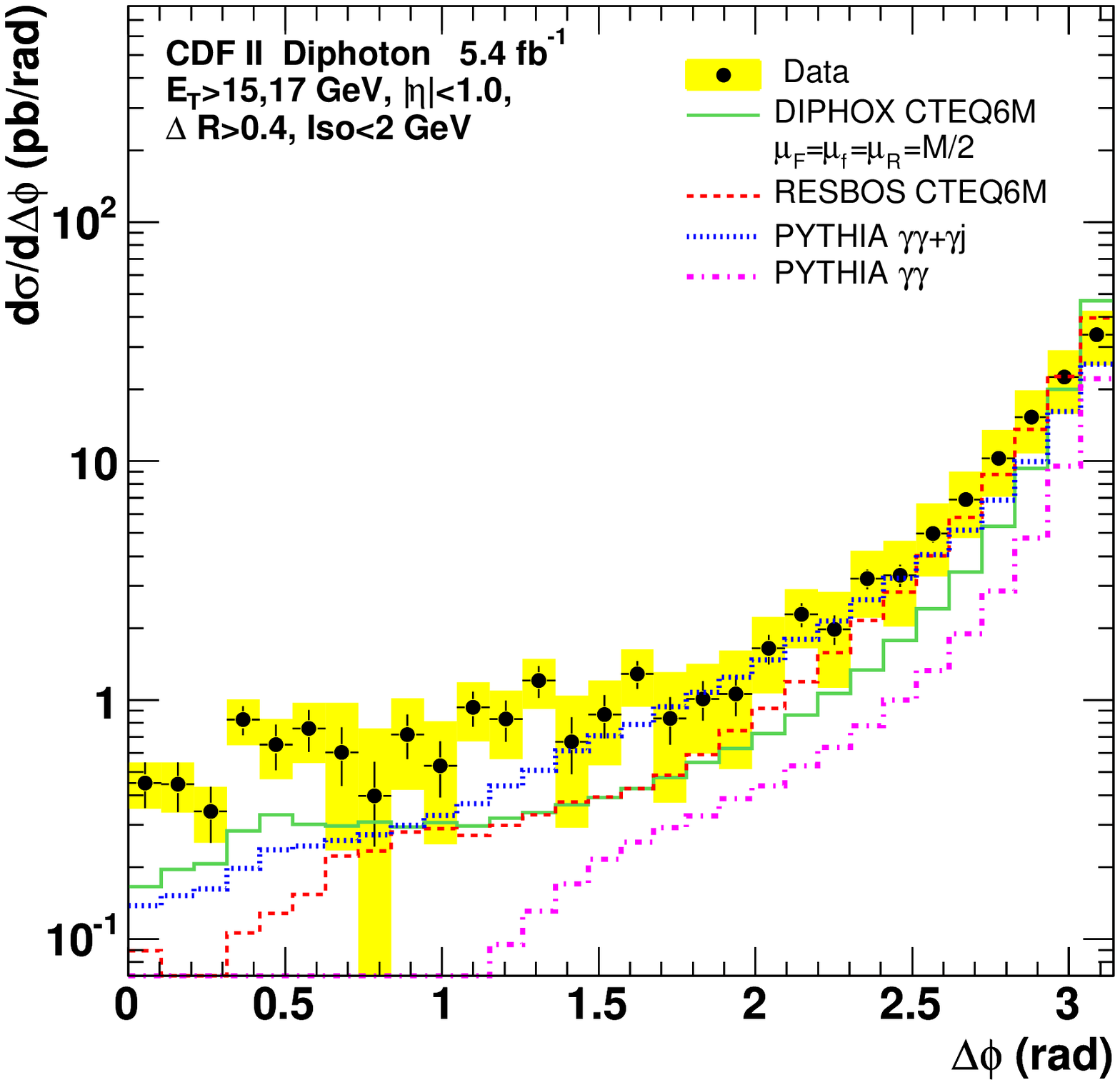}
\includegraphics[width=0.45\linewidth]{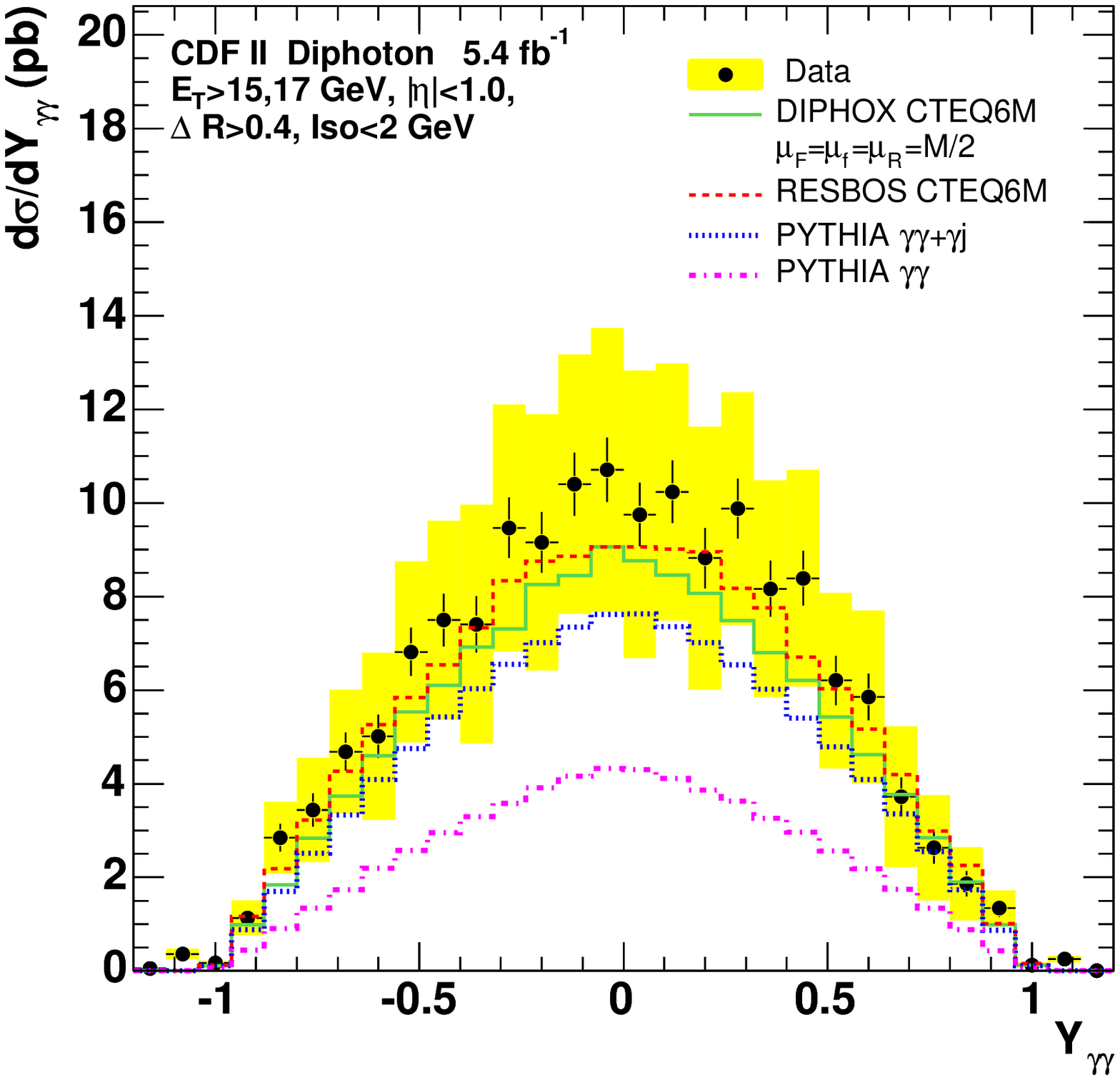}
\includegraphics[width=0.45\linewidth]{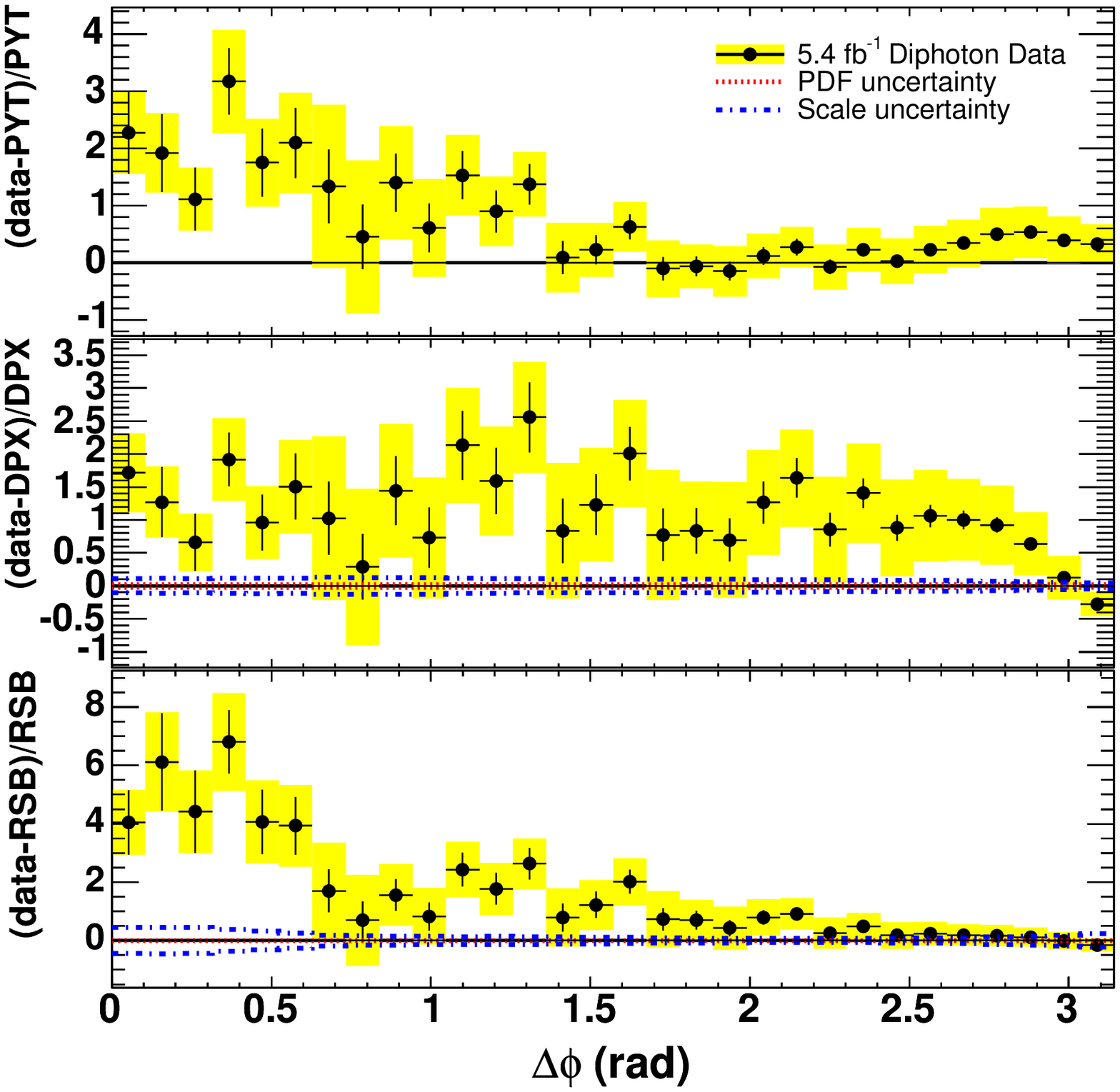}
\includegraphics[width=0.45\linewidth]{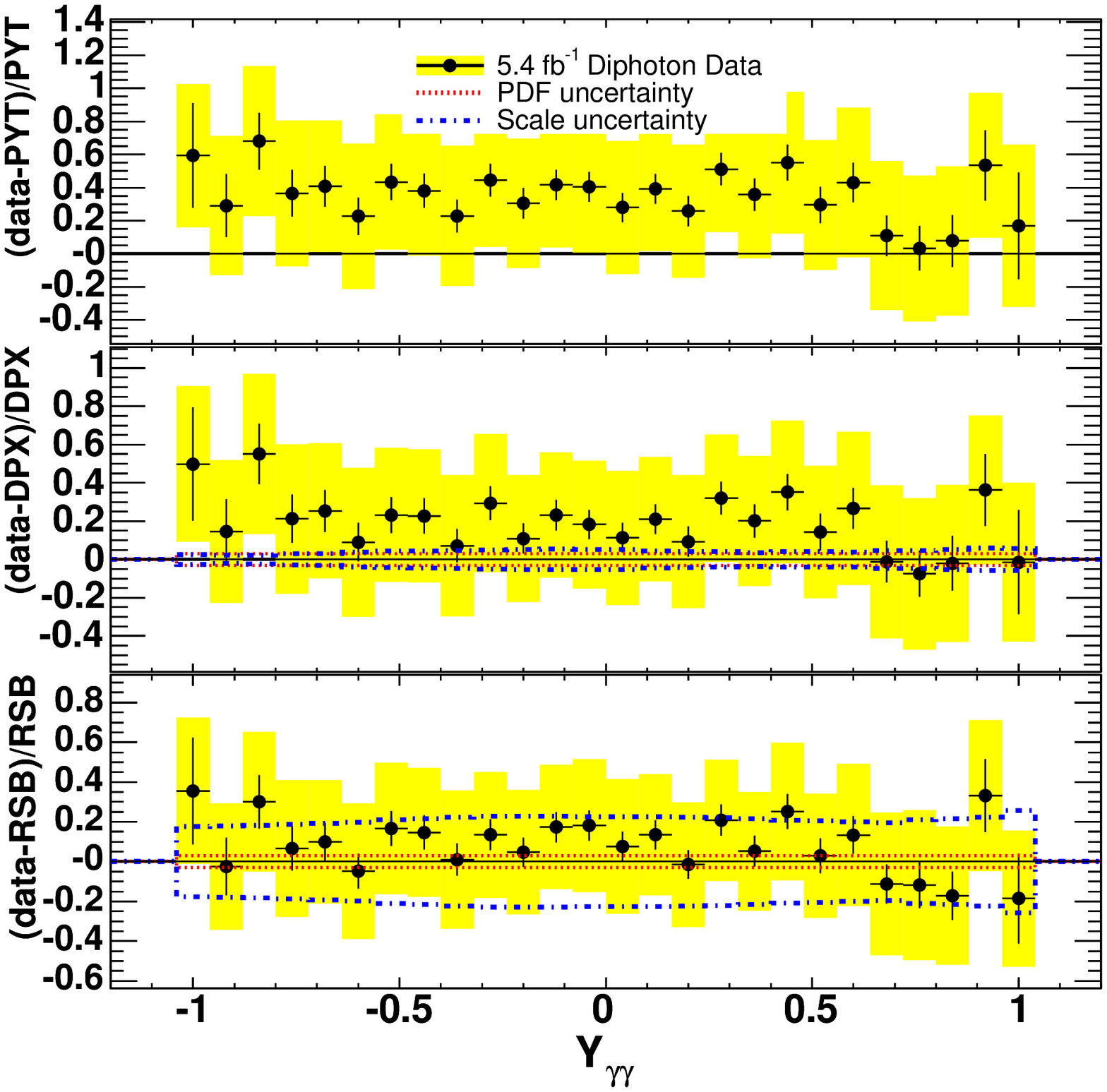}
\caption{The cross section as a function of the diphoton azimuthal distance
     (left) and of the diphoton rapidity (right). {\bf Top:} the absolute
     cross section values. {\bf Bottom:} the relative deviations of the data
     from the predictions. {\it Note:} the vertical axes scales differ between
     relative deviation plots. The shaded area is the total systematic
     uncertainty in the data.}
\label{fig:uncon_kin_xsec_angles}
\end{figure*}

Fig. \ref{fig:uncon_kin_xsec_angles} shows the results for
$d\sigma/d\Delta\phi$ and $d\sigma/dY_{\gamma\gamma}$. The $\Delta\phi$
spectrum peaks at $\Delta\phi$=$\pi$, corresponding to vanishing diphoton
$P_{\rm T}$, and the $Y_{\gamma\gamma}$ spectrum at $Y_{\gamma\gamma}$=0,
corresponding to vanishing diphoton momentum $P_{\rm z}$ parallel to the
proton beam. While all three predictions agree fairly well with the
measured $d\sigma/dY_{\gamma\gamma}$, within uncertainties, all three of
them underestimate the data in the low end of the $\Delta\phi$ spectrum.
This region is dominated by events with low mass and high $P_{\rm T}$.
The {\sc resbos} prediction provides the best description of the measured
$d\sigma/d\Delta\phi$ for $\Delta\phi$$>$2.5 rad, where soft gluon
resummation is important. {\sc pythia} provides the best description in
the region 1 rad$<$$\Delta\phi$$<$2.5 rad where the
$q\bar q\rightarrow g\gamma\gamma_{_{\rm ISR}}$ and
$gq\rightarrow q\gamma\gamma_{_{\rm ISR}}$ processes are the most important
(see the bottom plot of Fig. \ref{fig:pythia_breakdown}).

\begin{figure*}[!ht]
\centering
\includegraphics[width=0.45\linewidth]{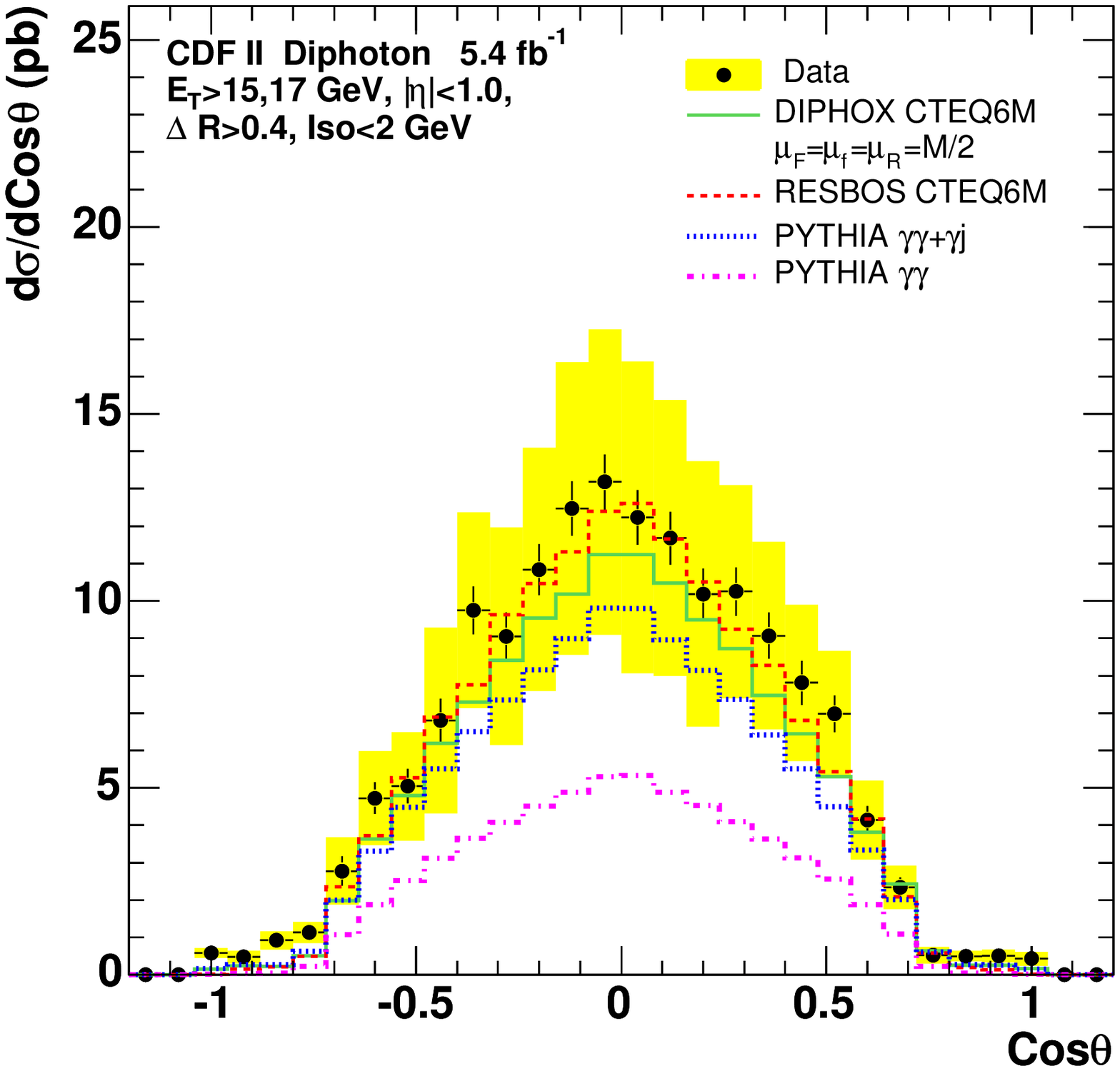}
\includegraphics[width=0.45\linewidth]{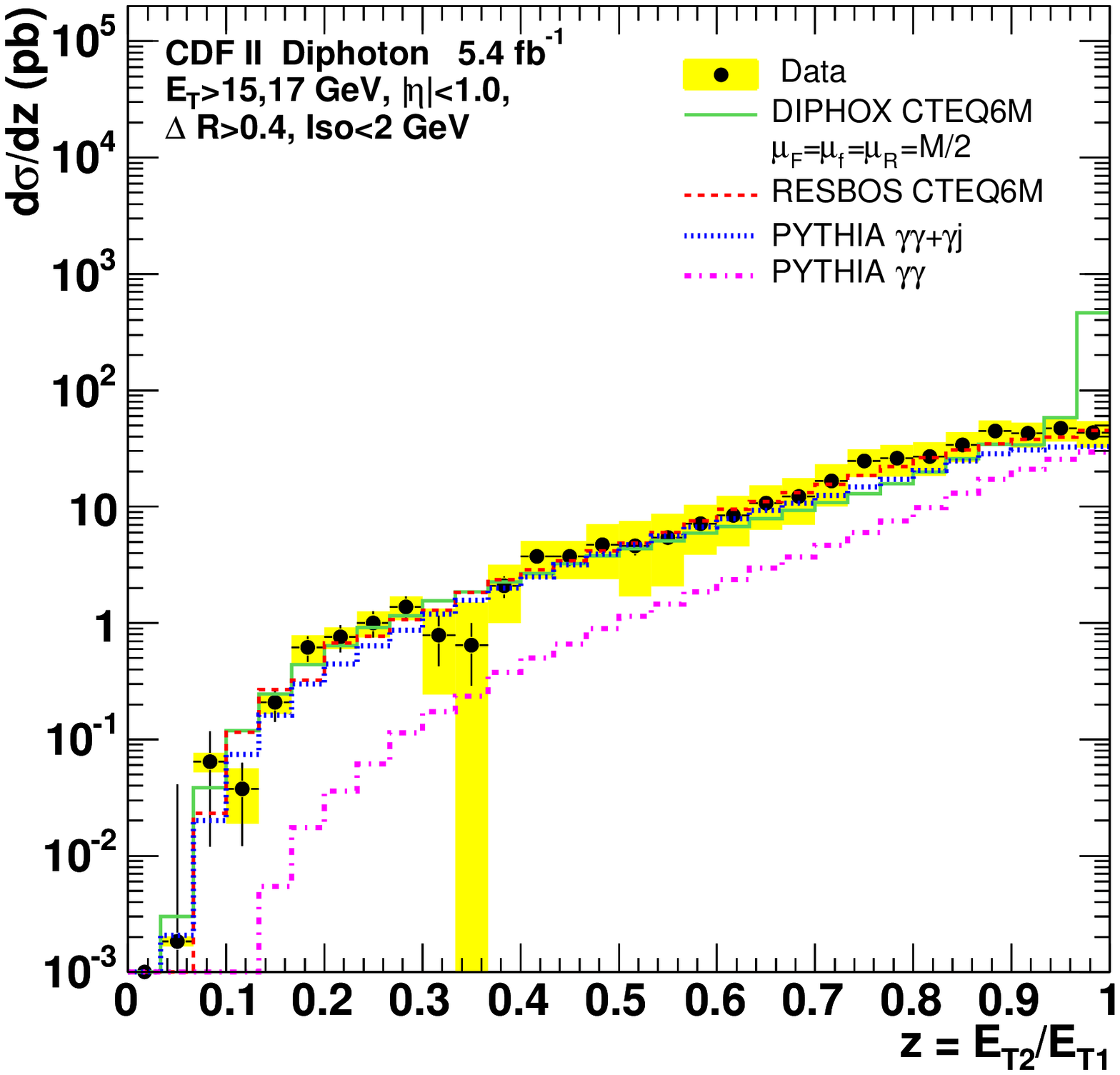}
\includegraphics[width=0.45\linewidth]{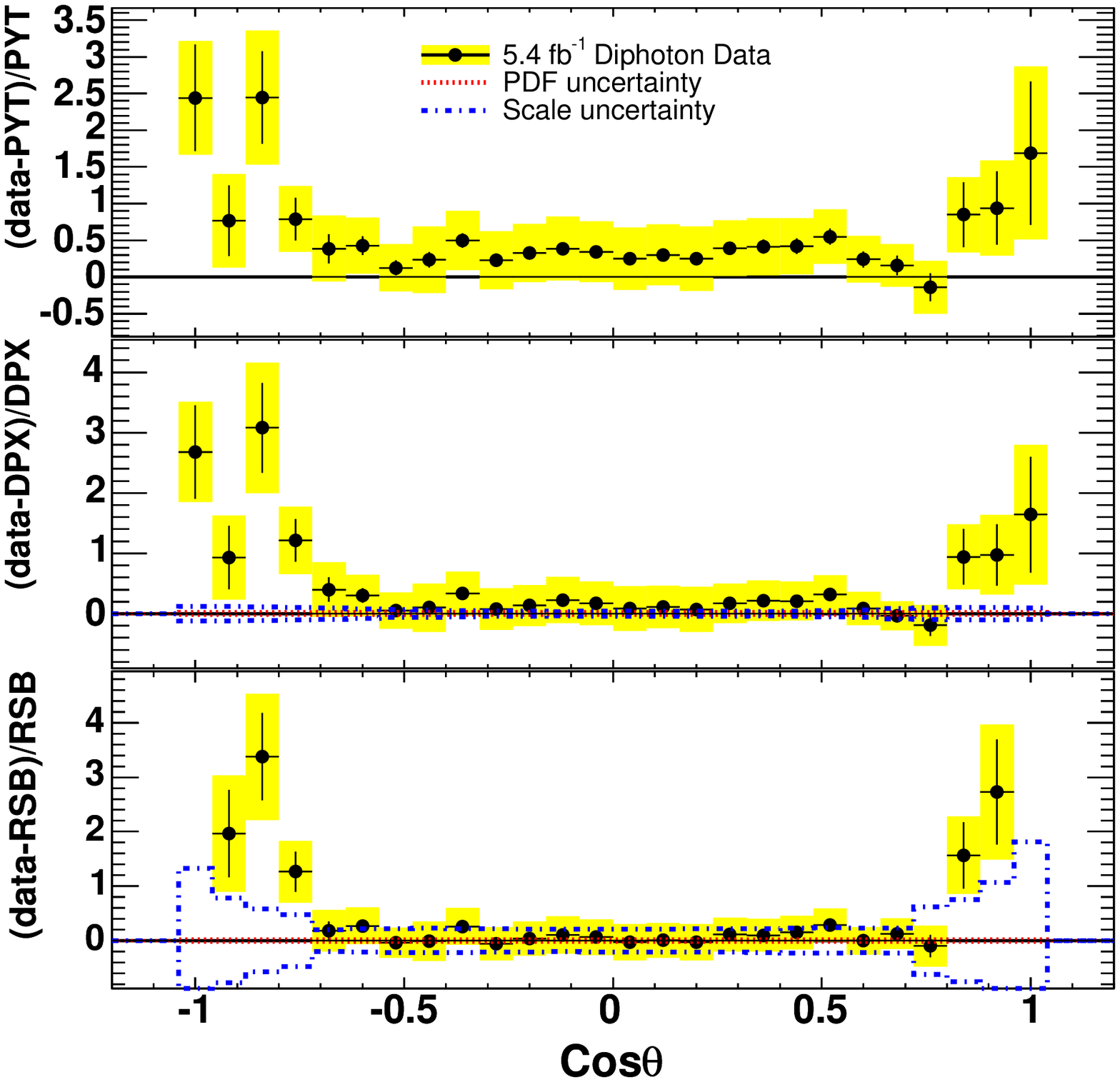}
\includegraphics[width=0.45\linewidth]{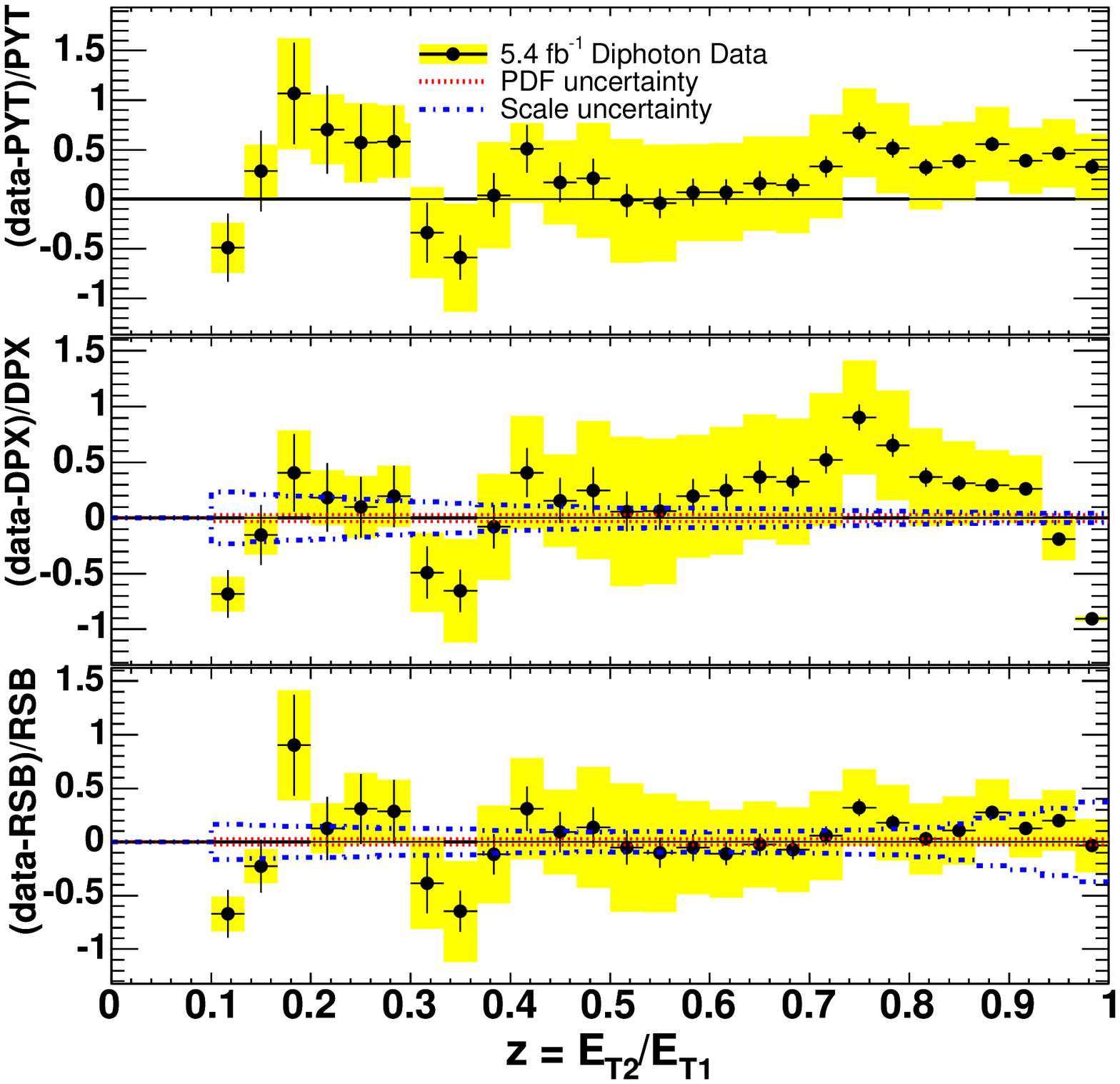}
\caption{The cross section as a function of the cosine of the polar angle in
     the Collins-Soper frame (left) and of the ratio of the subleading
     photon $E_{\rm T}$ to leading photon $E_{\rm T}$ (right). {\bf Top:}
     the absolute cross section values. {\bf Bottom:} the relative deviations
     of the data from the predictions. {\it Note:} the vertical axes scales
     differ between relative deviation plots. The shaded area is the total
     systematic uncertainty in the data.}
\label{fig:uncon_kin_xsec_cos}
\end{figure*}

Fig. \ref{fig:uncon_kin_xsec_cos} shows the results for $d\sigma/d\cos\theta$
and $d\sigma/dz$. All three predictions agree with the data, within
uncertainties. Exceptions are the predictions of all three calculations
underestimating the data in the two ends of the $\cos\theta$ spectrum, where
again gluon scattering processes and associated fragmentation are expected
to dominate \cite{resbos}.

In general, all three calculations reproduce most of the main features of
the data, as observed in the earlier diphoton cross section measurements
\cite{cdf6312,d0}. However, depending on their approximations, they display
differences with each other and with the data in certain kinematic regions.
There is a problem common to all three calculations in the description of
events with very low diphoton mass, low azimuthal distance and diphoton
transverse momentum in the region of the ``Guillet shoulder''. Such events
include fragmentation at a relatively high rate. The {\sc pythia}
$\gamma\gamma$ calculation fails completely to describe the data both in
the scale, where it is low by a factor of 2.5, and in the shape, particularly
of the $P_{\rm T}$, $\Delta\phi$ and $z$ distributions, where it predicts a
much softer spectrum than the data. This is in agreement with the conclusion
of Ref. \cite{cdf6312} which tested only {\sc pythia} $\gamma\gamma$ as a
parton showering MC prediction.

\subsection{Differential Cross Sections for $P_{\rm T}$$<$$M$ Kinematics}
\label{subsec:higgs-kin}

\begin{figure*}[!ht]
\centering
\includegraphics[width=0.45\linewidth]{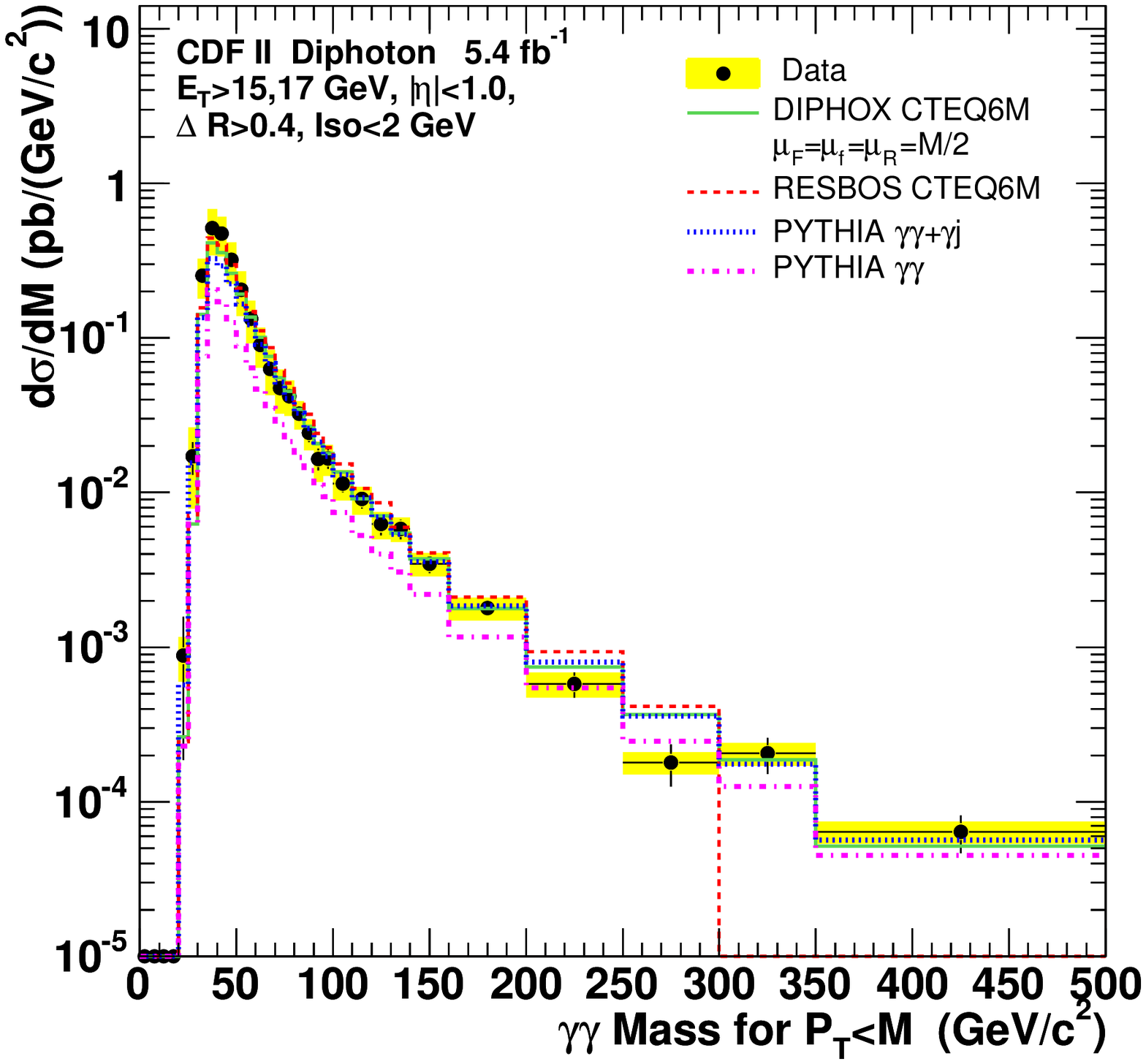}
\includegraphics[width=0.45\linewidth]{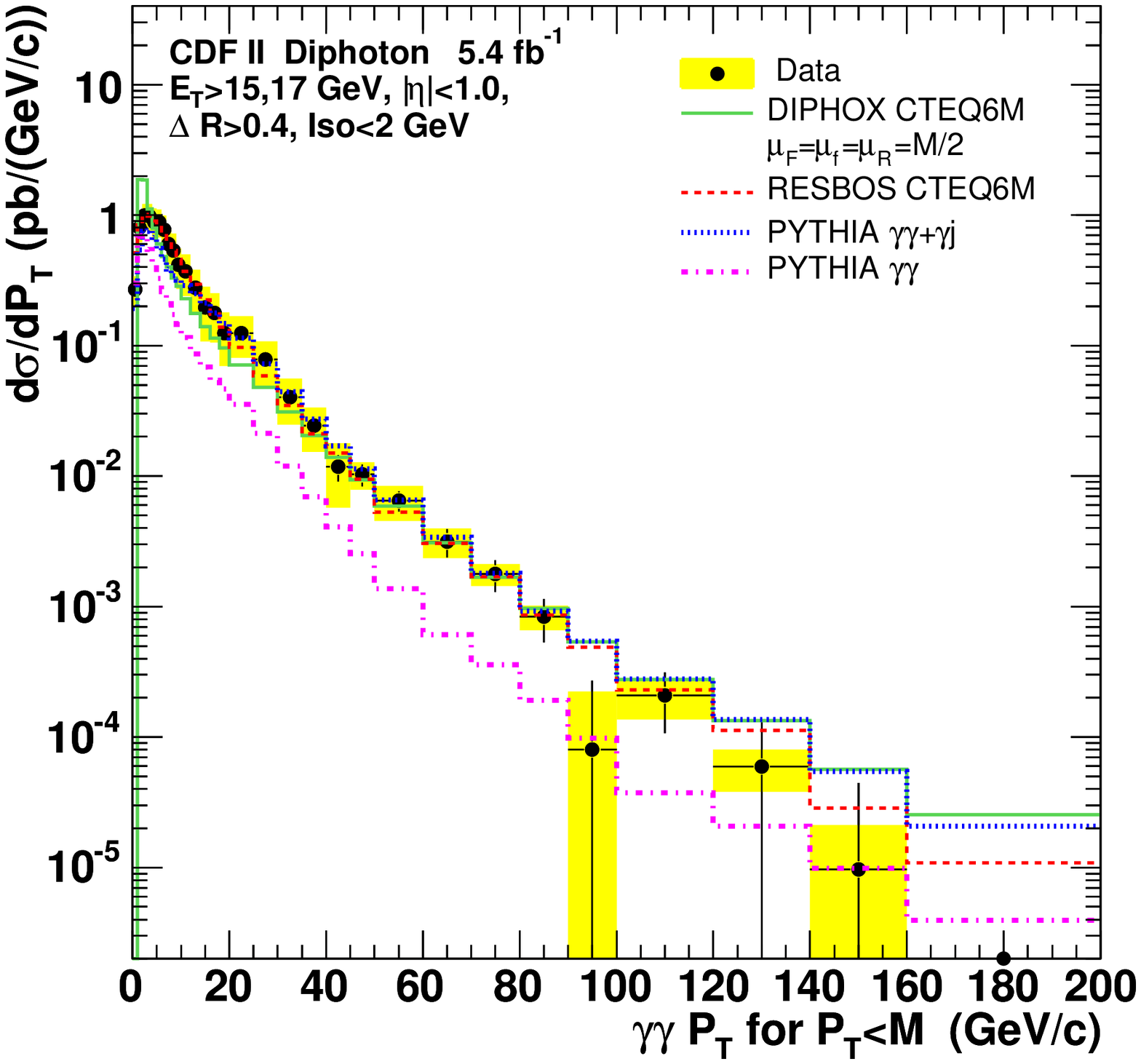}
\includegraphics[width=0.45\linewidth]
{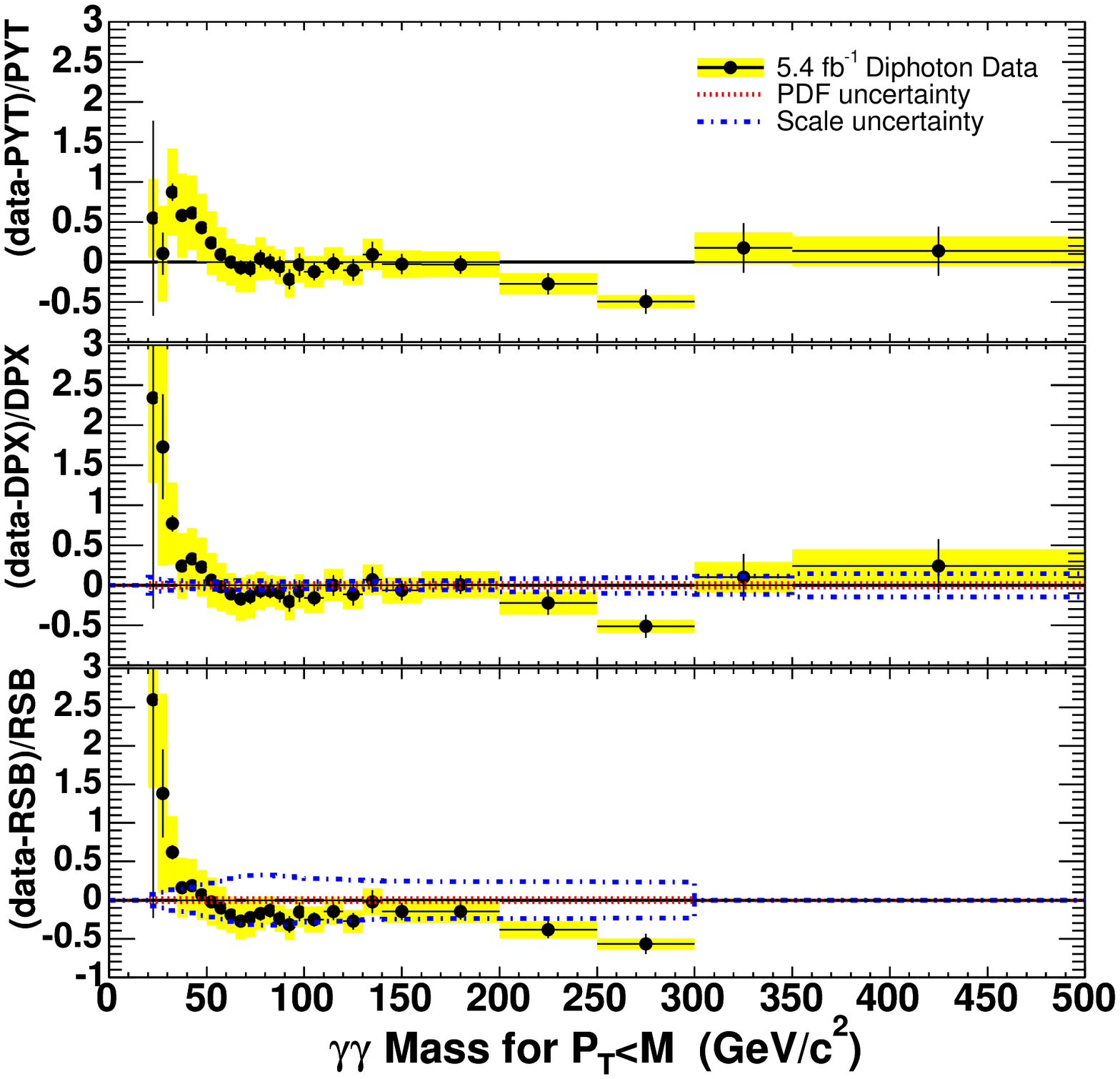}
\includegraphics[width=0.45\linewidth]
{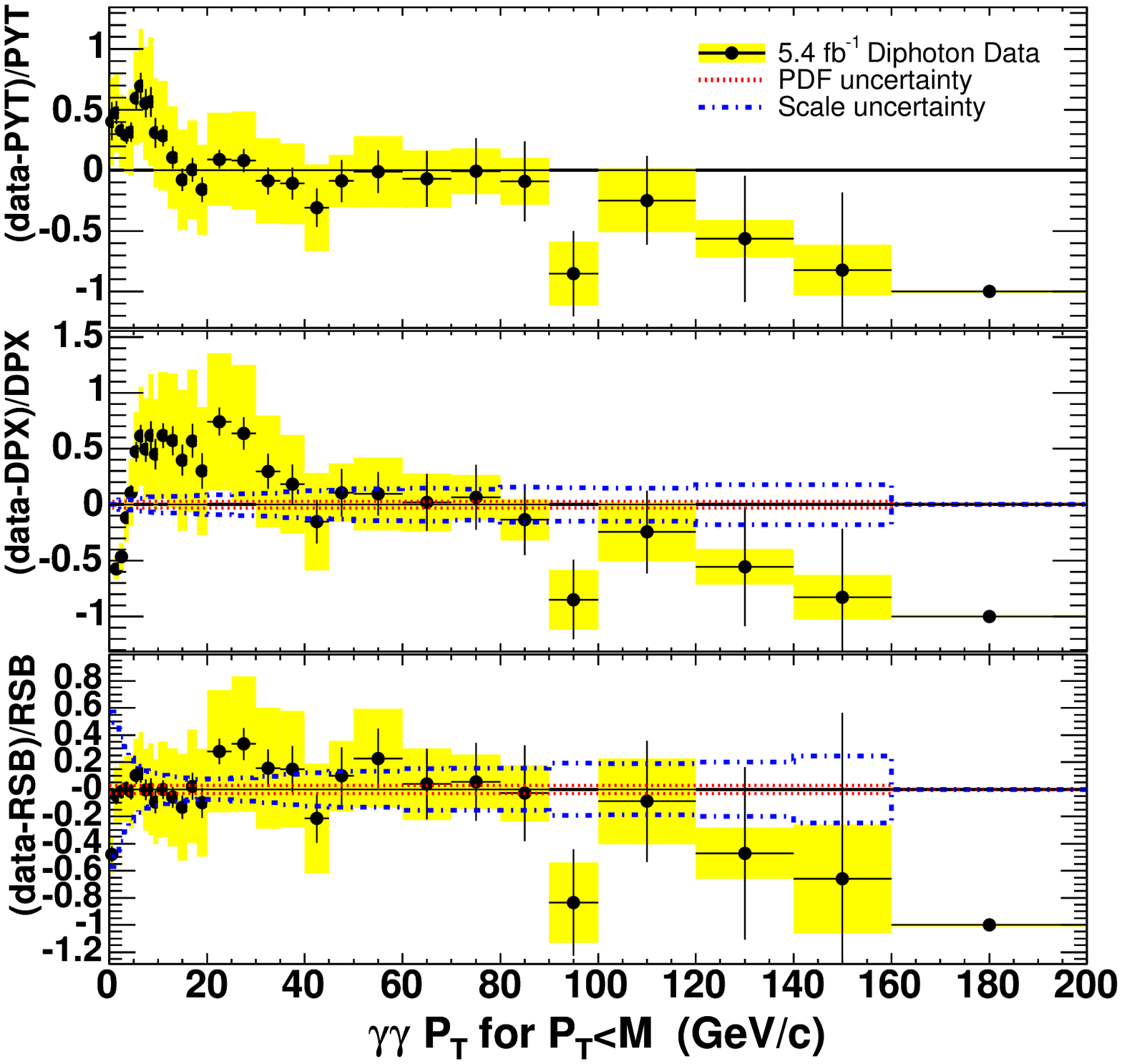}
\caption{The cross section as a function of the diphoton invariant mass (left)
     and transverse momentum (right) for $P_{\rm T}$$<$$M$. {\bf Top:} the
     absolute cross section values. {\bf Bottom:} the relative deviations of
     the data from the predictions. {\it Note:} the vertical axes scales
     differ between relative deviation plots. The shaded area is the total
     systematic uncertainty in the data.}
\label{fig:higgs_kin_xsec_mass}
\end{figure*}

Fig. \ref{fig:higgs_kin_xsec_mass} shows the results for $d\sigma/dM$ and
$d\sigma/dP_{\rm T}$ for $P_{\rm T}$$<$$M$. The low tail of the mass spectrum,
in the region 6 GeV/$c^2$$<$$M$$<$32 GeV/$c^2$, and the shoulder of the
$P_{\rm T}$ spectrum, in the region 20 GeV/$c$$<$$P_{\rm T}$$<$50 GeV/$c$,
are now eliminated. The agreement between the data and all three predictions
is improved in this case. However, {\sc diphox} still underestimates the data
for $P_{\rm T}$$<$40 GeV/$c$ and similarly {\sc pythia} still underestimates
the data for $P_{\rm T}$$<$10 GeV/$c$, thus showing the importance of NNLL
low--$P_{\rm T}$ resummation in this case as well.

\begin{figure*}[!ht]
\centering
\includegraphics[width=0.45\linewidth]{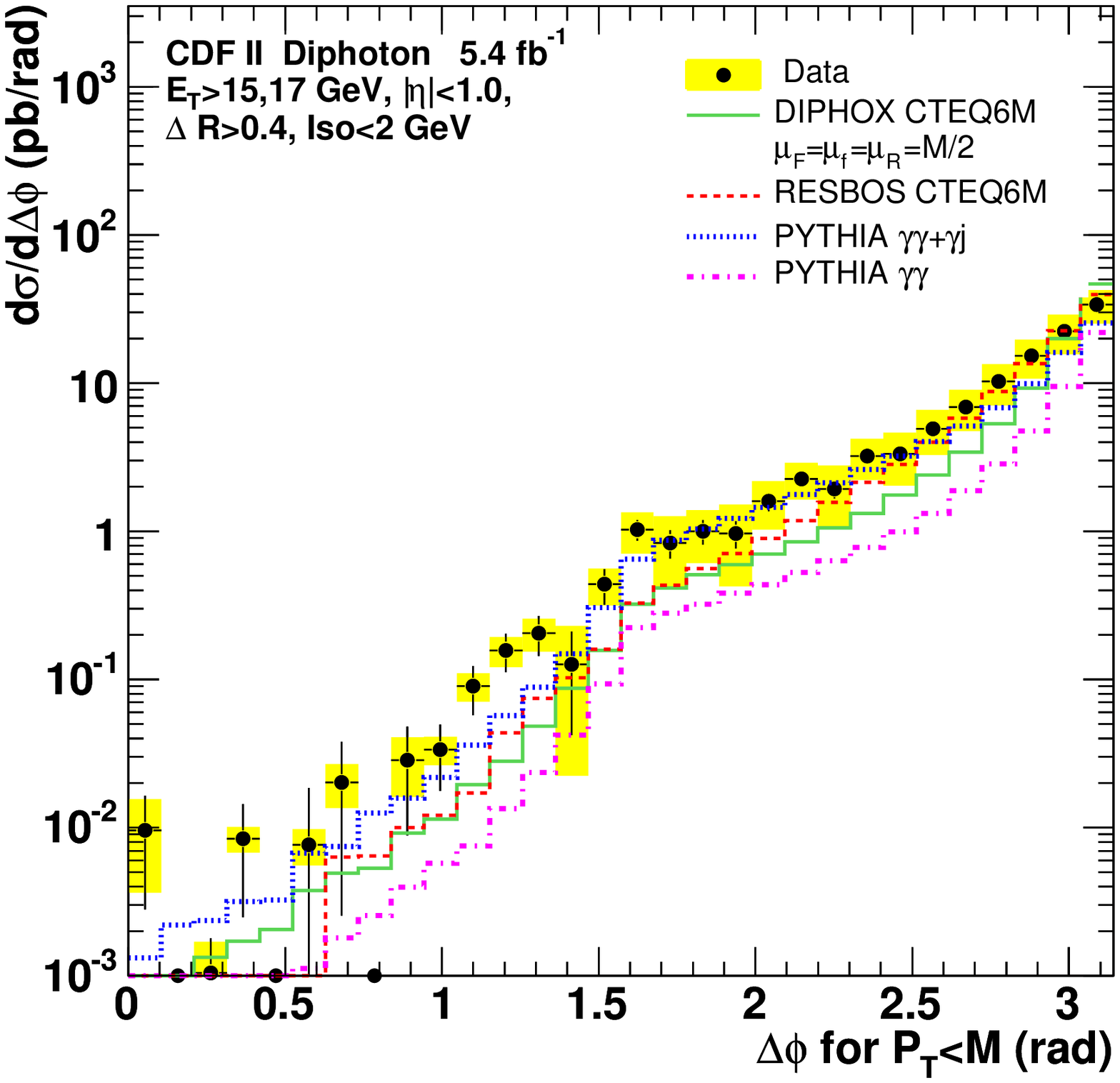}
\includegraphics[width=0.45\linewidth]{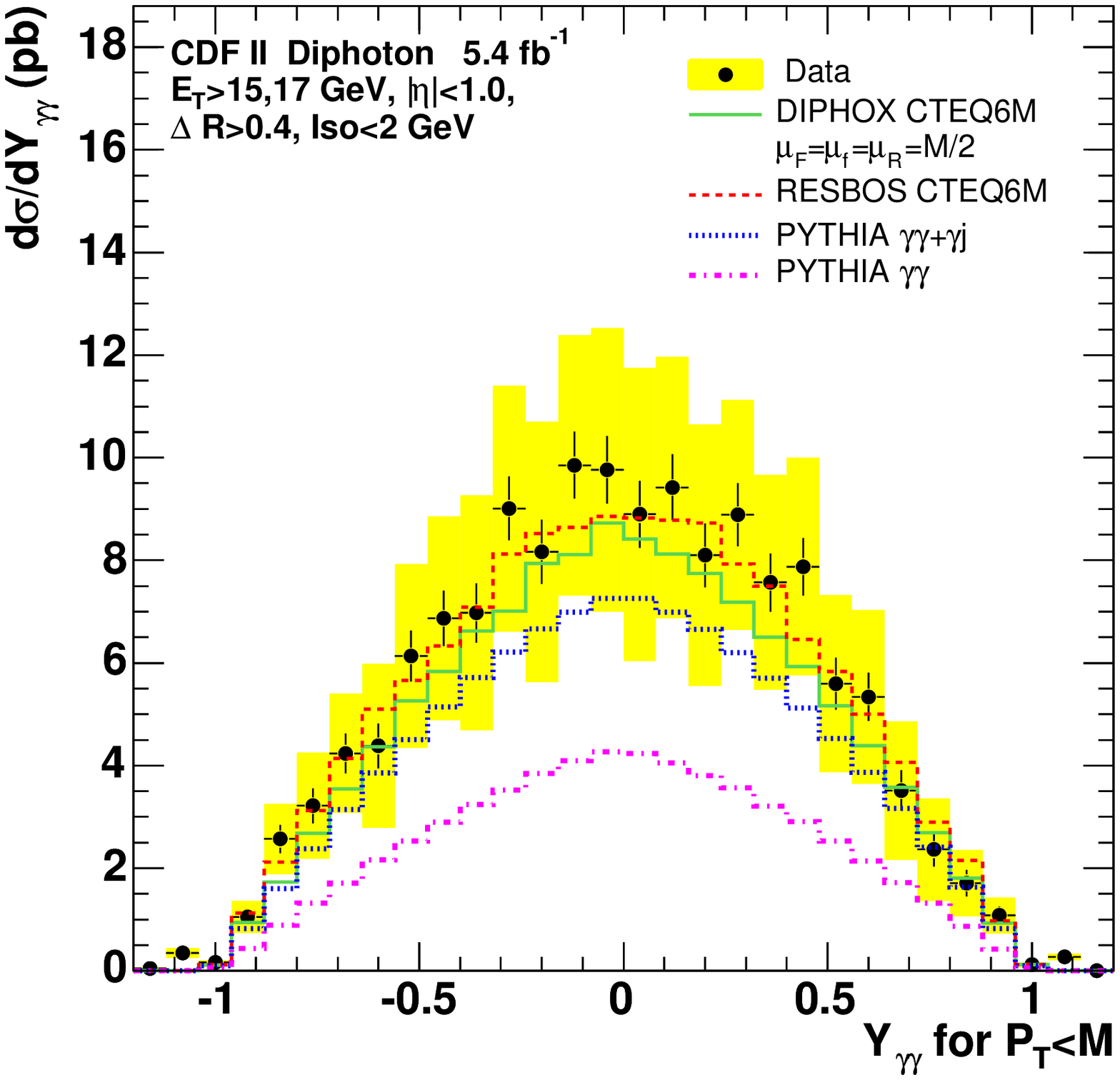}
\includegraphics[width=0.45\linewidth]
{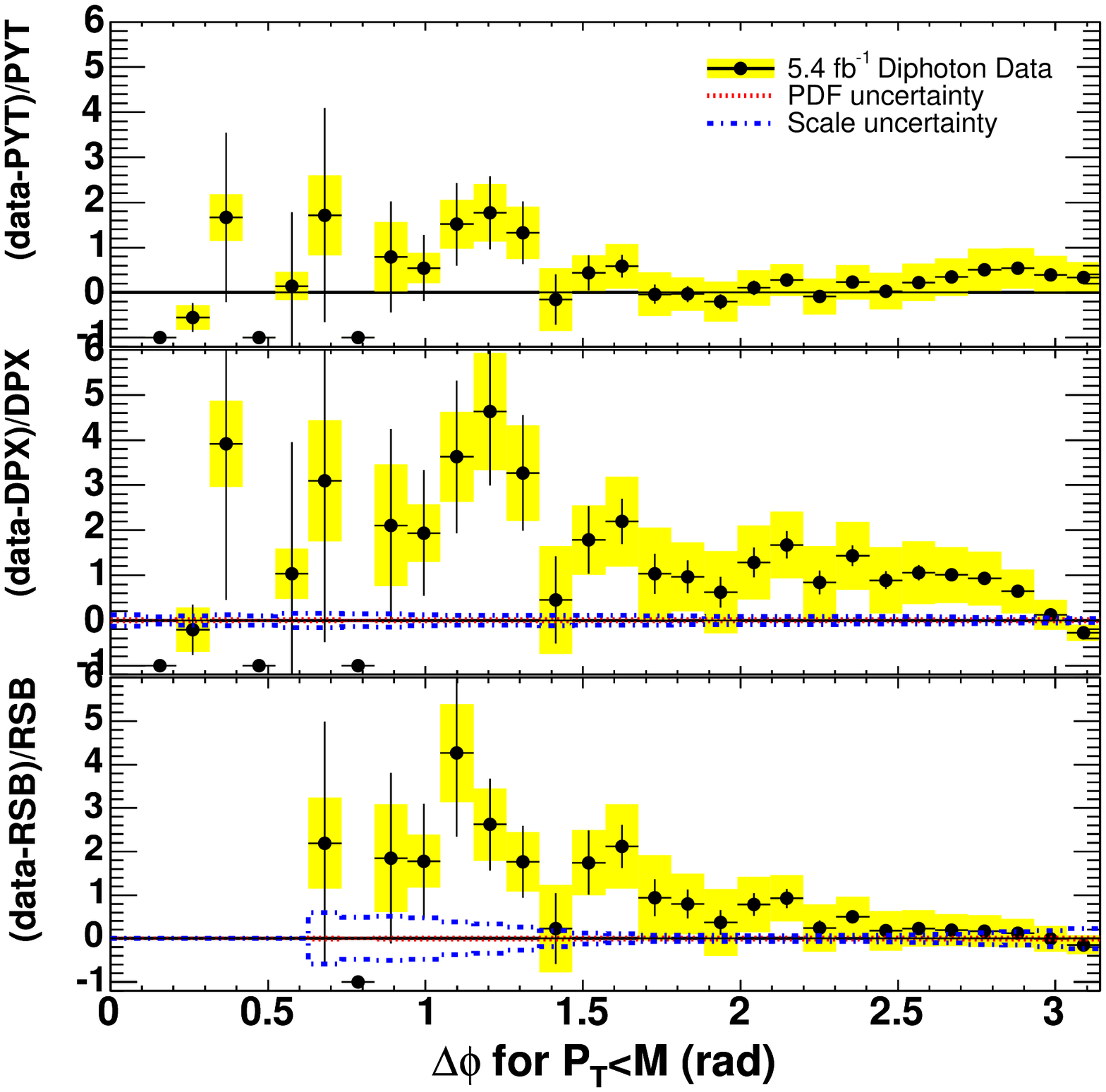}
\includegraphics[width=0.45\linewidth]
{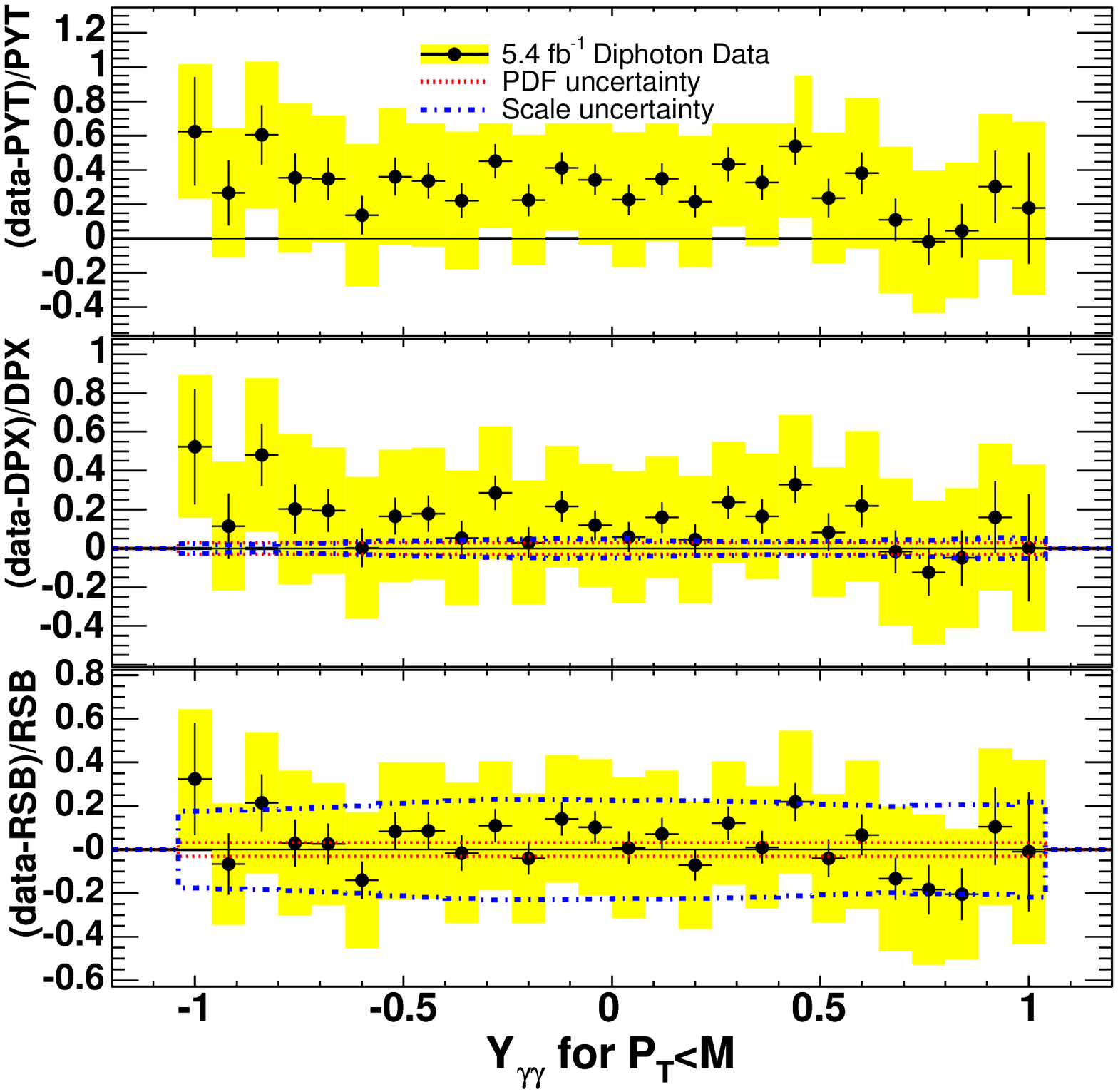}
\caption{The cross section as a function of the diphoton azimuthal distance
     (left) and of the diphoton rapidity (right) for $P_{\rm T}$$<$$M$.
     {\bf Top:} the absolute cross section values. {\bf Bottom:} the relative
     deviations of the data from the predictions. {\it Note:} the vertical
     axes scales differ between relative deviation plots. The shaded area is
     the total systematic uncertainty in the data.}
\label{fig:higgs_kin_xsec_angles}
\end{figure*}

Fig. \ref{fig:higgs_kin_xsec_angles} shows the results for
$d\sigma/d\Delta\phi$ and $d\sigma/dY_{\gamma\gamma}$ for $P_{\rm T}$$<$$M$. The
tail of the $\Delta\phi$ spectrum for $\Delta\phi$$<$$\pi$/2 is now weaker
but the measured cross section is underestimated by all three predictions,
as in the case of unconstrained kinematics.

\begin{figure*}[!ht]
\centering
\includegraphics[width=0.45\linewidth]{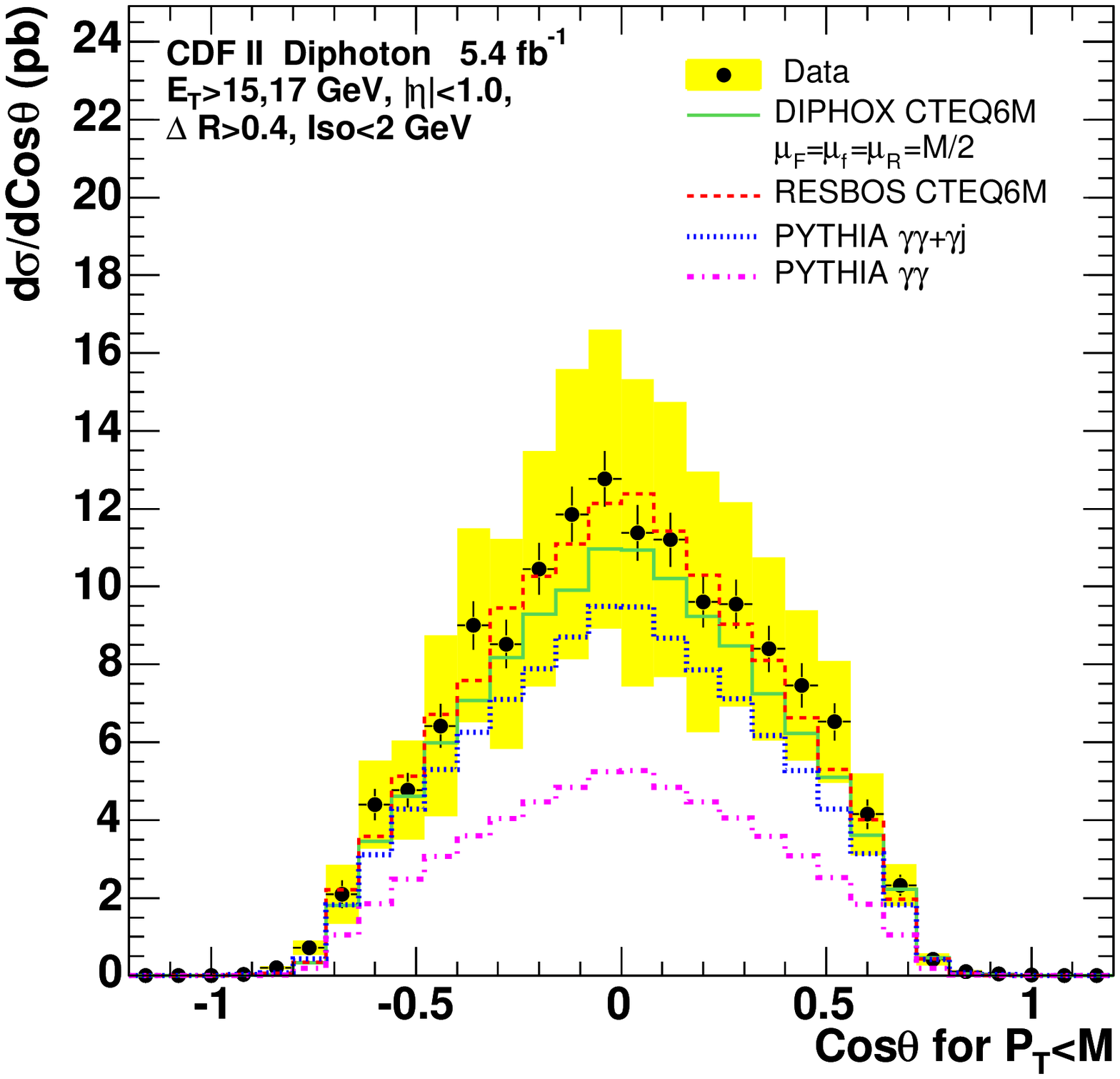}
\includegraphics[width=0.45\linewidth]{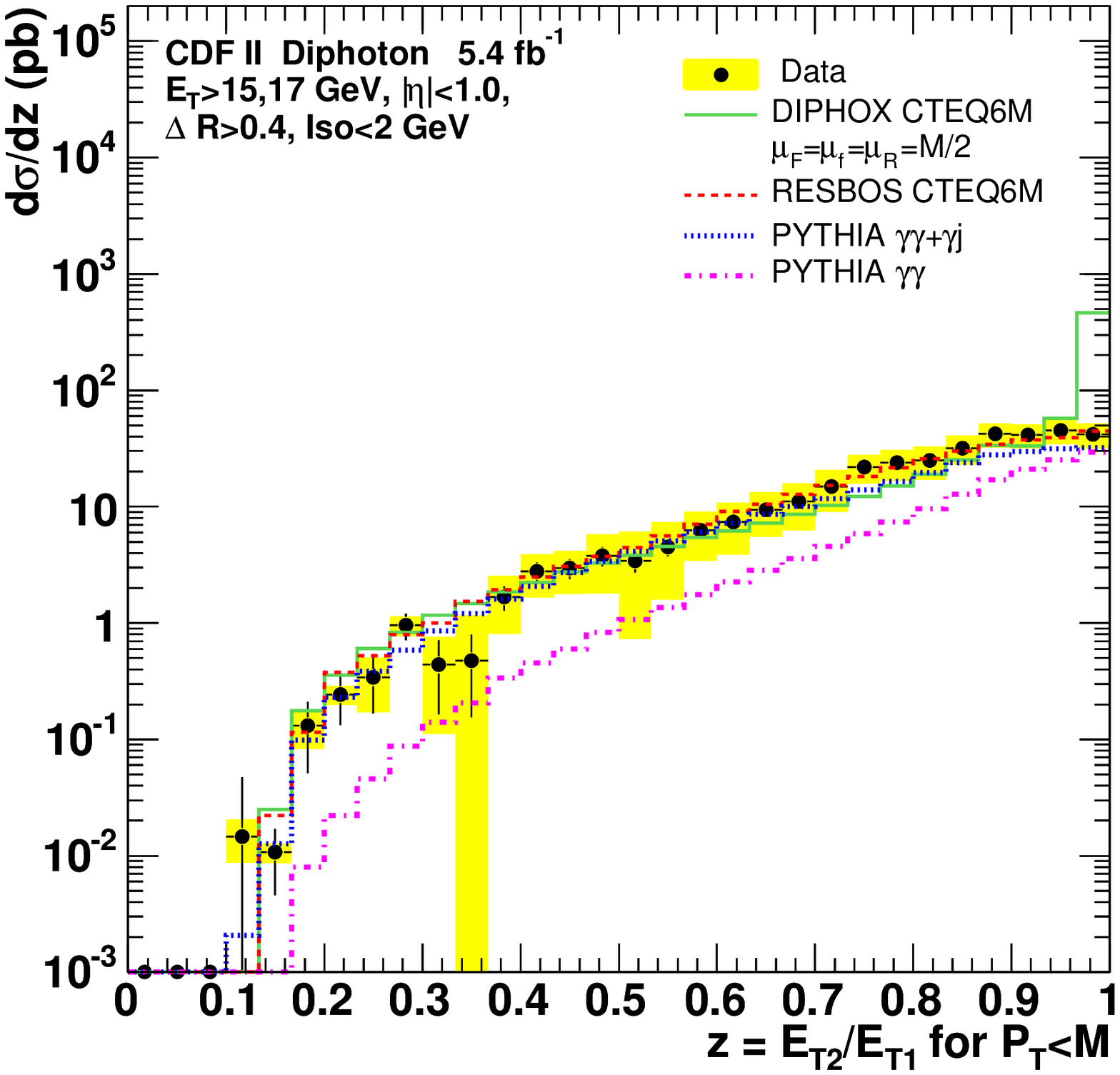}
\includegraphics[width=0.45\linewidth]
{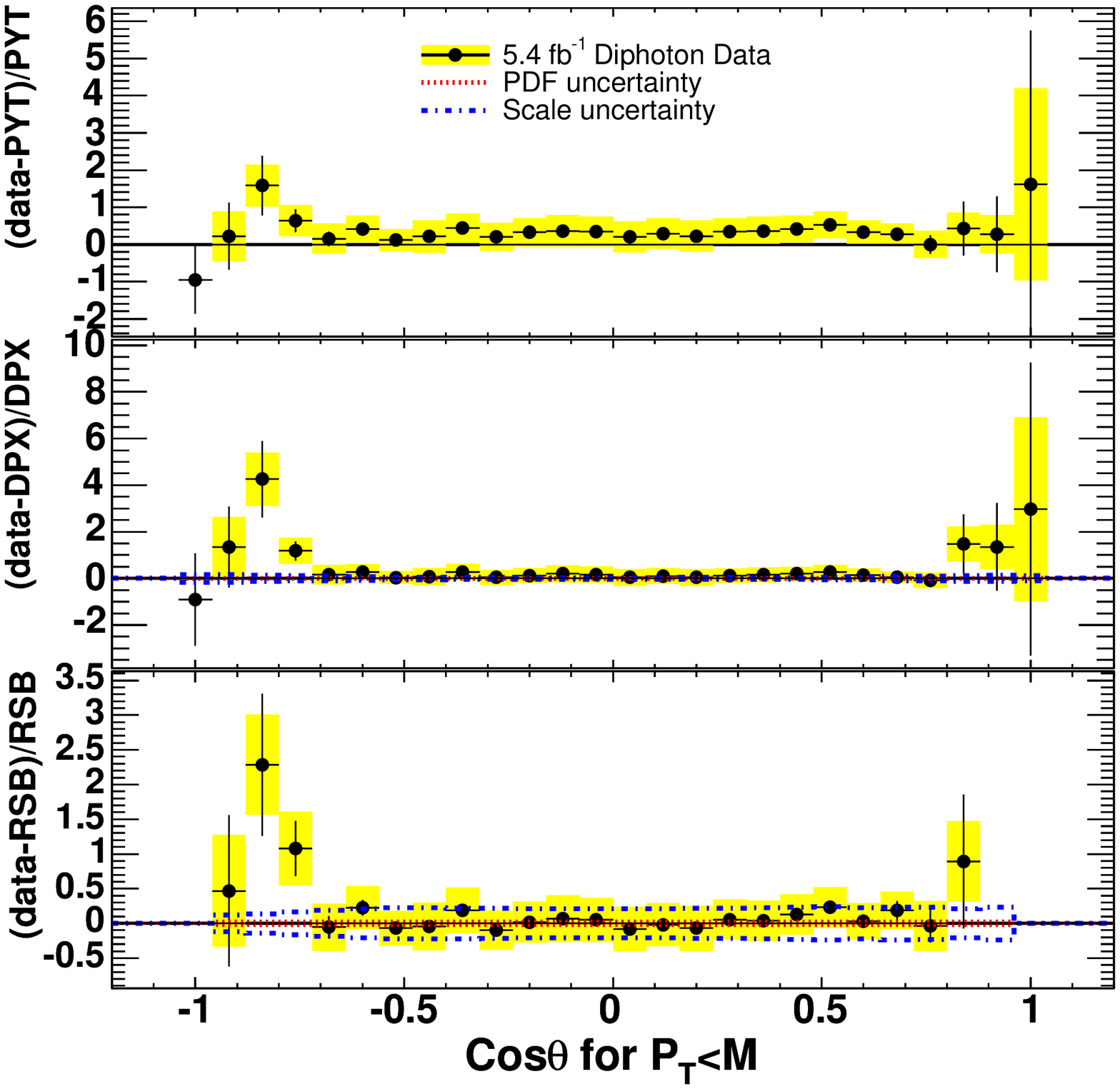}
\includegraphics[width=0.45\linewidth]{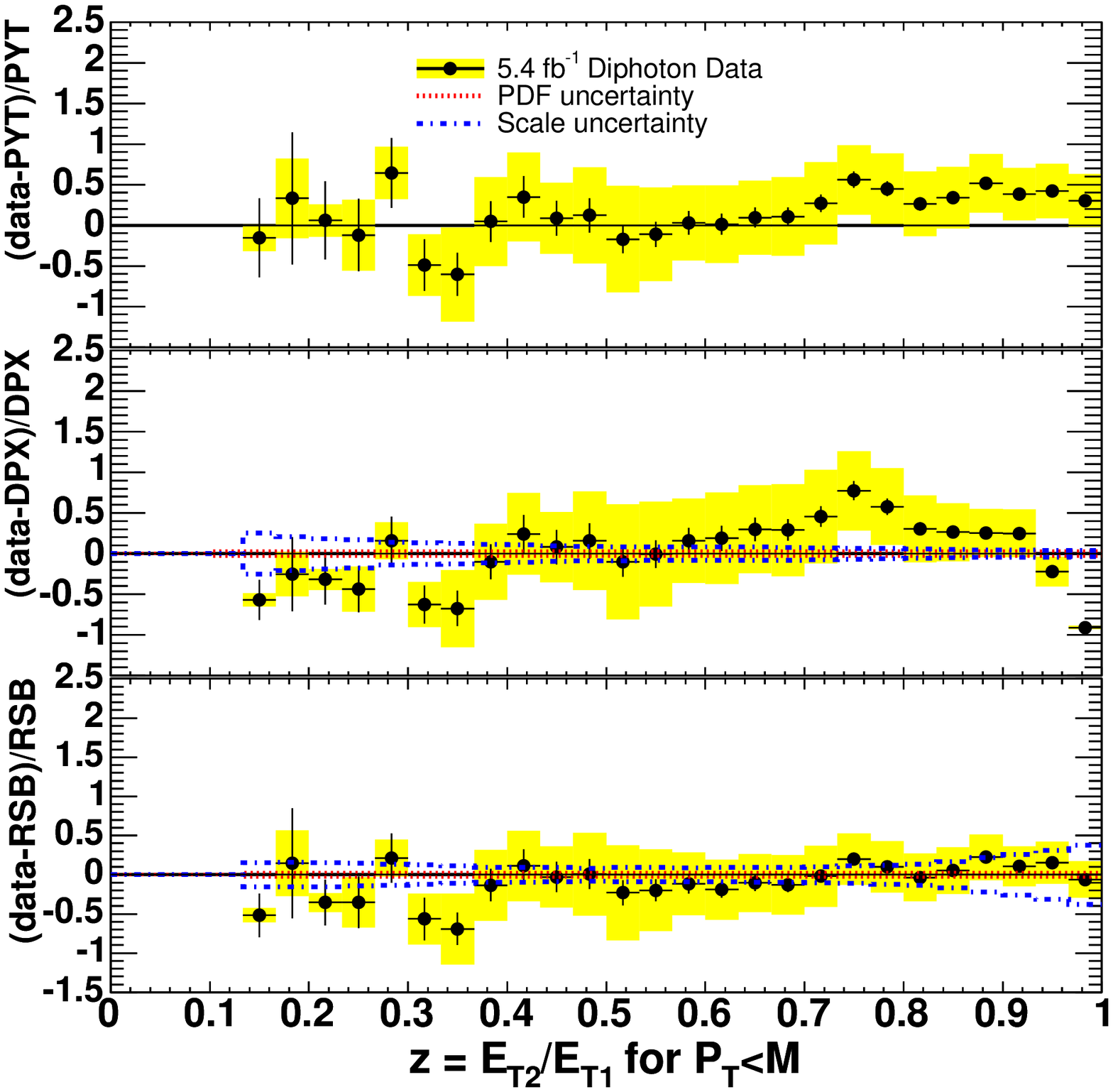}
\caption{The cross section as a function of the cosine of the polar angle in
     the Collins-Soper frame (left) and of the ratio of the subleading
     photon $E_{\rm T}$ to leading photon $E_{\rm T}$ (right) for
     $P_{\rm T}$$<$$M$. {\bf Top:} the absolute cross section values.
     {\bf Bottom:} the relative deviations of the data from the predictions.
     {\it Note:} the vertical axes scales differ between relative deviation
     plots. The shaded area is the total systematic uncertainty in the data.}
\label{fig:higgs_kin_xsec_cos}
\end{figure*}

Fig. \ref{fig:higgs_kin_xsec_cos} shows the results for $d\sigma/d\cos\theta$
and $d\sigma/dz$ for $P_{\rm T}$$<$$M$. The results are similar to the case of
unconstrained kinematics. Generally, all three calculations agree with
the data, within uncertainties. Exceptions are again the predictions of all
three calculations in the two ends of the $\cos\theta$ spectrum, where they
underestimate the data.

In general, events with kinematics similar to the decay of a heavy particle
with low transverse momentum into a photon pair, such as
gg$\to$H$\to$$\gamma\gamma$ production and decay, are better described by the
theory than events with low mass and high transverse momentum. This is also
observed in Ref. \cite{d0} which examines only the case of $P_{\rm T}$$<$$M$.
This observation is important for current searches of yet undiscovered
particles with a diphoton decay signature. The {\sc pythia}
$\gamma\gamma$ calculation again fails to describe the data both in the
scale and in the shape, in agreement with the conclusion of Ref. \cite{d0}
which tested only {\sc pythia} $\gamma\gamma$ as a parton showering MC
prediction.

\subsection{Differential Cross Sections for $P_{\rm T}$$>$$M$ Kinematics}
\label{subsec:box-kin}

\begin{figure*}[!ht]
\centering
\includegraphics[width=0.45\linewidth]{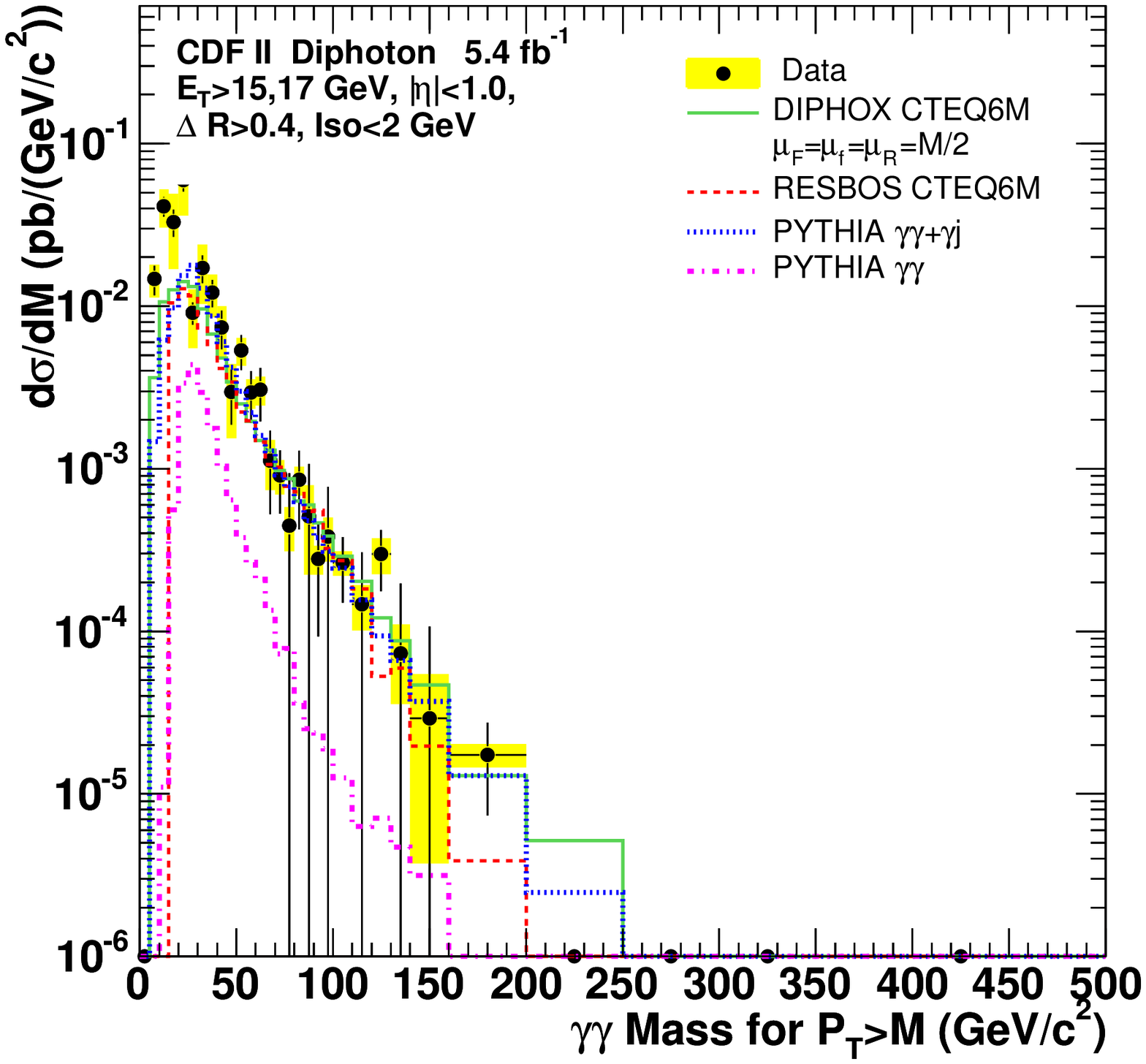}
\includegraphics[width=0.45\linewidth]{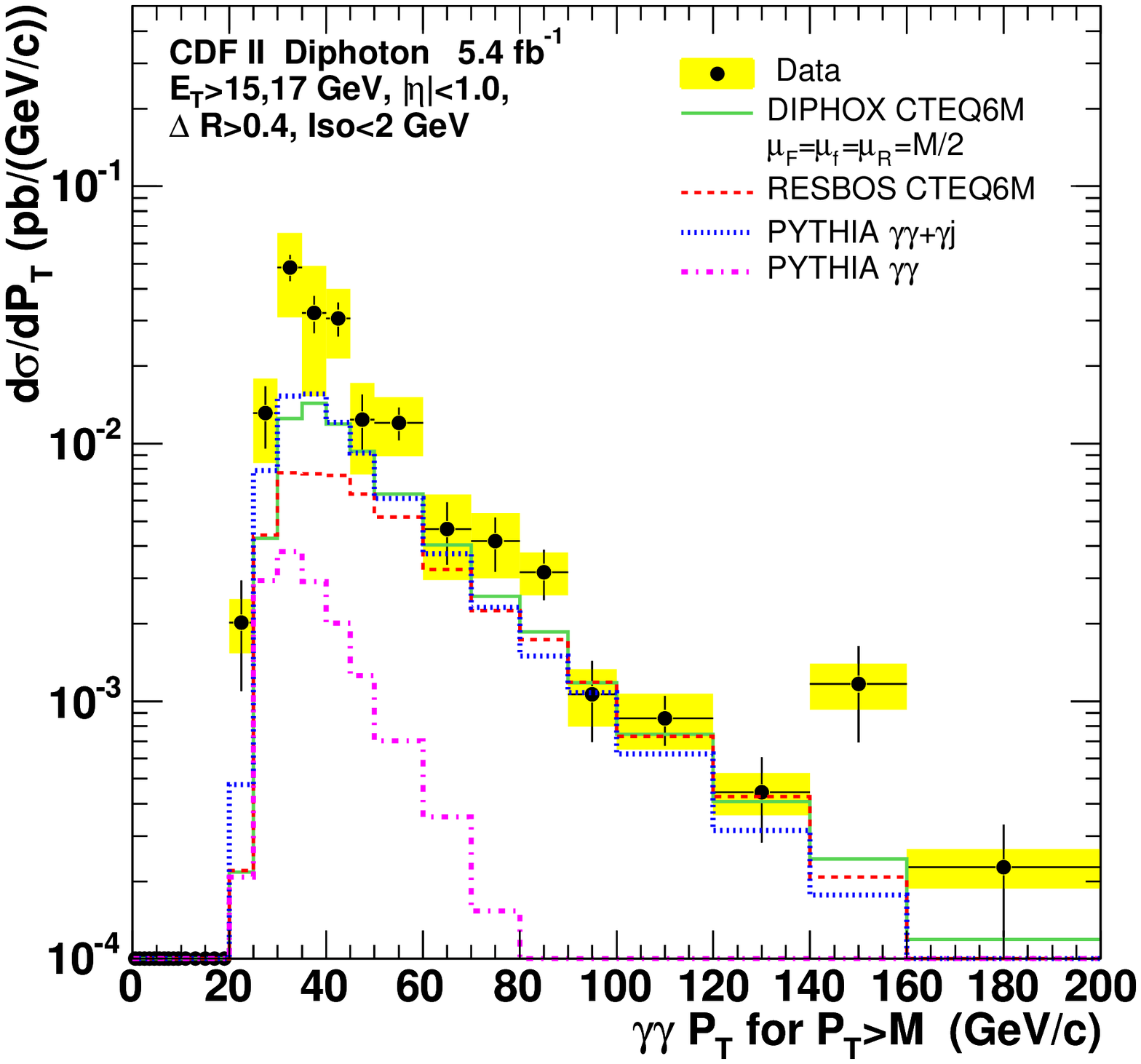}
\includegraphics[width=0.45\linewidth]
{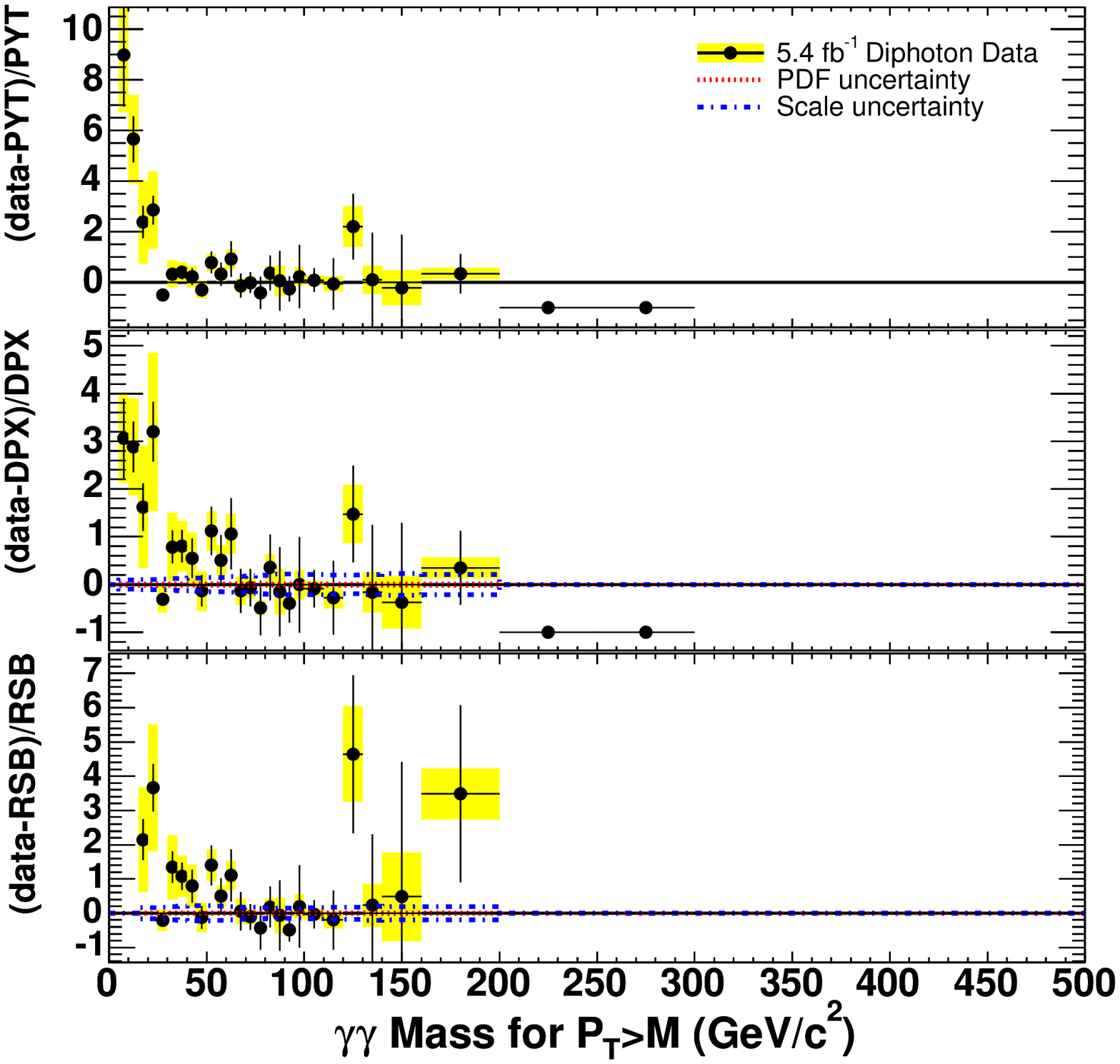}
\includegraphics[width=0.45\linewidth]
{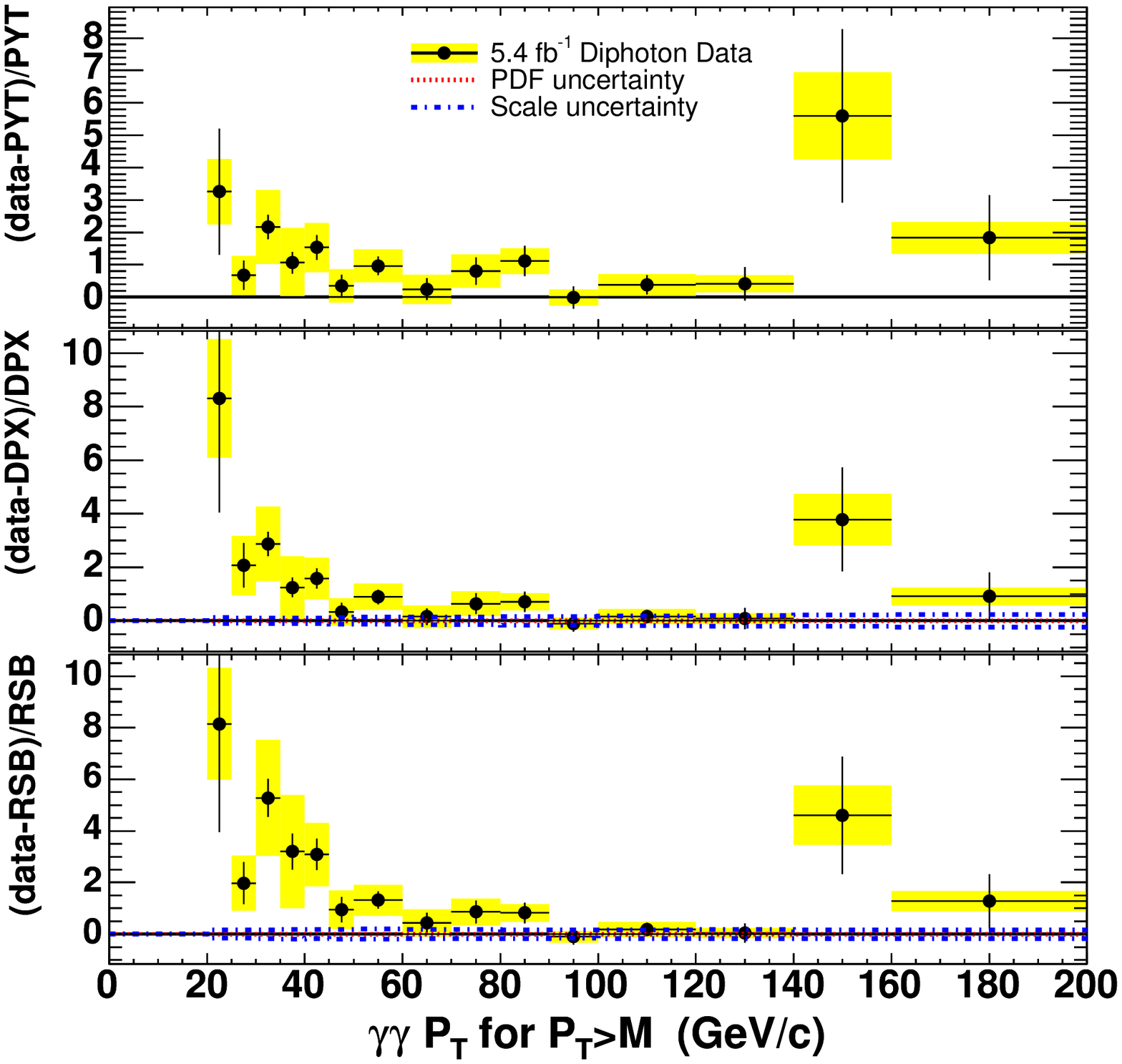}
\caption{The cross section as a function of the diphoton invariant mass (left)
     and transverse momentum (right) for $P_{\rm T}$$>$$M$. {\bf Top:} the
     absolute cross section values. {\bf Bottom:} the relative deviations of
     the data from the predictions. {\it Note:} the vertical axes scales
     differ between relative deviation plots. The shaded area is the total
     systematic uncertainty in the data.}
\label{fig:box_kin_xsec_mass}
\end{figure*}

Fig. \ref{fig:box_kin_xsec_mass} shows the results for $d\sigma/dM$ and
$d\sigma/dP_{\rm T}$ for $P_{\rm T}$$>$$M$. Both spectra are depleted in this
case:  the mass spectrum for $M$$>$200 GeV/$c^2$ and the transverse momentum
spectrum for $P_{\rm T}$$<$20 GeV/$c$. All three calculations underestimate
the data.

\begin{figure*}[!ht]
\centering
\includegraphics[width=0.45\linewidth]{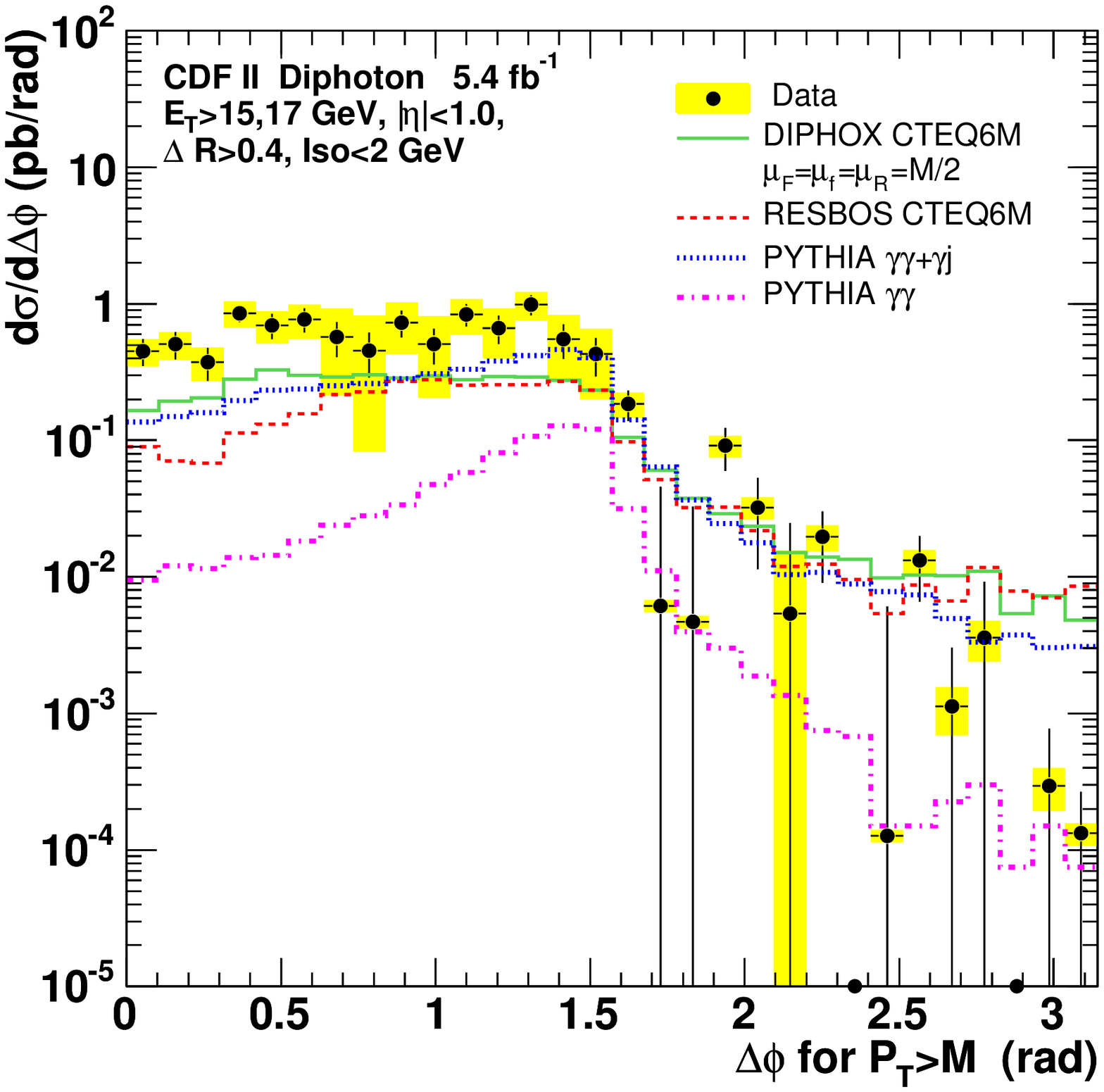}
\includegraphics[width=0.45\linewidth]{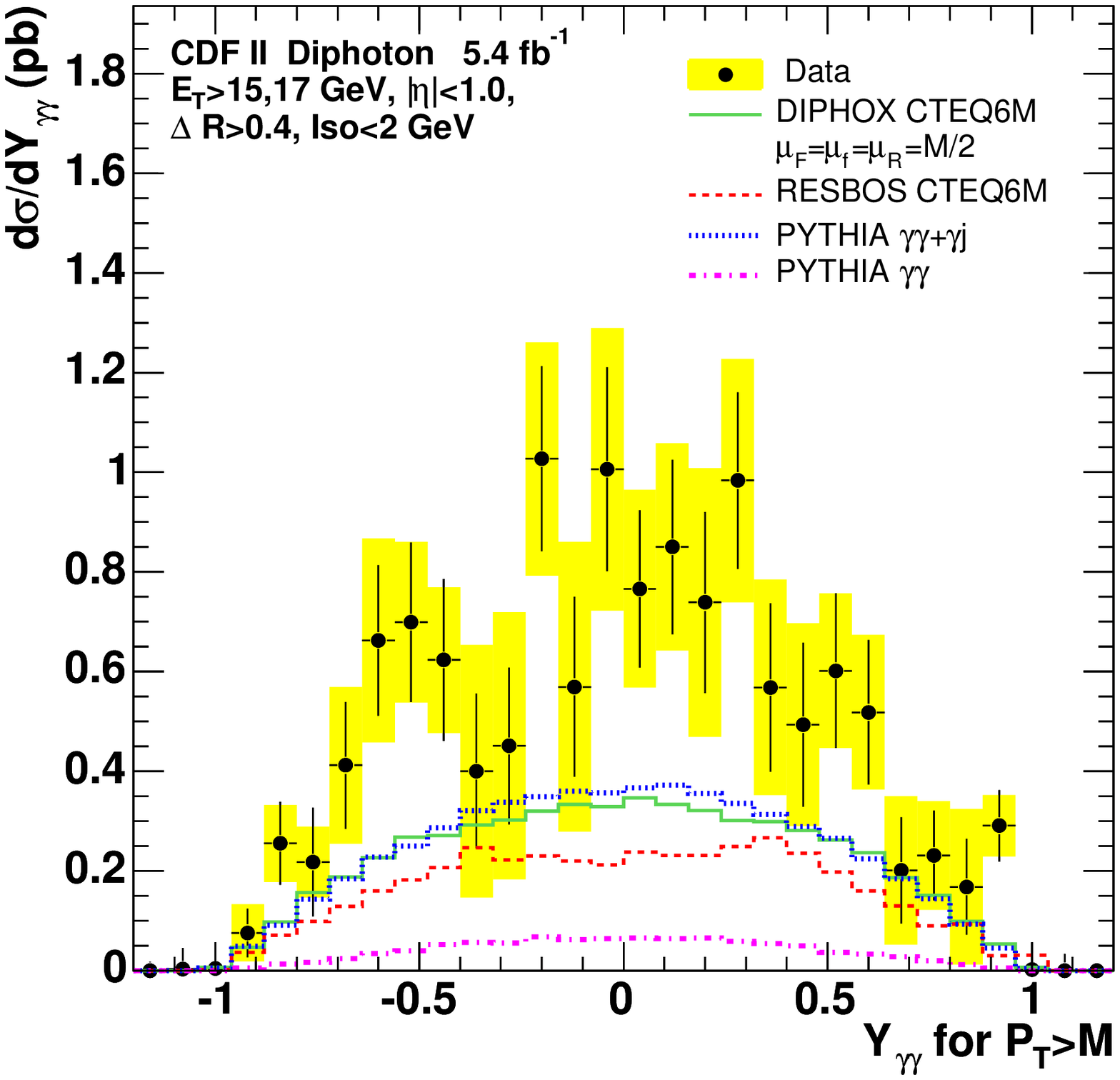}
\includegraphics[width=0.45\linewidth]
{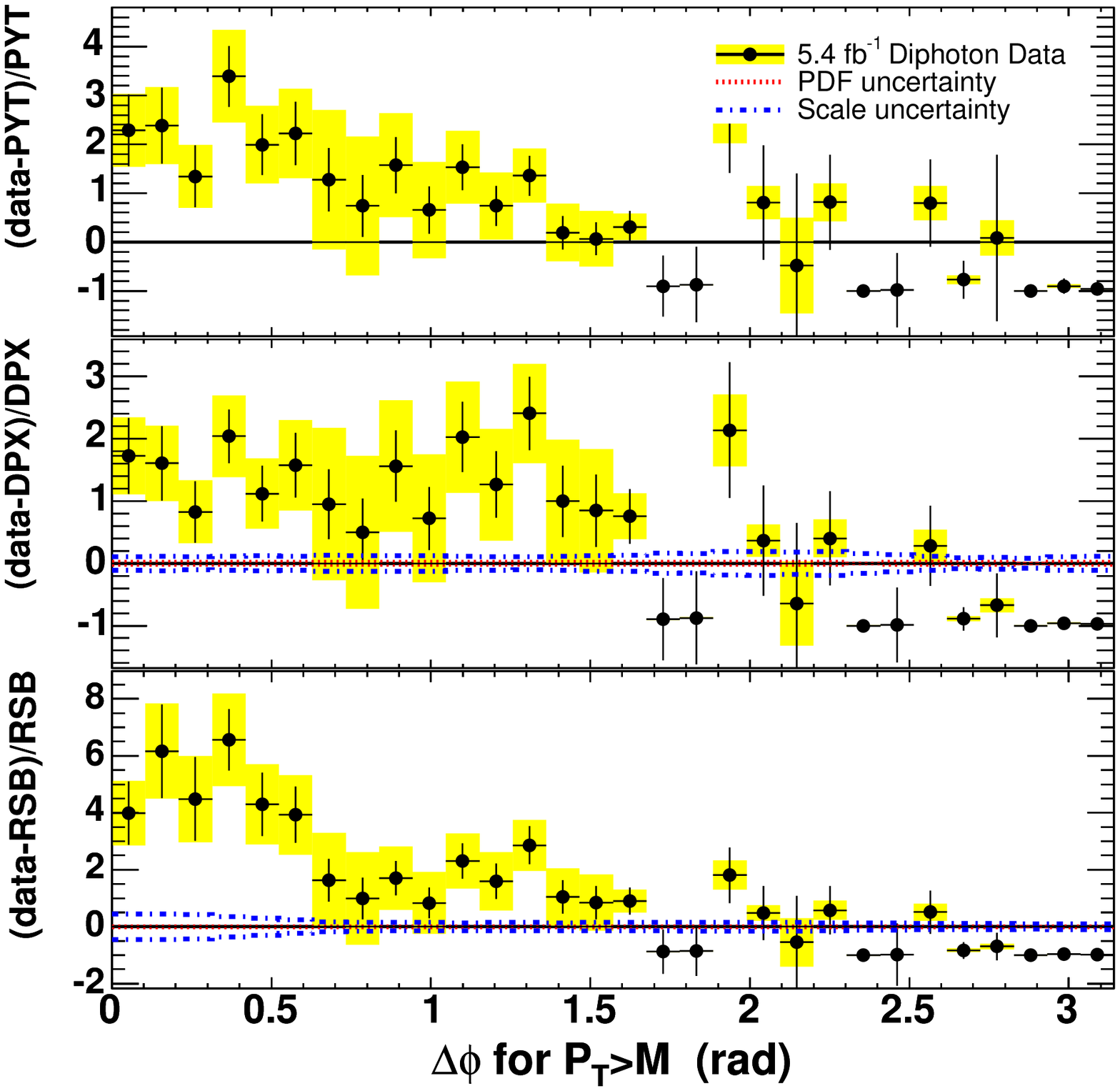}
\includegraphics[width=0.45\linewidth]
{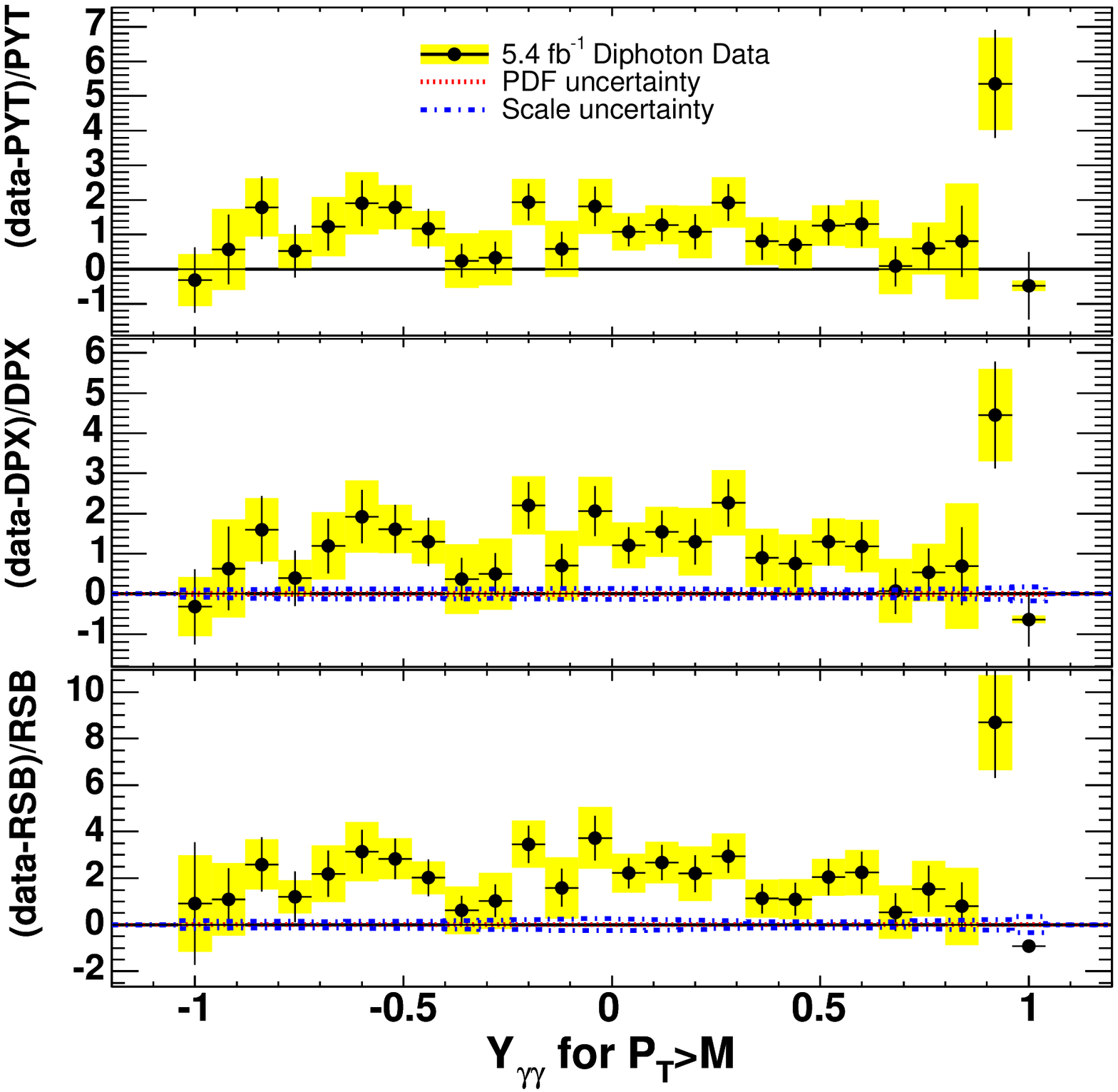}
\caption{The cross section as a function of the diphoton azimuthal distance
     (left) and of the diphoton rapidity (right) for $P_{\rm T}$$>$$M$.
     {\bf Top:} the absolute cross section values. {\bf Bottom:} the relative
     deviations of the data from the predictions. {\it Note:} the vertical
     axes scales differ between relative deviation plots. The shaded area is
     the total systematic uncertainty in the data.}
\label{fig:box_kin_xsec_angles}
\end{figure*}

Fig. \ref{fig:box_kin_xsec_angles} shows the results for $d\sigma/d\Delta\phi$
and $d\sigma/dY_{\gamma\gamma}$ for $P_{\rm T}$$>$$M$. The $\Delta\phi$ spectrum
is strongly suppressed for $\Delta\phi$$>$$\pi$/2. Again, the measured cross
section is underestimated by all three calculations.

\begin{figure*}[!ht]
\centering
\includegraphics[width=0.45\linewidth]{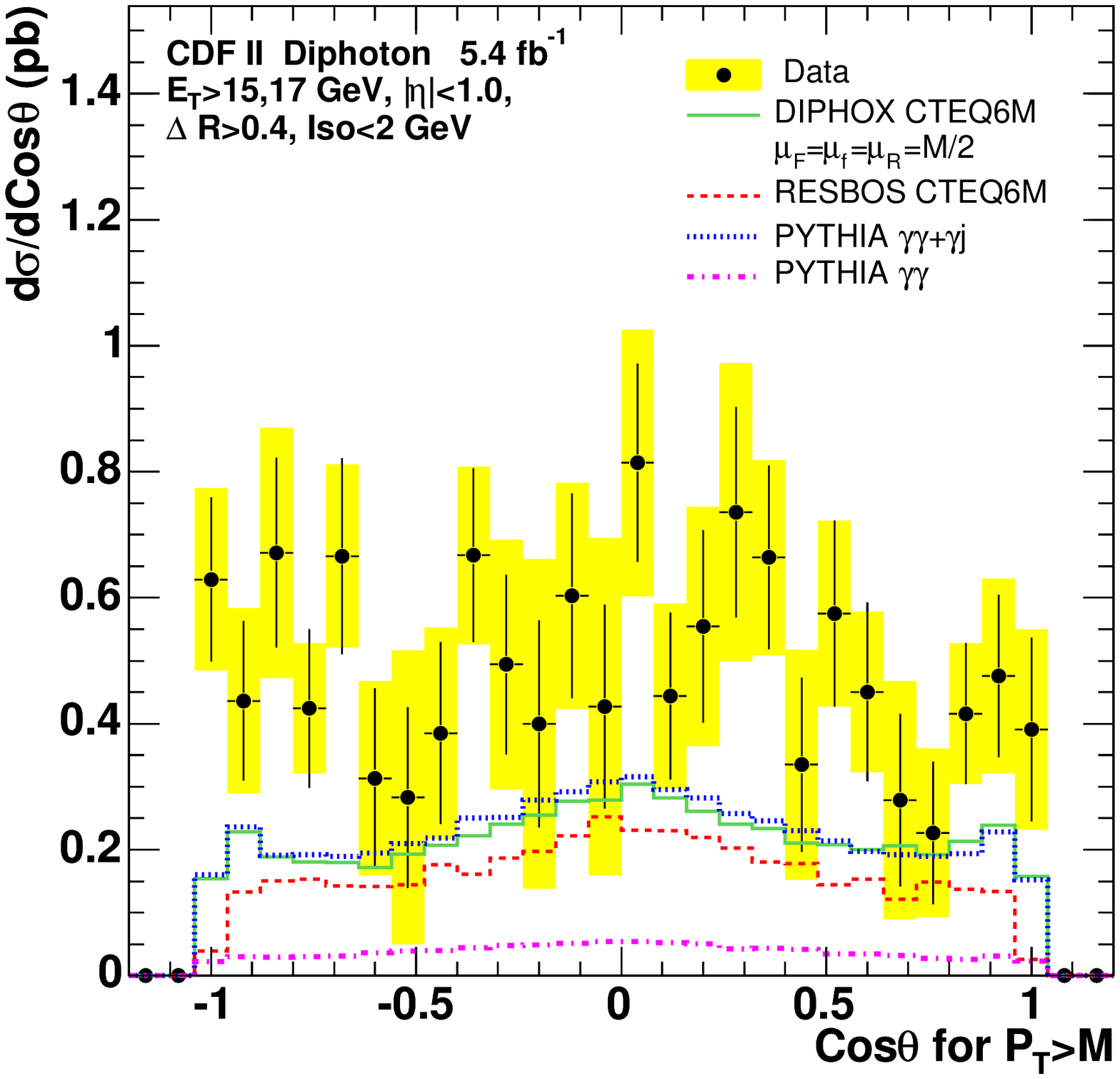}
\includegraphics[width=0.45\linewidth]{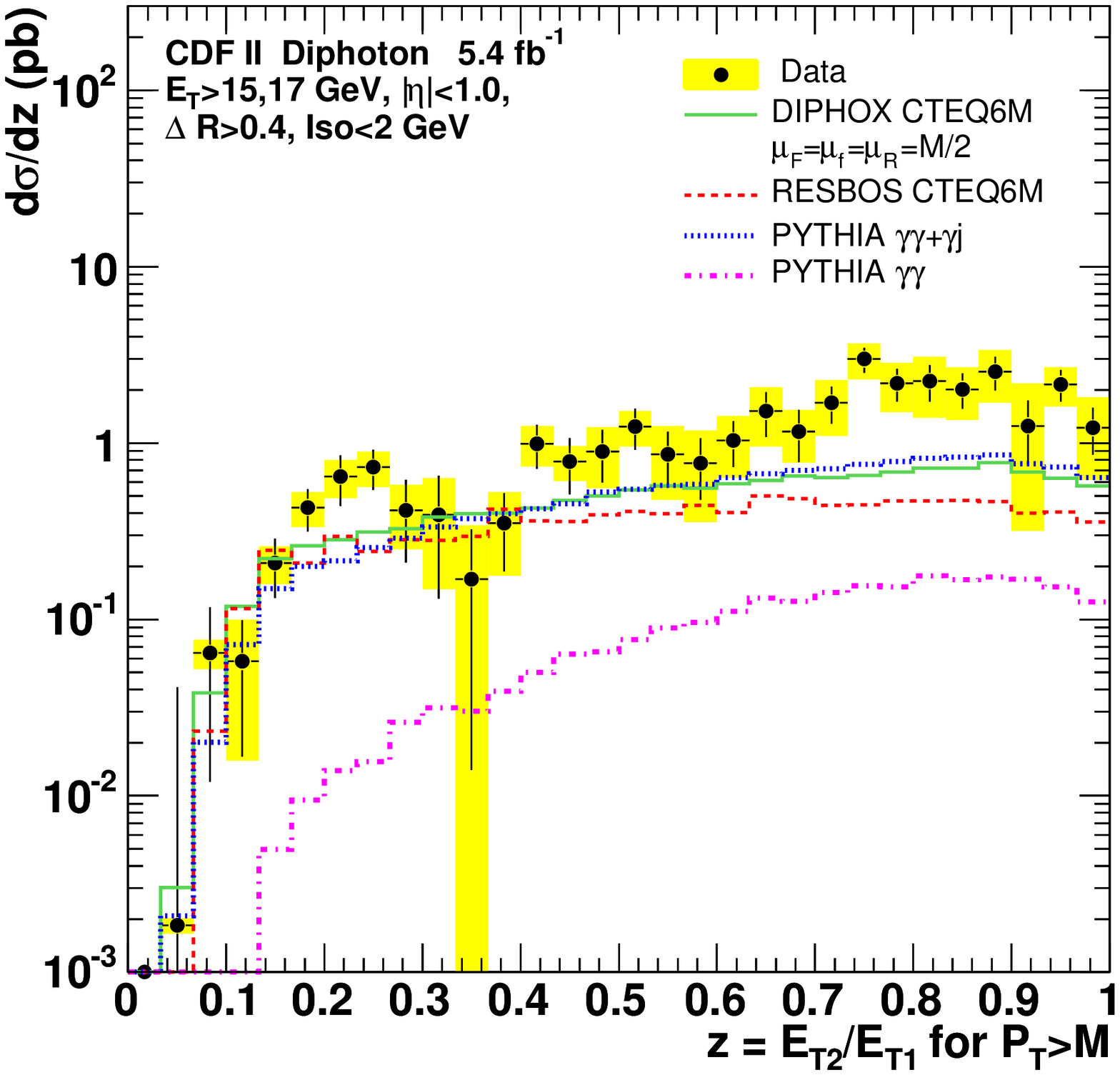}
\includegraphics[width=0.45\linewidth]
{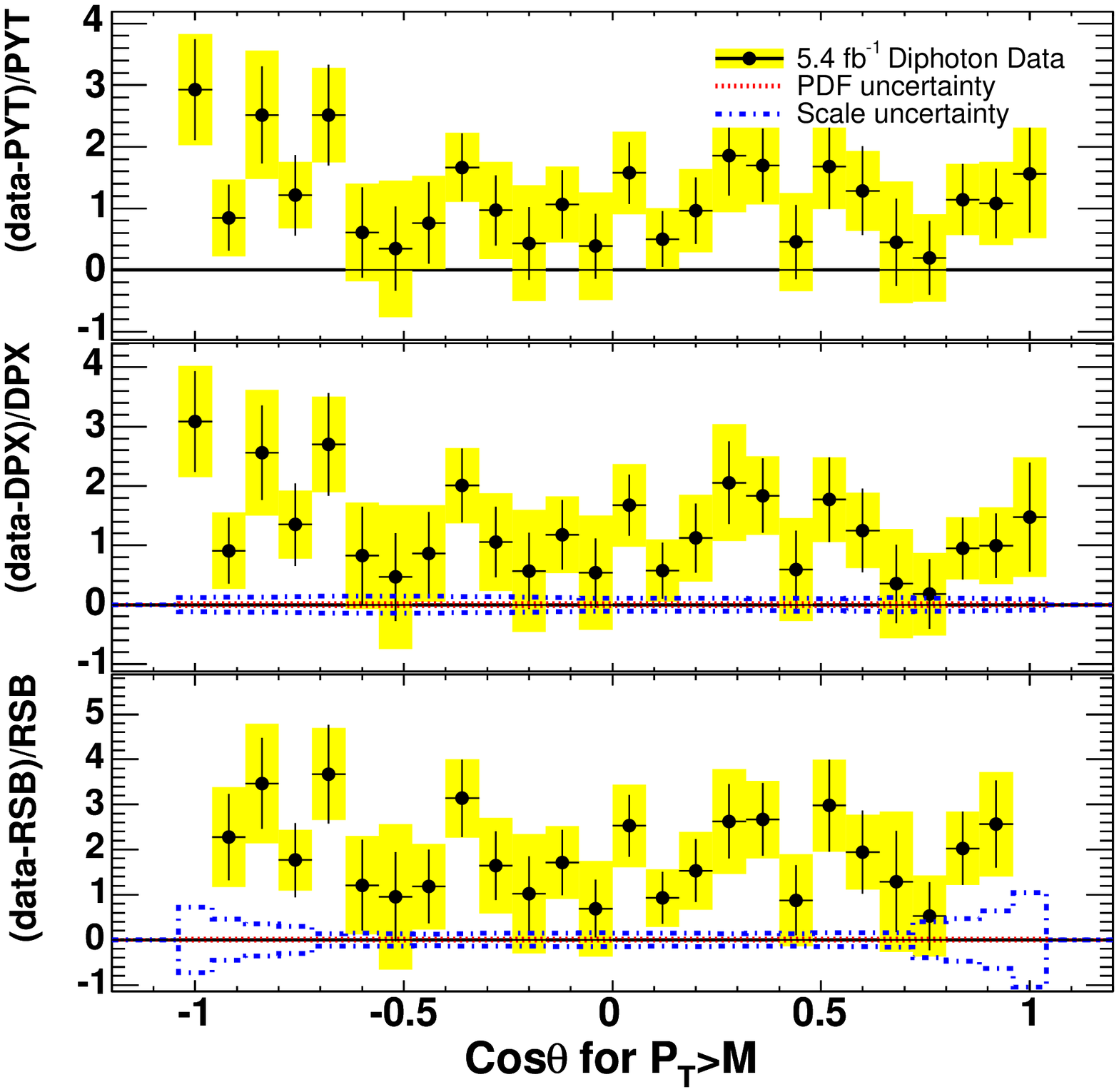}
\includegraphics[width=0.45\linewidth]{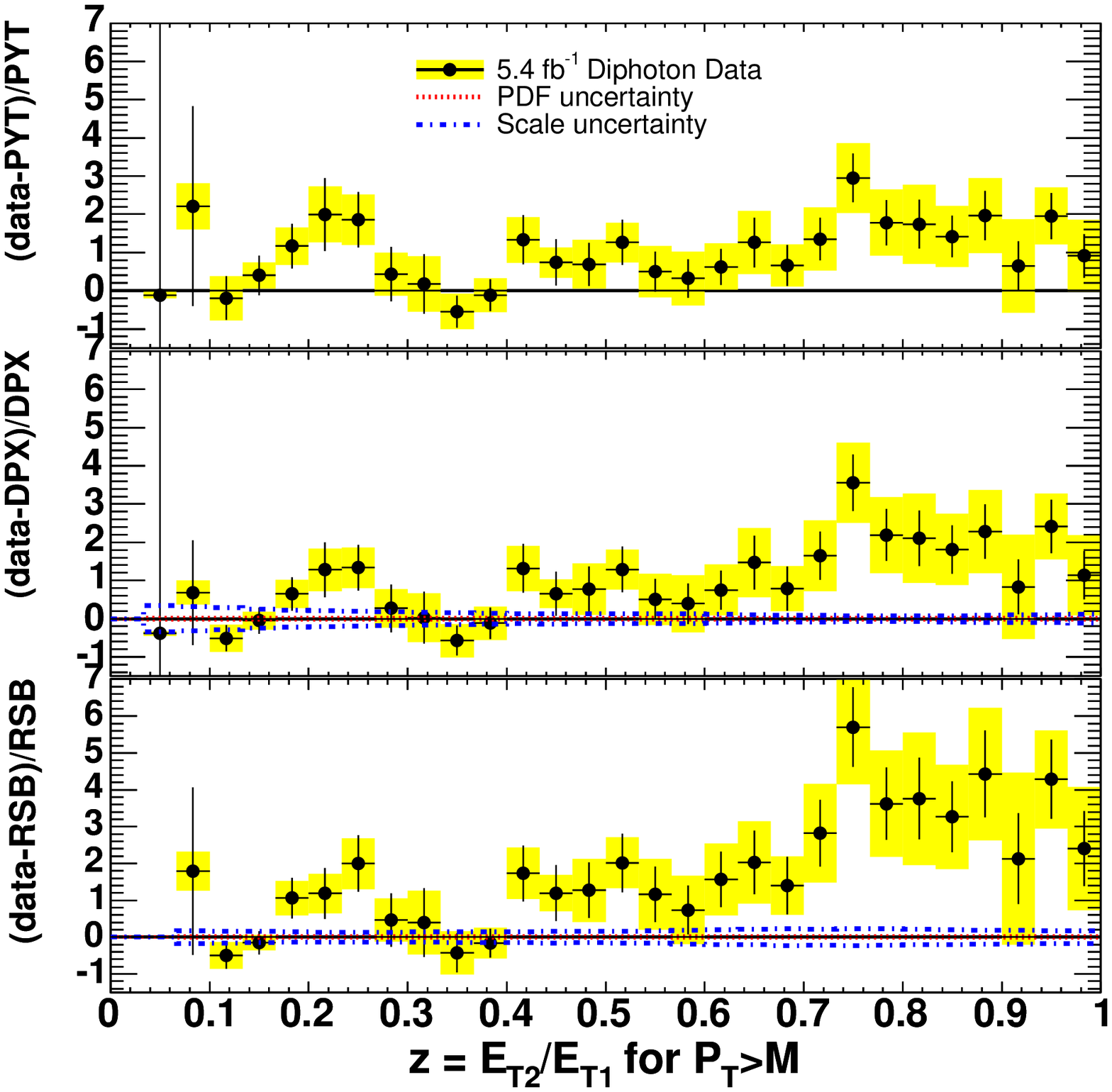}
\caption{The cross section as a function of the cosine of the polar angle in
     the Collins-Soper frame (left) and of the ratio of the subleading
     photon $E_{\rm T}$ to leading photon $E_{\rm T}$ (right) for
     $P_{\rm T}$$>$$M$. {\bf Top:} the absolute cross section values.
     {\bf Bottom:} the relative deviations of the data from the predictions.
     {\it Note:} the vertical axes scales differ between relative deviation
     plots. The shaded area is the total systematic uncertainty in the data.}
\label{fig:box_kin_xsec_cos}
\end{figure*}

Fig. \ref{fig:box_kin_xsec_cos} shows the results for $d\sigma/d\cos\theta$
and $d\sigma/dz$ for $P_{\rm T}$$>$$M$. In contrast with the unconstrained
kinematics and the $P_{\rm T}$$<$$M$ kinematics, in this case all three
calculations underestimate the data through the full ranges of the
$\cos\theta$ and $z$ spectra.

In general, events with low diphoton mass and high diphoton transverse
momentum, mainly coming from fragmentation, are not well described by the
examined calculations. This observation is important for measurements under
conditions where contributions from such events are strong, as in the LHC
\cite{higgs}.

\section{Conclusions}
\label{sec:summary}

In summary, the prompt diphoton production cross section, differential in
kinematic variables sensitive to the dynamics of the reaction mechanism,
is measured using data corresponding to an integrated of luminosity 5.36
fb$^{-1}$ collected with the CDF II detector. The large size of the data
sample allows for scanning a much more extended phase space and with a better
statistical precision than in earlier measurements. Using a novel technique
for the background subtraction, based on the track isolation, the overall
systematic uncertainty is limited to about 30$\%$ on average.

The results of the measurement are compared with three state-of-the-art
calculations, applying complementary techniques in modeling the reaction.
All three calculations describe events with large diphoton mass and small
diphoton transverse momentum fairly well, where the kinematics is similar
to the decay of a low-$P_{\rm T}$ heavy particle, such as the Higgs boson,
decaying into a photon pair. Exceptions are kinematic regions where gluon
interactions and the associated fragmentations of quarks into photons are
expected to be important, such as the low mass and azimuthal difference
regions and the region of the ``Guillet shoulder'' at moderate transverse
momentum. All three calculations underestimate the data in those regions.
Although the {\sc diphox} calculation explicitly includes a fragmentation
model, it fails to reproduce the data in those sensitive regions, possibly
because of the approximate nature of the requirement of photon isolation in
the {\sc diphox} framework. This requirement is mostly responsible for the
suppression of fragmentation contributions and is applied using hadron
variables in the data but using parton variables in {\sc diphox}. The low
transverse momentum and large azimuthal difference regions, where resummation
in the diphoton transverse momentum is important, are best described by
{\sc resbos}, as expected from the analytical resummation implemented in
this calculation. Photon radiation, especially from the initial state quarks,
in addition to the prompt photon production at the hard scattering, is for
the first time shown to play a very important role in the parton showering
{\sc pythia} calculation in order to bring the prediction into reasonable
agreement with the data.

\section{Acknowledgements}

We thank the Fermilab staff and the technical staffs of the participating
institutions for their vital contributions. We also thank P. Nadolsky, C.-P.
Yuan, Z. Li, J.-P. Guillet, C. Schmidt and S. Mrenna for their valuable help
in the theoretical calculations. This work was supported by the U.S.
Department of Energy and National Science Foundation; the Italian Istituto
Nazionale di Fisica Nucleare; the Ministry of Education, Culture, Sports,
Science and Technology of Japan; the Natural Sciences and Engineering Research
Council of Canada; the National Science Council of the Republic of China; the
Swiss National Science Foundation; the A.P. Sloan Foundation; the
Bundesministerium f\"ur Bildung und Forschung, Germany; the Korean World Class
University Program, the National Research Foundation of Korea; the Science and
Technology Facilities Council and the Royal Society, UK; the Institut National
de Physique Nucleaire et Physique des Particules/CNRS; the Russian Foundation
for Basic Research; the Ministerio de Ciencia e Innovaci\'{o}n, and Programa
Consolider-Ingenio 2010, Spain; the Slovak R\&D Agency; the Academy of
Finland; and the Australian Research Council (ARC). 

\appendix


\section{Content likelihood of the data sample}
\label{sec:likelihood}

The signal and background disentanglement is done in a maximum likelihood
framework. A likelihood of the composition of the baseline $\gamma\gamma$
sample is defined on the basis of binomial probabilities for the observed
photons to pass or fail the track isolation cut:

\begin{equation}
{\cal L}=
\prod_{ik}\mu_{i}(k)^{n_{i}(k)}{\left[1-\mu_{i}(k)\right]}^{1-n_{i}(k)}
\label{eq:lhood}
\end{equation}

\noindent
where $k=1,\ldots,N$ labels all events in the sample and $i=pp,pf,fp,ff$
labels the categories of events in which both photons pass, the leading
passes and the sub-leading fails, the leading fails and the sub-leading
passes, and both photons fail the track isolation cut, respectively.
$\mu_{i}(k)$ is the probability of the event $k$ to fall in the category
$i$ and $n_{i}(k)$ is the observation for the event $k$; i.e. $n_{i}(k)$
is 1 for one of the four categories $i$ and 0 for the other three. The
probabilities can be analyzed as follows:

\begin{equation}
\mu_{i}(k)=\sum_{j}\varepsilon_{ij}(k)p_{j}(k)
\label{eq:expct}
\end{equation}

\noindent
where $j=ss,sb,bs,bb$ labels the categories in which both photons are
signal, the leading is signal and the sub-leading background, the leading
is background and the sub-leading signal, and both photons are background,
respectively. $p_{j}(k)$ is the probability for an event $k$ to be in the
category $j$. $\varepsilon_{ij}(k)$ is the probability for an event $k$ of
the truth category $j$ to be observed in the observation category $i$. This
probability is directly related with the efficiencies of the two photons in
the event $k$ to pass the track isolation cut, since the efficiencies are
defined as the probabilities for the leading or sub-leading photon to pass
the cut. In a $4\times 4$ matrix notation,

\begin{widetext}
\begin{equation}
\varepsilon(k)=\left(\begin{array}{cccc}
\epsilon_{s1}(k)\epsilon_{s2}(k) & \epsilon_{s1}(k)\epsilon_{b2}(k) &
\epsilon_{b1}(k)\epsilon_{s2}(k) & \epsilon_{b1}(k)\epsilon_{b2}(k) \\
\epsilon_{s1}(k)(1-\epsilon_{s2}(k)) & \epsilon_{s1}(k)(1-\epsilon_{b2}(k)) &
\epsilon_{b1}(k)(1-\epsilon_{s2}(k)) & \epsilon_{b1}(k)(1-\epsilon_{b2}(k)) \\
(1-\epsilon_{s1}(k))\epsilon_{s2}(k) & (1-\epsilon_{s1}(k))\epsilon_{b2}(k) &
(1-\epsilon_{b1}(k))\epsilon_{s2}(k) & (1-\epsilon_{b1}(k))\epsilon_{b2}(k) \\
(1-\epsilon_{s1}(k))(1-\epsilon_{s2}(k)) &
(1-\epsilon_{s1}(k))(1-\epsilon_{b2}(k)) &
(1-\epsilon_{b1}(k))(1-\epsilon_{s2}(k)) &
(1-\epsilon_{b1}(k))(1-\epsilon_{b2}(k))\\
\end{array}\right)
\label{eq:mtrix}
\end{equation}
\end{widetext}

\noindent
where $\epsilon_{\alpha\beta}(k)$ ($\alpha=s$ or $b$, $\beta=1$ or 2)
are the efficiencies of the leading or sub-leading photon, coming from the
signal or from the background, to pass the track isolation cut.

The likelihood ${\cal L}$ is maximized or, equivalently, the opposite of
its natural logarithm

\begin{eqnarray}
L&=&-\ln{\cal L}\\
\nonumber
&=&\sum_{ik}\big\{n_{i}(k)\ln\mu_{i}(k)
+\left[1-n_{i}(k)\right]\ln\left[1-\mu_{i}(k)\right]\big\}
\label{eq:mnlog}
\end{eqnarray}

\noindent
is minimized with respect to the probabilities $p_{j}(k)$. The minimization
of the logarithm $L$ leads to the system of equations

\begin{equation}
n_{i}(k)=\mu_{i}(k)=\sum_{j}\varepsilon_{ij}(k)p_{j}(k)
\label{eq:obsvs}
\end{equation}

\noindent
with solutions

\begin{equation}
p_{j}(k)=\sum_{i}\varepsilon^{-1}_{ji}(k)n_{i}(k)
\label{eq:probs}
\end{equation}

\noindent
The choice of the track isolation cut at 1 GeV/$c$ gives the efficiencies
$\epsilon_{\alpha\beta}(k)$ sufficient discriminating power among the
truth categories $j$ for the matrix $\varepsilon(k)$ to be non-singular.
By summing the probabilties over all events in the baseline sample the
maximum likelihood composition of the sample is obtained:

\begin{equation}
w_{j}=\sum_{k}p_{j}(k)=\sum_{ik}\varepsilon^{-1}_{ji}(k)n_{i}(k)
\label{eq:fracs}
\end{equation}

\noindent
Eq. \ref{eq:fracs} for the truth category $j=ss$ provides the signal
fraction in the baseline $\gamma\gamma$ sample.

In general, the composition of a sample of events with $m$ photons each can be
resolved in a maximum likelihood framework by the inversion of a
$2^m\times 2^m$ matrix, constructed as in Eq. \ref{eq:mtrix}, which transforms
the probability $2^m$-vectors $p_{j}(k)$ to the observation $2^m$-vectors
$n_{i}(k)$. The generic matrix element $\varepsilon_{ij}(k)$ contains a factor
$\epsilon_{\alpha\beta}(k)$ or $1-\epsilon_{\alpha\beta}(k)$ for each photon
$\beta=1,~\! 2,~\!\ldots,~\! m$ with $\alpha=s$ $(b)$ if the photon is signal
(background) in the truth category $j$ and passes or fails, respectively, the
track isolation cut in the observation category $i$. In this context, Eq.
\ref{eq:phofr} for the single photon sample is derived from Eq. \ref{eq:fracs}
by inverting for each event the matrix

\begin{equation}
\varepsilon=\left(\begin{array}{cc}
\epsilon_{s} & \epsilon_{b} \\ 1-\epsilon_{s} & 1-\epsilon_{b} \\
\end{array}\right)\nonumber
\end{equation}
\begin{equation}
\Rightarrow\varepsilon^{-1}=\frac{1}{\epsilon_{s}-\epsilon_{b}}
\left(\begin{array}{cc}
1-\epsilon_{b} & -\epsilon_{b} \\ -1+\epsilon_{s} & \epsilon_{s} \\
\end{array}\right)
\label{eq:singl}
\end{equation}

\noindent
with observation vectors

\begin{equation}
n_{p}=\left(\begin{array}{c} 1 \\ 0 \\ \end{array}\right)~~~~~
n_{f}=\left(\begin{array}{c} 0 \\ 1 \\ \end{array}\right)
\label{eq:obsvec}
\end{equation}

\noindent
for photons passing $(n_{p})$ or failing $(n_{f})$ the track isolation cut.
Eq. \ref{eq:fracs} then gives for the signal fraction of the single photon
sample

\begin{eqnarray}
w_{s}&=&\sum_{p}\left(\frac{1-\epsilon_{b}}{\epsilon_{s}-\epsilon_{b}}\right)_{p}
+\sum_{f}\left(\frac{-\epsilon_{b}}{\epsilon_{s}-\epsilon_{b}}\right)_{f}
\nonumber \\ &=&
\sum_{i}\left(\frac{\epsilon-\epsilon_{b}}{\epsilon_{s}-\epsilon_{b}}\right)_{i}
\label{eq:purity}
\end{eqnarray}

\noindent
where, in the first line, the first sum runs over all photons passing the cut
and the second sum runs over all photons failing the cut. Eq. \ref{eq:purity}
is identical with Eq. \ref{eq:phofr}.


\section{Cross Section Tables}
\label{sec:xsec-tab}

This Appendix provides a set of Tables of the measured differential cross
section values as functions of the six kinematic variables selected in this
analysis. Each Table lists the bins of the selected variable, the values
of the cross section in the respective bins for the three examined cases of
no kinematic cut, $P_{\rm T}$$<$$M$, and $P_{\rm T}$$>$$M$, and the statistical and total
systematic uncertainties associated with each cross section value.

\begin{table*}[tbp]
\begin{center}
\caption{The diphoton production cross section differential in the diphoton
     invariant mass. The first error in the cross section is statistical and
     the second systematic.}
\label{tab:xsec-mass}
\begin{tabular}{cccc}
\hline\hline
Mass bin [GeV/$c^2$] & ~~~~~~Cross section without cut~~~~~~ & ~~~~~~Cross section for $P_{\rm T}$$<$$M$~~~~~~ & ~~~~~~Cross section for $P_{\rm T}$$>$$M$~~~~~~ \\
 & [pb/(GeV/$c^2$)] & [pb/(GeV/$c^2$)] & [pb/(GeV/$c^2$)] \\ \hline
0 - 5 & 0 & 0 & 0 \\
5 - 10 & $0.014679 \pm 0.002997 \pm 0.003369$ & 0 & $0.014679 \pm 0.002997 \pm 0.003301$ \\
10 - 15 & $0.004441 \pm 0.000607 \pm 0.001115$ & 0 & $0.036310 \pm 0.004965 \pm 0.009118$ \\
15 - 20 & $0.027568 \pm 0.005232 \pm 0.010986$ & 0 & $0.029520 \pm 0.005602 \pm 0.011764$ \\
20 - 25 & $0.037459 \pm 0.005482 \pm 0.012486$ & $0.000872 \pm 0.000686 \pm 0.000253$ & $0.039767 \pm 0.005921 \pm 0.013317$ \\
25 - 30 & $0.046105 \pm 0.006060 \pm 0.016932$ & $0.017214 \pm 0.004129 \pm 0.007523$ & $0.008772 \pm 0.001367 \pm 0.002915$ \\
30 - 35 & $0.263727 \pm 0.014488 \pm 0.072680$ & $0.242540 \pm 0.013891 \pm 0.066111$ & $0.016857 \pm 0.003275 \pm 0.005552$ \\
35 - 40 & $0.515524 \pm 0.020876 \pm 0.153195$ & $0.509522 \pm 0.021070 \pm 0.152077$ & $0.012408 \pm 0.002360 \pm 0.003196$ \\
40 - 45 & $0.478042 \pm 0.018964 \pm 0.128524$ & $0.467038 \pm 0.018734 \pm 0.125451$ & $0.007605 \pm 0.002024 \pm 0.002203$ \\
45 - 50 & $0.322442 \pm 0.015234 \pm 0.088299$ & $0.316623 \pm 0.015079 \pm 0.086365$ & $0.003078 \pm 0.001149 \pm 0.001156$ \\
50 - 55 & $0.207378 \pm 0.011824 \pm 0.057932$ & $0.201493 \pm 0.011730 \pm 0.056897$ & $0.005372 \pm 0.001299 \pm 0.001027$ \\
55 - 60 & $0.134243 \pm 0.009249 \pm 0.036287$ & $0.131094 \pm 0.009184 \pm 0.035695$ & $0.003014 \pm 0.001053 \pm 0.000599$ \\
60 - 65 & $0.092296 \pm 0.007243 \pm 0.023721$ & $0.089557 \pm 0.007169 \pm 0.023207$ & $0.002944 \pm 0.001071 \pm 0.000583$ \\
65 - 70 & $0.064259 \pm 0.006117 \pm 0.017943$ & $0.062955 \pm 0.006080 \pm 0.017584$ & $0.001183 \pm 0.000629 \pm 0.000328$ \\
70 - 75 & $0.049211 \pm 0.005194 \pm 0.013341$ & $0.047523 \pm 0.005150 \pm 0.012995$ & $0.000942 \pm 0.000396 \pm 0.000201$ \\
75 - 80 & $0.042325 \pm 0.004477 \pm 0.009863$ & $0.041887 \pm 0.004449 \pm 0.009750$ & $0.000480 \pm 0.000536 \pm 0.000126$ \\
80 - 85 & $0.033129 \pm 0.003469 \pm 0.006708$ & $0.032194 \pm 0.003436 \pm 0.006571$ & $0.000875 \pm 0.000443 \pm 0.000173$ \\
85 - 90 & $0.024546 \pm 0.003230 \pm 0.005261$ & $0.024111 \pm 0.003194 \pm 0.005127$ & $0.000522 \pm 0.000577 \pm 0.000253$ \\
90 - 95 & $0.016972 \pm 0.002675 \pm 0.004167$ & $0.016494 \pm 0.002657 \pm 0.004121$ & $0.000286 \pm 0.000191 \pm 0.000056$ \\
95 - 100 & $0.016820 \pm 0.002418 \pm 0.003531$ & $0.016531 \pm 0.002401 \pm 0.003486$ & $0.000402 \pm 0.000405 \pm 0.000122$ \\
100 - 110 & $0.011975 \pm 0.001421 \pm 0.002450$ & $0.011521 \pm 0.001407 \pm 0.002387$ & $0.000272 \pm 0.000118 \pm 0.000047$ \\
110 - 120 & $0.009187 \pm 0.001193 \pm 0.001782$ & $0.009037 \pm 0.001182 \pm 0.001747$ & $0.000145 \pm 0.000156 \pm 0.000037$ \\
120 - 130 & $0.006673 \pm 0.000968 \pm 0.001262$ & $0.006259 \pm 0.000953 \pm 0.001207$ & $0.000292 \pm 0.000119 \pm 0.000071$ \\
130 - 140 & $0.005805 \pm 0.000856 \pm 0.001031$ & $0.005742 \pm 0.000848 \pm 0.001017$ & $0.000065 \pm 0.000109 \pm 0.000029$ \\
140 - 160 & $0.003414 \pm 0.000448 \pm 0.000602$ & $0.003393 \pm 0.000444 \pm 0.000596$ & $0.000027 \pm 0.000072 \pm 0.000018$ \\
160 - 200 & $0.001801 \pm 0.000208 \pm 0.000303$ & $0.001757 \pm 0.000206 \pm 0.000296$ & $0.000015 \pm 0.000009 \pm 0.000002$ \\
200 - 250 & $0.000573 \pm 0.000107 \pm 0.000101$ & $0.000575 \pm 0.000107 \pm 0.000101$ & 0 \\
250 - 300 & $0.000182 \pm 0.000055 \pm 0.000030$ & $0.000182 \pm 0.000055 \pm 0.000030$ & 0 \\
300 - 350 & $0.000207 \pm 0.000055 \pm 0.000035$ & $0.000207 \pm 0.000055 \pm 0.000036$ & 0 \\
350 - 500 & $0.000064 \pm 0.000017 \pm 0.000011$ & $0.000064 \pm 0.000017 \pm 0.000011$ & 0 \\
\hline\hline
\end{tabular}
\end{center}
\end{table*}

\begin{table*}[tbp]
\begin{center}
\caption{The diphoton production cross section differential in the diphoton
     transverse momentum. The first error in the cross section is statistical
     and the second systematic.}
\label{tab:xsec-qT}
\begin{tabular}{cccc}
\hline\hline
$P_{\rm T}$ bin [GeV/$c$] & ~~~~~~Cross section without cut~~~~~~ & ~~~~~~Cross section for $P_{\rm T}$$<$$M$~~~~~~ & ~~~~~~Cross section for $P_{\rm T}$$>$$M$~~~~~~ \\
 & [pb/(GeV/$c$)] & [pb/(GeV/$c$)] & [pb/(GeV/$c$)] \\ \hline
 0 - 1 & $0.276549 \pm 0.030968 \pm 0.067613$ & $0.276542 \pm 0.030967 \pm 0.067612$ & 0 \\
 1 - 2 & $0.823081 \pm 0.051902 \pm 0.184444$ & $0.823132 \pm 0.051905 \pm 0.184456$ & 0 \\
 2 - 3 & $1.043320 \pm 0.056713 \pm 0.236673$ & $1.043373 \pm 0.056716 \pm 0.236685$ & 0 \\
 3 - 4 & $1.026369 \pm 0.055895 \pm 0.241145$ & $1.026574 \pm 0.055907 \pm 0.241193$ & 0 \\
 4 - 5 & $0.913944 \pm 0.055127 \pm 0.234272$ & $0.914185 \pm 0.055142 \pm 0.234334$ & 0 \\
 5 - 6 & $0.908628 \pm 0.053530 \pm 0.215222$ & $0.909148 \pm 0.053560 \pm 0.215345$ & 0 \\
 6 - 7 & $0.784121 \pm 0.049343 \pm 0.200779$ & $0.784624 \pm 0.049375 \pm 0.200908$ & 0 \\
 7 - 8 & $0.606111 \pm 0.045730 \pm 0.166151$ & $0.606435 \pm 0.045755 \pm 0.166240$ & 0 \\
 8 - 9 & $0.538172 \pm 0.042695 \pm 0.159898$ & $0.538651 \pm 0.042733 \pm 0.160040$ & 0 \\
 9 - 10 & $0.416951 \pm 0.039170 \pm 0.125580$ & $0.417071 \pm 0.039181 \pm 0.125616$ & 0 \\
10 - 12 & $0.371872 \pm 0.025701 \pm 0.113307$ & $0.372199 \pm 0.025723 \pm 0.113406$ & 0 \\
12 - 14 & $0.275282 \pm 0.022571 \pm 0.089464$ & $0.275797 \pm 0.022613 \pm 0.089631$ & 0 \\
14 - 16 & $0.196349 \pm 0.019572 \pm 0.072611$ & $0.196591 \pm 0.019597 \pm 0.072700$ & 0 \\
16 - 18 & $0.179301 \pm 0.017290 \pm 0.060538$ & $0.179549 \pm 0.017314 \pm 0.060622$ & 0 \\
18 - 20 & $0.125584 \pm 0.015368 \pm 0.045351$ & $0.126581 \pm 0.015490 \pm 0.045711$ & 0 \\
20 - 25 & $0.127625 \pm 0.009118 \pm 0.038128$ & $0.125027 \pm 0.009019 \pm 0.037597$ & $0.001932 \pm 0.000883 \pm 0.000457$ \\
25 - 30 & $0.092613 \pm 0.007839 \pm 0.028973$ & $0.078351 \pm 0.006931 \pm 0.024441$ & $0.013751 \pm 0.003737 \pm 0.004417$ \\
30 - 35 & $0.087187 \pm 0.007329 \pm 0.027382$ & $0.040507 \pm 0.004874 \pm 0.012880$ & $0.049065 \pm 0.005765 \pm 0.015309$ \\
35 - 40 & $0.056515 \pm 0.006341 \pm 0.020793$ & $0.024564 \pm 0.003619 \pm 0.007531$ & $0.033546 \pm 0.005550 \pm 0.014138$ \\
40 - 45 & $0.039824 \pm 0.004995 \pm 0.012246$ & $0.012094 \pm 0.002774 \pm 0.004861$ & $0.030706 \pm 0.004582 \pm 0.008315$ \\
45 - 50 & $0.022650 \pm 0.003604 \pm 0.006110$ & $0.010569 \pm 0.002027 \pm 0.002311$ & $0.012983 \pm 0.003290 \pm 0.004252$ \\
50 - 60 & $0.018204 \pm 0.002042 \pm 0.004416$ & $0.006578 \pm 0.001173 \pm 0.001669$ & $0.012472 \pm 0.001798 \pm 0.002971$ \\
60 - 70 & $0.007542 \pm 0.001410 \pm 0.002019$ & $0.003080 \pm 0.000766 \pm 0.000693$ & $0.004862 \pm 0.001318 \pm 0.001481$ \\
70 - 80 & $0.005717 \pm 0.001043 \pm 0.001271$ & $0.001773 \pm 0.000483 \pm 0.000331$ & $0.004388 \pm 0.001044 \pm 0.001080$ \\
80 - 90 & $0.003467 \pm 0.000656 \pm 0.000676$ & $0.000801 \pm 0.000292 \pm 0.000167$ & $0.003387 \pm 0.000749 \pm 0.000642$ \\
90 - 100 & $0.001355 \pm 0.000487 \pm 0.000377$ & $0.000079 \pm 0.000188 \pm 0.000103$ & $0.001242 \pm 0.000433 \pm 0.000279$ \\
100 - 120 & $0.001057 \pm 0.000212 \pm 0.000281$ & $0.000198 \pm 0.000097 \pm 0.000062$ & $0.000972 \pm 0.000212 \pm 0.000239$ \\
120 - 140 & $0.000655 \pm 0.000228 \pm 0.000135$ & $0.000059 \pm 0.000071 \pm 0.000017$ & $0.000516 \pm 0.000188 \pm 0.000094$ \\
140 - 160 & $0.000287 \pm 0.000122 \pm 0.000051$ & $0.000008 \pm 0.000030 \pm 0.000007$ & $0.001222 \pm 0.000496 \pm 0.000227$ \\
160 - 200 & $0.000100 \pm 0.000047 \pm 0.000021$ & 0 & $0.000227 \pm 0.000105 \pm 0.000039$ \\
\hline\hline
\end{tabular}
\end{center}
\end{table*}

\begin{table*}[tbp]
\begin{center}
\caption{The diphoton production cross section differential in the diphoton
     azimuthal difference. The first error in the cross section is statistical
     and the second systematic.}
\label{tab:xsec-Dphi}
\begin{tabular}{cccc}
\hline\hline
$\Delta\phi$ bin [radians] & ~~~~~~Cross section without cut~~~~~~ & ~~~~~~Cross section for $P_{\rm T}$$<$$M$~~~~~~ & ~~~~~~Cross section for $P_{\rm T}$$>$$M$~~~~~~ \\
 & [pb/rad] & [pb/rad] & [pb/rad] \\ \hline
 0.000 - 0.105 & $0.456343 \pm 0.099393 \pm 0.099826$ & $0.008959 \pm 0.006342 \pm 0.004135$ & $0.455591 \pm 0.101696 \pm 0.101433$ \\
 0.105 - 0.209 & $0.443854 \pm 0.103665 \pm 0.101738$ & 0 & $0.506682 \pm 0.116125 \pm 0.114737$ \\
 0.209 - 0.314 & $0.346811 \pm 0.090621 \pm 0.085583$ & $0.000964 \pm 0.000683 \pm 0.000446$ & $0.377053 \pm 0.101504 \pm 0.096562$ \\
 0.314 - 0.419 & $0.805494 \pm 0.112525 \pm 0.174161$ & $0.008695 \pm 0.006135 \pm 0.001692$ & $0.835398 \pm 0.118774 \pm 0.181144$ \\
 0.419 - 0.524 & $0.621352 \pm 0.134650 \pm 0.162677$ & 0 & $0.668658 \pm 0.139966 \pm 0.168432$ \\
 0.524 - 0.628 & $0.695012 \pm 0.138448 \pm 0.183259$ & $0.007809 \pm 0.011151 \pm 0.001958$ & $0.712287 \pm 0.143289 \pm 0.188042$ \\
 0.628 - 0.733 & $0.553260 \pm 0.152492 \pm 0.264377$ & $0.023257 \pm 0.020326 \pm 0.006509$ & $0.528230 \pm 0.151279 \pm 0.258421$ \\
 0.733 - 0.838 & $0.375583 \pm 0.144617 \pm 0.261528$ & 0 & $0.433524 \pm 0.158014 \pm 0.270138$ \\
 0.838 - 0.942 & $0.671584 \pm 0.143112 \pm 0.228641$ & $0.029115 \pm 0.020067 \pm 0.010039$ & $0.679242 \pm 0.151863 \pm 0.231038$ \\
 0.942 - 1.047 & $0.522112 \pm 0.137889 \pm 0.219688$ & $0.038178 \pm 0.018111 \pm 0.008311$ & $0.496178 \pm 0.145545 \pm 0.230738$ \\
 1.047 - 1.152 & $0.864793 \pm 0.144971 \pm 0.220201$ & $0.087654 \pm 0.032029 \pm 0.018939$ & $0.781381 \pm 0.145224 \pm 0.207021$ \\
 1.152 - 1.257 & $0.798674 \pm 0.154962 \pm 0.224791$ & $0.150640 \pm 0.043789 \pm 0.033689$ & $0.638090 \pm 0.150851 \pm 0.210793$ \\
 1.257 - 1.361 & $1.124851 \pm 0.168956 \pm 0.257447$ & $0.188806 \pm 0.056573 \pm 0.044931$ & $0.939944 \pm 0.162249 \pm 0.212925$ \\
 1.361 - 1.466 & $0.639906 \pm 0.169993 \pm 0.284205$ & $0.118937 \pm 0.079298 \pm 0.074899$ & $0.527203 \pm 0.151112 \pm 0.210203$ \\
 1.466 - 1.571 & $0.834419 \pm 0.171606 \pm 0.270175$ & $0.421297 \pm 0.113865 \pm 0.107330$ & $0.408768 \pm 0.128128 \pm 0.171398$ \\
 1.571 - 1.676 & $1.250665 \pm 0.169418 \pm 0.311391$ & $1.007357 \pm 0.159164 \pm 0.278003$ & $0.230173 \pm 0.057530 \pm 0.047005$ \\
 1.676 - 1.780 & $0.832275 \pm 0.186990 \pm 0.364452$ & $0.826622 \pm 0.181796 \pm 0.334001$ & $0.006883 \pm 0.044420 \pm 0.000755$ \\
 1.780 - 1.885 & $1.001287 \pm 0.187680 \pm 0.340539$ & $0.998464 \pm 0.185177 \pm 0.326921$ & $0.004982 \pm 0.030035 \pm 0.000548$ \\
 1.885 - 1.990 & $1.076470 \pm 0.209360 \pm 0.442750$ & $0.985068 \pm 0.206856 \pm 0.435764$ & $0.097539 \pm 0.033960 \pm 0.017947$ \\
 1.990 - 2.094 & $1.667315 \pm 0.233681 \pm 0.501973$ & $1.628501 \pm 0.232554 \pm 0.498599$ & $0.030895 \pm 0.019985 \pm 0.005693$ \\
 2.094 - 2.199 & $2.275603 \pm 0.257154 \pm 0.585295$ & $2.271233 \pm 0.256406 \pm 0.580308$ & $0.005091 \pm 0.018432 \pm 0.007031$ \\
 2.199 - 2.304 & $1.982347 \pm 0.275434 \pm 0.703905$ & $1.945552 \pm 0.274931 \pm 0.703233$ & $0.021202 \pm 0.011392 \pm 0.004164$ \\
 2.304 - 2.409 & $3.180244 \pm 0.298472 \pm 0.883152$ & $3.185517 \pm 0.298001 \pm 0.878647$ & 0 \\
 2.409 - 2.513 & $3.344205 \pm 0.354571 \pm 1.099021$ & $3.345126 \pm 0.354498 \pm 1.097423$ & $0.000142 \pm 0.006657 \pm 0.000016$ \\
 2.513 - 2.618 & $4.913562 \pm 0.403046 \pm 1.447020$ & $4.883607 \pm 0.402930 \pm 1.445024$ & $0.010602 \pm 0.005301 \pm 0.001996$ \\
 2.618 - 2.723 & $6.787434 \pm 0.476941 \pm 1.881799$ & $6.788491 \pm 0.477385 \pm 1.881996$ & $0.000988 \pm 0.001690 \pm 0.000302$ \\
 2.723 - 2.827 & $9.949192 \pm 0.584235 \pm 2.773900$ & $9.948025 \pm 0.584526 \pm 2.773673$ & $0.001445 \pm 0.002278 \pm 0.000396$ \\
 2.827 - 2.932 & $14.781949 \pm 0.694936 \pm 3.966263$ & $14.791632 \pm 0.695391 \pm 3.968861$ & 0 \\
 2.932 - 3.037 & $21.597660 \pm 0.861897 \pm 5.826928$ & $21.602392 \pm 0.862322 \pm 5.828225$ & $0.000218 \pm 0.000352 \pm 0.000061$ \\
 3.037 - 3.142 & $33.827076 \pm 0.998489 \pm 8.203793$ & $33.825920 \pm 0.998661 \pm 8.204532$ & $0.000133 \pm 0.000133 \pm 0.000024$ \\
\hline\hline
\end{tabular}
\end{center}
\end{table*}

\begin{table*}[tbp]
\begin{center}
\caption{The diphoton production cross section differential in the diphoton
     rapidity. The first error in the cross section is statistical and the
     second systematic.}
\label{tab:xsec-Y}
\begin{tabular}{cccc}
\hline\hline
~~~~~~~$Y_{gg}$ bin~~~~~~~ & ~~~~~~Cross section without cut~~~~~~ & ~~~~~~Cross section for $P_{\rm T}$$<$$M$~~~~~~ & ~~~~~~Cross section for $P_{\rm T}$$>$$M$~~~~~~ \\
 & [pb] & [pb] & [pb] \\ \hline
$-$1.20 - $-$1.12 & $0.047796 \pm 0.032826 \pm 0.014355$ & $0.047253 \pm 0.027906 \pm 0.017471$ & $0.000543 \pm 0.017286 \pm 0.000059$ \\
$-$1.12 - $-$1.04 & $0.357094 \pm 0.120949 \pm 0.102530$ & $0.353126 \pm 0.113517 \pm 0.089113$ & $0.003968 \pm 0.041744 \pm 0.000434$ \\
$-$1.04 - $-$0.96 & $0.187965 \pm 0.037254 \pm 0.047292$ & $0.179976 \pm 0.034997 \pm 0.041721$ & $0.004941 \pm 0.006801 \pm 0.004111$ \\
$-$0.96 - $-$0.88 & $1.216128 \pm 0.179918 \pm 0.350105$ & $1.123226 \pm 0.169654 \pm 0.303864$ & $0.087621 \pm 0.056102 \pm 0.049798$ \\
$-$0.88 - $-$0.80 & $2.934750 \pm 0.300535 \pm 0.734394$ & $2.634538 \pm 0.284117 \pm 0.653983$ & $0.284037 \pm 0.092706 \pm 0.076637$ \\
$-$0.80 - $-$0.72 & $3.523381 \pm 0.364743 \pm 1.002344$ & $3.275366 \pm 0.344374 \pm 0.929432$ & $0.243582 \pm 0.121183 \pm 0.071961$ \\
$-$0.72 - $-$0.64 & $4.727782 \pm 0.415361 \pm 1.224628$ & $4.251420 \pm 0.388978 \pm 1.079371$ & $0.442221 \pm 0.136633 \pm 0.142010$ \\
$-$0.64 - $-$0.56 & $5.043031 \pm 0.463117 \pm 1.552137$ & $4.391841 \pm 0.437811 \pm 1.375512$ & $0.704898 \pm 0.160548 \pm 0.190566$ \\
$-$0.56 - $-$0.48 & $6.849294 \pm 0.521381 \pm 1.783643$ & $6.156165 \pm 0.496178 \pm 1.631273$ & $0.704821 \pm 0.161752 \pm 0.160419$ \\
$-$0.48 - $-$0.40 & $7.515287 \pm 0.565874 \pm 1.957807$ & $6.894050 \pm 0.542207 \pm 1.818195$ & $0.612712 \pm 0.159704 \pm 0.141450$ \\
$-$0.40 - $-$0.32 & $7.432176 \pm 0.605607 \pm 2.230531$ & $7.009432 \pm 0.582405 \pm 2.029948$ & $0.387476 \pm 0.149935 \pm 0.191711$ \\
$-$0.32 - $-$0.24 & $9.473290 \pm 0.644927 \pm 2.438148$ & $9.018377 \pm 0.624024 \pm 2.249574$ & $0.448143 \pm 0.156451 \pm 0.207981$ \\
$-$0.24 - $-$0.16 & $9.171682 \pm 0.650452 \pm 2.482947$ & $8.178396 \pm 0.625007 \pm 2.276978$ & $1.030157 \pm 0.186527 \pm 0.232884$ \\
$-$0.16 - $-$0.08 & $10.391827 \pm 0.673024 \pm 2.604154$ & $9.833859 \pm 0.649550 \pm 2.409783$ & $0.581114 \pm 0.183635 \pm 0.235125$ \\
$-$0.08 - 0 & $10.685762 \pm 0.683909 \pm 2.779428$ & $9.737009 \pm 0.655963 \pm 2.536669$ & $1.011320 \pm 0.205173 \pm 0.258684$ \\
 0 - 0.08 & $9.778326 \pm 0.680032 \pm 2.746355$ & $8.910962 \pm 0.656693 \pm 2.541139$ & $0.779653 \pm 0.160122 \pm 0.189297$ \\
 0.08 - 0.16 & $10.245283 \pm 0.664715 \pm 2.572113$ & $9.428201 \pm 0.642795 \pm 2.382707$ & $0.844021 \pm 0.173754 \pm 0.196508$ \\
 0.16 - 0.24 & $8.898950 \pm 0.650517 \pm 2.522506$ & $8.174833 \pm 0.625682 \pm 2.299986$ & $0.740208 \pm 0.181740 \pm 0.229029$ \\
 0.24 - 0.32 & $9.873399 \pm 0.639878 \pm 2.378116$ & $8.881087 \pm 0.614339 \pm 2.143591$ & $0.980913 \pm 0.177119 \pm 0.232361$ \\
 0.32 - 0.40 & $8.208516 \pm 0.599002 \pm 2.150160$ & $7.608601 \pm 0.571739 \pm 1.962087$ & $0.575259 \pm 0.170775 \pm 0.186692$ \\
 0.40 - 0.48 & $8.310832 \pm 0.582531 \pm 2.114900$ & $7.805289 \pm 0.557393 \pm 1.950114$ & $0.489608 \pm 0.162967 \pm 0.168822$ \\
 0.48 - 0.56 & $6.158322 \pm 0.525197 \pm 1.682342$ & $5.551052 \pm 0.501466 \pm 1.536281$ & $0.602316 \pm 0.155006 \pm 0.146833$ \\
 0.56 - 0.64 & $5.764018 \pm 0.484449 \pm 1.615546$ & $5.254852 \pm 0.463164 \pm 1.480387$ & $0.515143 \pm 0.143704 \pm 0.136823$ \\
 0.64 - 0.72 & $3.746496 \pm 0.413343 \pm 1.272077$ & $3.516958 \pm 0.394670 \pm 1.144314$ & $0.219857 \pm 0.116288 \pm 0.126414$ \\
 0.72 - 0.80 & $2.638815 \pm 0.344297 \pm 0.937298$ & $2.365470 \pm 0.327398 \pm 0.833140$ & $0.236348 \pm 0.092090 \pm 0.090185$ \\
 0.80 - 0.88 & $1.908901 \pm 0.278836 \pm 0.661709$ & $1.745827 \pm 0.263054 \pm 0.558774$ & $0.178584 \pm 0.101977 \pm 0.123069$ \\
 0.88 - 0.96 & $1.342328 \pm 0.185391 \pm 0.348576$ & $1.092484 \pm 0.174767 \pm 0.311890$ & $0.266332 \pm 0.065150 \pm 0.055395$ \\
 0.96 - 1.04 & $0.125836 \pm 0.034791 \pm 0.043927$ & $0.121207 \pm 0.033430 \pm 0.042727$ & $0.002452 \pm 0.004579 \pm 0.000700$ \\
 1.04 - 1.12 & $0.251782 \pm 0.113387 \pm 0.068974$ & $0.278853 \pm 0.108155 \pm 0.064602$ & 0 \\
 1.12 - 1.20 & $0.001304 \pm 0.029359 \pm 0.000143$ & $0.001304 \pm 0.029359 \pm 0.000143$ & 0 \\
\hline\hline
\end{tabular}
\end{center}
\end{table*}

\begin{table*}[tbp]
\begin{center}
\caption{The diphoton production cross section differential in the cosine of
     the polar angle in Collins-Soper frame. The first error in the cross
     section is statistical and the second systematic.}
\label{tab:xsec-cos}
\begin{tabular}{cccc}
\hline\hline
~~~~~~~$\cos\theta$ bin~~~~~~~ & ~~~~~~Cross section without cut~~~~~~ & ~~~~~~Cross section for $P_{\rm T}$$<$$M$~~~~~~ & ~~~~~~Cross section for $P_{\rm T}$$>$$M$~~~~~~ \\
 & [pb] & [pb] & [pb] \\ \hline
$-$1.20 - $-$1.12 & 0 & 0 & 0 \\
$-$1.12 - $-$1.04 & 0 & 0 & 0 \\
$-$1.04 - $-$0.96 & $0.696711 \pm 0.146723 \pm 0.155742$ & $0.000604 \pm 0.012602 \pm 0.000066$ & $0.679356 \pm 0.141129 \pm 0.153197$ \\
$-$0.96 - $-$0.88 & $0.593058 \pm 0.160701 \pm 0.183348$ & $0.043900 \pm 0.032514 \pm 0.018749$ & $0.534910 \pm 0.155592 \pm 0.157684$ \\
$-$0.88 - $-$0.80 & $0.901409 \pm 0.165027 \pm 0.221503$ & $0.214033 \pm 0.066821 \pm 0.046916$ & $0.679008 \pm 0.152504 \pm 0.179300$ \\
$-$0.80 - $-$0.72 & $0.964840 \pm 0.155297 \pm 0.234473$ & $0.686284 \pm 0.131504 \pm 0.167953$ & $0.394443 \pm 0.116872 \pm 0.094604$ \\
$-$0.72 - $-$0.64 & $2.322500 \pm 0.330788 \pm 0.674056$ & $1.895647 \pm 0.321399 \pm 0.603007$ & $0.702685 \pm 0.163956 \pm 0.152732$ \\
$-$0.64 - $-$0.56 & $3.911811 \pm 0.353577 \pm 0.984306$ & $3.971193 \pm 0.364919 \pm 0.976839$ & $0.308842 \pm 0.140763 \pm 0.123241$ \\
$-$0.56 - $-$0.48 & $4.225161 \pm 0.385762 \pm 1.118926$ & $4.381881 \pm 0.405094 \pm 1.105399$ & $0.225221 \pm 0.114141 \pm 0.141418$ \\
$-$0.48 - $-$0.40 & $5.878916 \pm 0.500103 \pm 1.841742$ & $5.998004 \pm 0.523626 \pm 1.862583$ & $0.328751 \pm 0.123479 \pm 0.118609$ \\
$-$0.40 - $-$0.32 & $8.447426 \pm 0.558218 \pm 2.115941$ & $8.476556 \pm 0.589264 \pm 2.168045$ & $0.611133 \pm 0.126341 \pm 0.129931$ \\
$-$0.32 - $-$0.24 & $8.006947 \pm 0.563629 \pm 2.281457$ & $8.205863 \pm 0.595600 \pm 2.316283$ & $0.424552 \pm 0.122576 \pm 0.141255$ \\
$-$0.24 - $-$0.16 & $9.621518 \pm 0.610748 \pm 2.601253$ & $10.100368 \pm 0.646808 \pm 2.669680$ & $0.352094 \pm 0.144908 \pm 0.177738$ \\
$-$0.16 - $-$0.08 & $11.031545 \pm 0.647516 \pm 3.074353$ & $11.400319 \pm 0.685597 \pm 3.182760$ & $0.544025 \pm 0.146581 \pm 0.146555$ \\
$-$0.08 - 0 & $11.701358 \pm 0.652059 \pm 3.229385$ & $12.252276 \pm 0.687876 \pm 3.313652$ & $0.409344 \pm 0.154987 \pm 0.199857$ \\
 0 - 0.08 & $10.744885 \pm 0.639661 \pm 3.180184$ & $10.786094 \pm 0.672314 \pm 3.242347$ & $0.808290 \pm 0.156327 \pm 0.195042$ \\
 0.08 - 0.16 & $10.215599 \pm 0.620273 \pm 2.867342$ & $10.599528 \pm 0.657294 \pm 2.971375$ & $0.475211 \pm 0.141936 \pm 0.137377$ \\
 0.16 - 0.24 & $8.894859 \pm 0.591847 \pm 2.681245$ & $9.066109 \pm 0.621564 \pm 2.735215$ & $0.593223 \pm 0.163517 \pm 0.176269$ \\
 0.24 - 0.32 & $8.996937 \pm 0.573034 \pm 2.308823$ & $9.083449 \pm 0.602347 \pm 2.317747$ & $0.814932 \pm 0.184770 \pm 0.227355$ \\
 0.32 - 0.40 & $8.131862 \pm 0.550158 \pm 2.082427$ & $8.160151 \pm 0.578638 \pm 2.111536$ & $0.687574 \pm 0.151101 \pm 0.156505$ \\
 0.40 - 0.48 & $7.136364 \pm 0.535862 \pm 1.790615$ & $7.337457 \pm 0.558791 \pm 1.797939$ & $0.363666 \pm 0.150402 \pm 0.157323$ \\
 0.48 - 0.56 & $6.314780 \pm 0.443943 \pm 1.479071$ & $6.414981 \pm 0.468845 \pm 1.498226$ & $0.642568 \pm 0.164994 \pm 0.156587$ \\
 0.56 - 0.64 & $4.206630 \pm 0.370591 \pm 1.022774$ & $4.311754 \pm 0.395680 \pm 1.044270$ & $0.438339 \pm 0.137911 \pm 0.112795$ \\
 0.64 - 0.72 & $2.507368 \pm 0.287244 \pm 0.605107$ & $2.372216 \pm 0.277399 \pm 0.544206$ & $0.242620 \pm 0.118928 \pm 0.127240$ \\
 0.72 - 0.80 & $0.723595 \pm 0.163114 \pm 0.254653$ & $0.508954 \pm 0.128116 \pm 0.162397$ & $0.253995 \pm 0.126498 \pm 0.118790$ \\
 0.80 - 0.88 & $0.553629 \pm 0.131739 \pm 0.141735$ & $0.117661 \pm 0.060000 \pm 0.032288$ & $0.422761 \pm 0.113453 \pm 0.106068$ \\
 0.88 - 0.96 & $0.486210 \pm 0.125296 \pm 0.141378$ & $0.046081 \pm 0.036977 \pm 0.015266$ & $0.420675 \pm 0.114459 \pm 0.120587$ \\
 0.96 - 1.04 & $0.389773 \pm 0.141480 \pm 0.142265$ & $0.014859 \pm 0.023566 \pm 0.010889$ & $0.372354 \pm 0.138736 \pm 0.128626$ \\
 1.04 - 1.12 & 0 & 0 & 0 \\
 1.12 - 1.20 & 0 & 0 & 0 \\
\hline\hline
\end{tabular}
\end{center}
\end{table*}

\begin{table*}[tbp]
\begin{center}
\caption{The diphoton production cross section differential in the
     sub-leading to leading photon transverse momentum ratio. The first
     error in the cross section is statistical and the second systematic.}
\label{tab:xsec-z}
\begin{tabular}{cccc}
\hline\hline
~~~~~~~$z$ bin~~~~~~~ & ~~~~~~Cross section without cut~~~~~~ & ~~~~~~Cross section for $P_{\rm T}$$<$$M$~~~~~~ & ~~~~~~Cross section for $P_{\rm T}$$>$$M$~~~~~~ \\
 & [pb] & [pb] & [pb] \\ \hline
 0 - 0.033 & 0 & 0 & 0 \\
 0.033 - 0.067 & $0.001845 \pm 0.039308 \pm 0.000183$ & 0 & $0.001845 \pm 0.039308 \pm 0.000183$ \\
 0.067 - 0.100 & $0.064683 \pm 0.052631 \pm 0.011964$ & 0 & $0.064683 \pm 0.052631 \pm 0.011541$ \\
 0.100 - 0.133 & $0.040160 \pm 0.027228 \pm 0.018852$ & $0.014680 \pm 0.032433 \pm 0.004628$ & $0.059705 \pm 0.042528 \pm 0.042287$ \\
 0.133 - 0.167 & $0.208681 \pm 0.066194 \pm 0.042837$ & $0.011290 \pm 0.006518 \pm 0.002107$ & $0.219178 \pm 0.080746 \pm 0.052940$ \\
 0.167 - 0.200 & $0.581526 \pm 0.144204 \pm 0.158584$ & $0.130976 \pm 0.080148 \pm 0.047058$ & $0.421745 \pm 0.112886 \pm 0.095122$ \\
 0.200 - 0.233 & $0.687775 \pm 0.179968 \pm 0.138709$ & $0.220290 \pm 0.099866 \pm 0.041656$ & $0.624676 \pm 0.198432 \pm 0.138087$ \\
 0.233 - 0.267 & $0.882504 \pm 0.219745 \pm 0.200850$ & $0.311307 \pm 0.158139 \pm 0.120736$ & $0.662516 \pm 0.169350 \pm 0.152908$ \\
 0.267 - 0.300 & $1.254394 \pm 0.290529 \pm 0.272821$ & $0.891524 \pm 0.232063 \pm 0.173525$ & $0.380905 \pm 0.188202 \pm 0.125645$ \\
 0.300 - 0.333 & $0.742542 \pm 0.342084 \pm 0.399154$ & $0.427234 \pm 0.266842 \pm 0.247371$ & $0.362702 \pm 0.241421 \pm 0.173907$ \\
 0.333 - 0.367 & $0.633573 \pm 0.347426 \pm 0.629057$ & $0.465991 \pm 0.313708 \pm 0.505393$ & $0.171920 \pm 0.157777 \pm 0.130833$ \\
 0.367 - 0.400 & $2.120037 \pm 0.452224 \pm 0.852624$ & $1.715373 \pm 0.410393 \pm 0.695311$ & $0.360338 \pm 0.169736 \pm 0.140800$ \\
 0.400 - 0.433 & $3.736916 \pm 0.590222 \pm 1.123369$ & $2.770300 \pm 0.524436 \pm 0.913390$ & $1.008748 \pm 0.280388 \pm 0.241464$ \\
 0.433 - 0.467 & $3.631943 \pm 0.628078 \pm 1.112420$ & $2.889061 \pm 0.570550 \pm 0.962623$ & $0.784327 \pm 0.273687 \pm 0.172160$ \\
 0.467 - 0.500 & $4.521544 \pm 0.752621 \pm 1.788419$ & $3.644498 \pm 0.692765 \pm 1.516931$ & $0.862816 \pm 0.286054 \pm 0.268761$ \\
 0.500 - 0.533 & $4.504285 \pm 0.765007 \pm 2.208000$ & $3.370450 \pm 0.701471 \pm 2.016244$ & $1.208186 \pm 0.316267 \pm 0.270012$ \\
 0.533 - 0.567 & $5.323637 \pm 0.834206 \pm 2.560178$ & $4.458064 \pm 0.775756 \pm 2.247832$ & $0.852710 \pm 0.299524 \pm 0.308661$ \\
 0.567 - 0.600 & $7.041574 \pm 0.923763 \pm 2.652894$ & $6.228433 \pm 0.870212 \pm 2.309346$ & $0.758477 \pm 0.289013 \pm 0.321800$ \\
 0.600 - 0.633 & $8.403337 \pm 1.002178 \pm 3.176046$ & $7.326917 \pm 0.951026 \pm 2.835386$ & $1.050768 \pm 0.308693 \pm 0.336094$ \\
 0.633 - 0.667 & $10.720245 \pm 1.130024 \pm 3.698758$ & $9.382812 \pm 1.061512 \pm 3.277018$ & $1.561377 \pm 0.447592 \pm 0.490694$ \\
 0.667 - 0.700 & $12.252908 \pm 1.215009 \pm 4.376297$ & $11.087163 \pm 1.154250 \pm 4.032151$ & $1.190425 \pm 0.387427 \pm 0.352079$ \\
 0.700 - 0.733 & $16.415794 \pm 1.325045 \pm 5.439192$ & $14.773602 \pm 1.265468 \pm 4.943280$ & $1.709796 \pm 0.404441 \pm 0.516840$ \\
 0.733 - 0.767 & $24.203773 \pm 1.475829 \pm 6.110600$ & $21.437672 \pm 1.403616 \pm 5.494815$ & $2.989357 \pm 0.484161 \pm 0.680585$ \\
 0.767 - 0.800 & $25.442360 \pm 1.568095 \pm 6.926982$ & $23.305996 \pm 1.500378 \pm 6.332095$ & $2.170811 \pm 0.461754 \pm 0.605827$ \\
 0.800 - 0.833 & $26.262505 \pm 1.639982 \pm 7.542237$ & $24.221176 \pm 1.570060 \pm 6.897027$ & $2.164903 \pm 0.503890 \pm 0.696416$ \\
 0.833 - 0.867 & $33.239754 \pm 1.758692 \pm 8.786769$ & $31.182541 \pm 1.696572 \pm 8.184977$ & $2.019600 \pm 0.455356 \pm 0.595763$ \\
 0.867 - 0.900 & $43.423592 \pm 1.926264 \pm 10.230121$ & $41.122257 \pm 1.860713 \pm 9.587927$ & $2.478693 \pm 0.538976 \pm 0.732571$ \\
 0.900 - 0.933 & $41.876434 \pm 1.940630 \pm 9.835004$ & $40.736725 \pm 1.885258 \pm 9.372191$ & $1.262268 \pm 0.497901 \pm 0.732053$ \\
 0.933 - 0.967 & $46.414635 \pm 2.069529 \pm 10.663040$ & $44.152313 \pm 2.017412 \pm 10.112022$ & $2.116757 \pm 0.431365 \pm 0.515146$ \\
 0.967 - 1.000 & $43.381981 \pm 2.094066 \pm 10.489670$ & $41.883457 \pm 2.042450 \pm 9.953175$ & $1.217864 \pm 0.363204 \pm 0.483372$ \\
\hline\hline
\end{tabular}
\end{center}
\end{table*}


\begin{thebibliography}{99}

\bibitem{higgs} T. Aaltonen {\it et al.} (CDF Collaboration), Phys. Rev. Lett.
{\bf 103}, 061803 (2009); V. M. Abazov {\it et al.} (D0 Collaboration),
Phys. Rev. Lett. {\bf 102}, 231801 (2009); G. Aad {\it et al.}
(ATLAS Collaboration), arXiv:0901.0512 [hep-ex] (2009); G. L. Bayatian
{\it et al.} (CMS Collaboration), J. Phys. G {\bf 34}, 995 (2007).
\bibitem{grav} T. Aaltonen {\it et al.} (CDF Collaboration), Phys. Rev. Lett.
{\bf 99}, 171801 (2007); T. Aaltonen {\it et al.} (CDF Collaboration), Phys.
Rev. D {\bf 83}, 011102 (2011); The CMS Collaboration, J. High Energy Phys. 05
(2011) 085; The ATLAS Collaboration, arXiv:1107.0561v2 [hep-ex] (2011).
\bibitem{pythia} T. Sj{\"o}strand, Comput. Phys. Commun. {\bf 82}, 74 (1994);
S. Mrenna, Comput. Phys. Commun. {\bf 101}, 232 (1997).
\bibitem{herwig} G. Corcella, I. G. Knowles, G. Marchesini, S. Moretti,
K. Odagiri, P. Richardson, M. H. Seymour and B. R. Webber, J. High Energy Phys.
01 (2001) 010.
\bibitem{sherpa} S. H{\"o}che, S. Schumann and F. Siegert, Phys. Rev.
{\bf D81}, 034026 (2010).
\bibitem{diphox} T. Binoth, J. P. Guillet, E. Pilon and M. Werlen,
Eur. Phys. J. {\bf C16}, 311 (2000); T. Binoth, J. P. Guillet, E. Pilon
and M. Werlen, Phys. Rev. {\bf D63}, 114016 (2001).
\bibitem{g2mc} Z. Bern, L. Dixon and C. Schmidt, Phys. Rev. {\bf D66},
074018 (2002).
\bibitem{resbos} C. Balazs, E. L. Berger, P. Nadolsky and C.-P. Yuan,
Phys. Lett. {\bf D637}, 235 (2006); C. Balazs, E. L. Berger, P. Nadolsky
and C.-P. Yuan, Phys. Rev. {\bf D76}, 013009 (2007); C. Balazs, E. L. Berger,
P. Nadolsky and C.-P. Yuan, Phys. Rev. {\bf D76}, 013008 (2007).
\bibitem{frag} L. Bourhis, M. Fontannaz and J. P. Guillet,
Eur. Phys. J. {\bf C2}, 529 (1998).
\bibitem{cdf6312} D. Acosta {\it et al.} (CDF Collaboration),
Phys.\ Rev.\ Lett.\ {\bf 95}, 022003 (2005).
\bibitem{d0} V. M. Abazov {\it et al.} (D0 Collaboration), Phys. Lett. B
{\bf 690}, 108 (2010).
\bibitem{eta} The CDF II detector uses a cylindrical coordinate system
in which $\phi$ is the azimuthal angle, $\theta$ is the polar angle with
respect to the proton beam, $r$ is the radius from the nominal beam line,
and $z$ points in the proton beam direction, with the origin at the center
of the detector. The transverse $r$-$\phi$, or $x$-$y$ plane, is the plane
perpendicular to the $z$ axis. The pseudorapidity, $\eta$, is defined as
$-\ln(\tan(\theta/2))$. For the photons, which have zero mass, this is
identical to the rapidity $y$=$\tanh^{-1}(p_{\rm z}/E)$, where
$p_{\rm z}$=$p\cdot\cos\theta$ is the momentum parallel to the beam \cite{PDG}.
The transverse energy of a particle is $E_{\rm T}$=$E\cdot\sin\theta$. The
transverse momentum of a particle is defined as $p_{\rm T}$=$p\cdot\sin\theta$. 
\bibitem{CDF} D. Acosta {\it et al.} (CDF Collaboration), Phys.\ Rev.\ D
{\bf 71}, 032001 (2005). 
\bibitem{L00} C.S. Hill {\it et al.}, Nucl. Instrum. Methods Phys. Res.,
Sect. A {\bf 511}, 118 (2003).
\bibitem{SVX} A. Sill {\it et al.}, Nucl. Instrum. Methods Phys. Res.,
Sect. A {\bf 447}, 1 (2000).
\bibitem{ISL} A. Affolder {\it et al.} (CDF Collaboration), Nucl. Instrum.
Methods Phys. Res., Sect. A {\bf 453}, 84 (2000).
\bibitem{COT} A. Affolder {\it et al.} (CDF Collaboration), Nucl. Instrum.
Methods Phys. Res., Sect. A {\bf 526}, 249 (2004).
\bibitem{cem} L. Balka {\it et al.}, Nucl. Instrum. Methods Phys. Res.,
Sect. A {\bf 267}, 272 (1988); S.R.~Hahn {\it et al.}, {\it ibid.}
{\bf 267}, 351 (1988). 
\bibitem{cwha} S. Bertolucci {\it et al.}, Nucl. Instrum. Methods Phys. Res.,
Sect. A {\bf 267}, 301 (1988).
\bibitem{EMtiming} M. Goncharov {\it et al.}, Nucl. Instrum. Methods Phys.
Res., Sect. A {\bf 565}, 543 (2006).
\bibitem{CLC} D. Acosta {\it et al.}, Nucl. Instrum. Methods Phys. Res.,
Sect. A {\bf 494}, 57 (2002).
\bibitem{trigger} F. Abe {\it et al.} (CDF Collaboration), Nucl. Instrum.
Methods Phys. Res., Sect. A {\bf 271}, 387 (1988).
\bibitem{CESfid} For the central calorimeter, the fiducial region covers
$\sim$87$\%$ of the total area.
\bibitem{diphoAnom} T. Aaltonen {\it et al.} (CDF Collaboration),
Phys. Rev. D {\bf 82}, 052005 (2010).
\bibitem{cdfsim} E. Gerchtein and M. Paulini, CHEP-2003-TUMT005,
arXiv:physics/0306031 (2003); the version of {\sc geant} used for the
detector simulation is 3.21, see the CERN Program Library Long Writeup W5013.
\bibitem{jetsfrag} T. Aaltonen {\it et al.} (CDF Collaboration), Phys. Rev D
{\bf 77}, 092001 (2008); Phys. Rev. Lett. {\bf 102}, 232002 (2009).
\bibitem{cteq5l} H. L. Lai, J. Huston, S. Kuhlmann, J. Morfin, F. Olness,
J.F. Owens, J. Pumplin and W. K. Tung, Eur. Phys. J. C {\bf 12}, 375 (2000).
\bibitem{tuneA} R. Field and C. Group, arXiv:0510198 [hep-ph] (2005).
\bibitem{cdf9590} T. Aaltonen {\it et al.} (CDF Collaboration),
Phys.\ Rev.\ D {\bf 80}, 111106 (2009).
\bibitem{PDG} W. M. Yao {\it et al.}, ``Review of Particle Physics'',
J. Phys. G {\bf 33}, 2006.
\bibitem{dy} D. Acosta {\it et al.} (CDF Collaboration), Phys. Rev. Lett.
{\bf 94}, 091803 (2005).
\bibitem{cteq6m} J. Pumplin, D. R. Stump, J. Huston, H. L. Lai, P. Nadolsky
and W. K. Tung, J. HEP {\bf 0207}, 012 (2002).
\bibitem{CS} J. C. Collins and D. E. Soper, Phys. Rev D {\bf 16}, 2219 (1977).

\end{thebibliography}
\end{document}